\documentclass{svjour2}                    
\smartqed  
\usepackage{amsmath}
\usepackage{amssymb}
\usepackage{graphicx}
\usepackage{array}
\begin{document}

\newcommand{\D}{\displaystyle} 
\newcommand{\be}{\begin{eqnarray*}}
\newcommand{\ee}{\end{eqnarray*}}
\newcommand{\ben}{\begin{eqnarray}}
\newcommand{\een}{\end{eqnarray}}
\newcommand{\ud}{\underline}
\newcommand{\nd}{\noindent}
\newcommand{\nn}{\nonumber}
\newcommand{\tl}{\triangleleft}
\newcommand{\tr}{\triangleright}
\newcommand{\llangle}{\left\langle}
\newcommand{\rrangle}{\right\rangle}
\newcommand{\lel}{\left\langle}
\newcommand{\rir}{\right\rangle}
\newcommand{\bs}{\begin{subequations}}
\newcommand{\es}{\end{subequations}}

\title{Renormalized kinetic theory of classical fluids in and out of equilibrium}



\author{J\'er\^ome Daligault}


\institute{J. Daligault \at
              Theoretical Division, Los Alamos National Laboratory, Los Alamos, NM 87545\\
              Tel.: +1-505-6610132\\
              \email{daligaul@lanl.gov}
}

\date{Received: date / Accepted: date}
\date{\today}

\maketitle

\begin{abstract}
We present a theory for the construction of renormalized kinetic equations to describe the dynamics of classical systems of particles in or out of equilibrium.
A {\it closed, self-consistent} set of evolution equations is derived for the single-particle phase-space distribution function $f$, the correlation function $C=\langle \delta f\delta f \rangle$, the retarded and advanced density response functions $\chi^{R,A}=\delta f/\delta\phi$ to an external potential $\phi$, and the associated memory functions $\Sigma^{R,A,C}$.
The basis of the theory is an effective action functional $\Omega$ of external potentials $\phi$ that contains all information about the dynamical properties of the system.
In particular, its functional derivatives generate successively the single-particle phase-space density $f$ and all the correlation and density response functions, which are coupled through an infinite hierarchy of evolution equations.
Traditional renormalization techniques (involving Legendre transform and vertex functions) are then used to perform the closure of the hierarchy through memory functions.
The latter satisfy functional equations that can be used to devise systematic approximations.
The present formulation can be equally regarded as (i) a generalization to dynamical problems of the density functional theory of fluids in equilibrium and (ii) as the classical mechanical counterpart of the theory of non-equilibrium Green's functions in quantum field theory.
It unifies and encompasses previous results for classical Hamiltonian systems with any initial conditions.
For equilibrium states, the theory reduces to the equilibrium memory function approach used in the kinetic theory of fluids in thermal equilibrium.
For non-equilibrium fluids, popular closures of the BBGKY hierarchy (e.g. Landau, Boltzmann, Lenard-Balescu-Guernsey) are simply recovered and we discuss the correspondence with the seminal approaches of Martin-Siggia-Rose and of Rose.and we discuss the correspondence with the seminal approaches of Martin-Siggia-Rose and of Rose.
\keywords{Kinetic theory \and Closure \and Renormalization \and Effective action \and Schwinger closed-time contour}
\PACS{05.20.Dd,05.20.-y}
\end{abstract}


\maketitle

\clearpage

\section{Introduction}

The present paper elaborates the foundations of a unifying approach for the construction of renormalized kinetic equations of classical systems of particles in and out of equilibrium.
The statistical dynamics is formulated in terms of self-consistently determined single-particle phase-space distribution function, density response, correlation and memory functions.
The memory functions satisfy a functional equation that can be used to devise systematic renormalized perturbative approximations.
Important topical applications include the systematic generalization of the Boltzmann equation to dense gases and liquids, the kinetic theory of strongly coupled plasmas and Hamiltonian systems in general, the proper macroscopic description of fluids beyond the Navier-Stokes equations, and the physics of plasma turbulence.

Before embarking on the detailed description of the theory, we recall a few of the most striking developments and unresolved issues in kinetic theory that motivated this work.

\subsection{Background.}  Kinetic theory attempts to carry out a complete statistical description of the macroscopic dynamics of fluids in terms of the underlying microscopic interactions of its constituents.
The foundation of modern kinetic theory dates back to 1872 when Boltzmann published his famous equation for the single-particle distribution function for dilute gases \cite{Boltzmann1872,Brush2003,Cercignani2006}.
The solution of the Boltzmann equation, derived independently by Chapman and Enskog around 1915, achieved two basic goals of kinetic theory.
That is, it (i) established the connection between the microscopic dynamics and the macroscopic equations of hydrodynamics and (ii) provided explicit expressions for the transport coefficients in terms of the molecular parameters.
Unfortunately, since it is based on the assumption of uncorrelated binary collisions, the Boltzmann equation gives a satisfactory description of the behavior of sufficiently low density gases with short-range interactions only, and is inadequate to describe dense gases, liquids and plasmas.

The generalization of the Boltzmann equation to more dense fluids and to plasmas confronts the full complexity of a highly correlated many-body problem.
This became very clear when, in the early 1960's  \cite{Surveys}, it was realized that all the remarkable systematic developments made until then, which were based on some form of perturbation theory of the Liouville equation or the BBGKY hierarchy, were plagued by unphysical divergences \cite{Montgomery1967,Balescu1997,history1}.
Those findings, together with others permitted by the advent of computer simulations, shifted attention towards the development of a renormalization (regularization) procedure, which would encompass the ``theoretical'' divergences and take proper account of the correlated collisions, whereby particles interact via the effective (renormalized) potential that integrates the average effects of the medium on the bare interactions \cite{history2,history3}.
Additionaly, such a theory should yet retain the desirable properties of the Boltzmann equation and be tractable for practical calculations.

Such a renormalized kinetic theory was first presented by Mazenko in 1974 using the memory function approach \cite{Mazenko1974}.
His so-called ``fully renormalized'' theory applies only to fluids in or near thermal equilibrium by means of the fluctuation-dissipation theorem.
At equilibrium, the distribution function $f$ is known (the Maxwell-Boltzmann distribution) and the statistical dynamics is encoded in the equilibrium two-point time-correlation function $C_{eq}(1,1')$ of the microscopic phase-space density.
It was shown in the 1960's \cite{equilibriumtheory} that an exact evolution equation for $C_{eq}$ can be written in the form of a generalized Langevin equation in terms of a memory function kernel $\Sigma_{eq}$.
Mazenko devised an iterative scheme for its calculation that can be used to develop systematic renormalized approximations of $\Sigma_{eq}$.

The development of a renormalization procedure that can describe non-equilibrium states faces additional challenges \cite{Krommes2002}.
A first major development in that direction was done in 1973 by Martin, Siggia and Rose (MSR) \cite{MartinSiggiaRose}.
MSR elaborated a procedure to derive self-consistent approximations for calculating the statistical dynamical properties of a classical random variable or field whose time dependence is governed by a nonlinear differential evolution equation.
In other words, they devised a systematic approach to the ubiquitous statistical closure problem, which occurs as a consequence of the statistical averages and the non-linearity of the fundamental evolution equation.
Noting that satisfactory approaches to the closure problem had been successfully developed in quantum field theory, MSR recast the classical problem into a quantum-mechanical-looking problem and borrowed techniques of quantum field theory to derive a renormalized perturbation theory.
To accomplish the recast, MSR introduced an additional field operator that does not commute with the fundamental field and allows the calculation of both correlation functions and response functions simultaneously, where the latter describe the response of the fluid to the injection of particles.

Although the basic problem in kinetic theory also amounts to a closure problem, it was then recognized that its specificities did not fit into the framework of MSR theory; in particular, its natural field variable, the microscopic single-particle phase-space density, is discrete, singular and its statistics is strongly non-Gaussian.
In 1979, Rose \cite{Rose1979} developed another approach specifically for dealing with kinetic theory.
Like MSR, Rose also recast the basic problem in a quantum-mechanical form but this time using an occupation number representation in classical phase-space \cite{Doi}.
The response to the injection of particles included in MSR is naturally incorporated here.
The implications of Rose's paper to kinetic theory have not yet been fully explored.
In addition, Rose points out that the steps necessary to reduce his general formalism to Mazenko's theory of fluids in thermal equilibrium are not apparent.

Finally, very recently, Mazenko \cite{Mazenko2010} published a new fundamental ``theory of statistical particle dynamics that solves the chronic problem of self-consistency''.
In that paper, the theory is presented for the Smoluchowski dynamics and the author announces that its extension to Newtonian dynamics is under way.
From the information available in \cite{Mazenko2010}, it is likely that the approach presented here is still different from Mazenko's new theory.

\subsection{Present work.} 

In this paper, we present an alternative, arguably simpler and elegant approach to the derivation of renormalized kinetic equations for classical fluids that applies to {\em both} equilibrium and non-equilibrium states and encompasses the previous works on this topic.
For simplicity of the exposition, we shall consider a system of identical point particles mutually interacting via a pair-wise additive central potential; however, no peculiar assumption is made on its range and strength, so the theory can be applied to investigate systems ranging from neutral liquids with short-range interactions to classical plasmas with the Coulomb interaction.
Moreover, the theory is derived independently of the nature of initial state of the system, and the influence of initial correlations on the system dynamics, usually neglected or difficult to include in the previous works, is treated explicitly.

A {\it closed, self-consistent} set of evolution equations for the phase-space density $f({\bf r},{\bf p},t)$, the two-point correlation function $C$, and the retarded and advanced response functions $\chi^{R,A}$ is derived in terms of time history integrals that involve three memory functions $\Sigma^{R,A,C}$.
Loosely speaking, the correlation and response functions describe the dynamics of emission and absorption of phase-space density fluctuations in the fluid, which in turn determine the effect of collisions on the dynamics of the distribution function $f$.
The memory response functions describe how many-body effects affect, i.e. renormalize, the time-dependent propagation and lifetime of density flucutations, and as such play a role similar to self-energies in quantum field-theory.

To obtain this closure, we construct an action functional $\tilde\Omega[\phi]$ of external potentials $\phi$ that contains all information about the dynamical properties of the fluid.
In particular, its derivatives with respect to $\phi$ successively generate the phase-space density $f$ and all the correlation and response functions.
The latter are coupled through an infinite hierarchy of evolution equations, which subsumes the traditional BBGKY equations.
The hierarchy is formally closed using traditional closure (renormalization) techniques involving the Legendre transform $\tilde\Gamma[f]=-\tilde\Omega[\phi]+\int{\phi f}$ of the effective action $\Omega[\phi]$.

The basis of the present approach is the extension to classical systems of the closed-time contour idea originally introduced in 1961 by Schwinger \cite{Schwinger1961} in his seminal paper on the Brownian motion of a quantum particle.
Schwinger's idea was later fruitfully used to develop the theory of non-equilibrium Green's functions in quantum field theory \cite{RammerSmith1986,CalzettaHu2008}.
Two fully equivalent formulations of the theory are presented.
In the first formulation, the theory is expressed in terms of time-dependent quantities taking their values on the closed-time contour.
The effective action $\tilde\Omega[\phi]$ is a functional of external potentials $\phi$ that linearly couples  along the closed-time contour to the single-particle phase-space density.
The first functional derivative with respect to $\phi$ generates the phase-space distribution function $f=\delta\tilde\Omega/\delta\phi$ and the second derivative $\delta^{2}\tilde\Omega/\delta\phi^2$ equals a quantity that combines both the two-point correlation function $C$ and the density response functions $\chi^{R,A}$.
In the second formulation, the theory is recast into a form that {\it directly} generates the physical quantities of interest.
The effective action is written as a functional of two external potentials, $\Omega[\phi_p,\phi_\Delta]$, of the physical time variable $t$ and has the remarkable derivatives $f=\frac{\delta\Omega}{\delta \phi_\Delta}$, $C=\frac{\delta^{2}\Omega}{\delta \phi_\Delta\delta \phi_\Delta}$, $\chi^{R}=\frac{\delta^{2}\Omega}{\delta \phi_\Delta\delta \phi_p}$ and $\chi^{A}=\frac{\delta^{2}\Omega}{\delta \phi_p\delta \phi_\Delta}$.

The main purpose of this paper is to lay down the general foundations of the theory without reference to a specific physical system or model.
Its application to physical problems would require substantial additional work, some of it being under way.
The paper is organized as follows.
In Sec. \ref{sectionII}, the basic definitions and notations are introduced.
To provide a self-contained presentation of the work, we recall several basic results of classical statistical mechanics that are used thereafter.
The original contributions of this work really begin into Sec. \ref{sectionIII} where we develop the closed-time formulation of our theory and continue in Sec. \ref{sectionIV} and \ref{sectionV} where we recast the results into equations that involve quantities depending on the physical time.
For clarity, the details of the proofs of the calculations are given in the appendices.
Throughout these sections, we discuss the implications and merits of the new approach presented.
We study the special case of a fluid at equilibrium, make contact with Mazenko's renormalized theory, show the correspondence with popular kinetic equations (Landau, Boltzmann, Lenard-Balescu-Guernsey), and study the correspondence with MSR and Rose's theories.
Also the present approach naturally shed light on analogies and differences between quantum and classical kinetic theory, which we highlight throughout the paper and in appendices.
The comparison allows us (i) to comprehend the longer resistance of classical statistical mechanics to renormalization, and (ii) to justify some of the ingenious approaches used by previous works like MSR.
For convenience, a summary of the main components of the theory is given in Sec.\ref{sectionVIA}.

\section{Basic definitions} \label{sectionII}

In this section, we introduce the physical systems under consideration.
For completeness, we also recall several basic notions of statistical dynamics used thereafter such as the correlation functions, the linear response functions to external perturbations, and the closure problem.
Most of the material is discussed in textbooks, e.g. \cite{Balescu1997,MazenkoBook}, and is recalled here for completeness.
The reader already familiar with these notions may wish to skip ahead to Sec.\ref{sectionIII} where the original contribution of this paper begins.

\subsection{Physical system}

We consider a system consisting of $N$ identical point-particles of mass $m$ enclosed in a volume $V$ of the $d$-dimensional space $\mathbb{R}^{d}$.
We assume that the system dynamics is governed by the laws of classical mechanics under the total, possibly time-dependent Hamiltonian $H_{tot}$.
Thus, the position ${\bf r}_{j}(t)$ and momentum ${\bf p}_{j}(t)$ of the $j$-th particle at a time $t$ evolve according to the Hamilton equations,
\ben
\frac{d{\bf r}_{j}}{dt}=\frac{\partial H_{tot}(x,t)}{\partial {\bf p}_{j}}\quad,\quad\frac{d{\bf p}_{j}}{dt}=-\frac{\partial H_{tot}(x,t)}{\partial {\bf r}_{j}}\,. \label{realHamiltonequations}
\een
where $x=({\bf r}_{1},\dots,{\bf r}_{N};{\bf p}_{1},\dots,{\bf p}_{N})\in V^{N}\times\mathbb{R}^{dN}$.
These equations (\ref{realHamiltonequations}) are to be solved subject of $2dN$ initial conditions on the coordinates and momenta $x(t_{0})=x_{0}$ at an initial time $t_0$.

We assume that the initial state $x_0$ of the system is imperfectly known but can be characterized by a distribution $F_0(x_0)$ defined such that $F_{0}(x_{0})dx_{0}$ is the probability that the system is initially in a microscopic state represented by a phase-space point of volume $dx_{0}$ around $x_{0}$; in the following, $F_0$ is assumed to be normalized, i.e. $\int{dx_{0}F_{0}(x_{0})}=1$.
Most importantly, the present theory is derived independently of the nature of initial state $F_0$ of the system.

Thus we are interested in the dynamics at time $t\geq t_0$ of a statistical ensemble of independent systems, each of which is a replica of the system defined above, and initially distributed according to $F_0$; the dynamics prior to the initial time $t_{0}$ is not of interest to us.
The uncertainty on the initial conditions will be the only element of statistics in the present theory.
Once the non-equilibrium initial state is specified, the time-evolution is deterministic and completely determined by the Hamiltonian.

As time increases, the initial distribution of states $F_0$ evolves into the distribution $F(x,t)$ at $t\geq t_0$ according to Liouville equation,
\ben
\frac{\partial}{\partial t}F(x,t)&=&\left[H_{tot},F\right]_{PB}(x,t)\,, 
\label{Liouvilleequation}
\een
where $[A,B]_{PB}$ denotes the $N$-particle Poisson bracket
\be
[A,B]_{PB}=\sum_{j=1}^{N}{\left(\frac{\partial A}{\partial {\bf r}_{j}}\cdot\frac{\partial B}{\partial {\bf p}_{j}}-\frac{\partial A}{\partial {\bf p}_{j}}\cdot\frac{\partial B}{\partial {\bf r}_{j}}\right)}\/.
\ee

We assume that the total Hamiltonian can be written as
\ben
H_{tot}(x,t)=H_{N}(x)+H_{ext}(x,t)\/. \label{totalHamiltonian}
\een
The time-independent term $H_N$ characterizes the system in the absence of external, time-dependent perturbations, and is of the form,
\ben
H_{N}(x)&=&\sum_{j=1}^{N}{\left(\frac{{\bf p}_{i}^{2}}{2m}+v_{0}({\bf r}_{i})\right)}+\frac{1}{2}\sum_{i\neq j=1}^{N}{v(|{\bf r}_{i}-{\bf r}_{j}|)}\nn\\
&=&\sum_{j=1}^{N}{h_{0}({\bf r}_{j},{\bf p}_{j})}+\frac{1}{2}\sum_{i\neq j=1}^{N}{v(|{\bf r}_{i}-{\bf r}_{j}|)}\/. \label{HN}
\een
The particles interact according to a pair-wise additive central potential $v(r)$ and the whole system is possibly confined by a static potential $v_{0}({\bf r})$.
Throughout the paper, no peculiar assumption is made on the range and strength of the interaction potential $v$ other than being differentiable for $r>0$.
In particular, the theory applies to both short-range potentials as encountered in neutral liquids \cite{HansenMcDonald3rd} and also to the long-range, Coulomb potential of importance to plasma physics; in the latter case, a background potential may be included in $v_{0}$ to ensure electrical neutrality \cite{BausHansen}.

The term $H_{ext}$ in Eq.(\ref{totalHamiltonian}) represents the possible interaction to an external, time-dependent perturbation $\phi_{0}({\bf r},{\bf p},t)$,
\ben
H_{ext}(x,t)&=&\sum_{j=1}^{N}{\phi_{0}({\bf r}_{j},{\bf p}_{j},t)} 
=\iint{d{\bf r}d{\bf p}\,N({\bf r},{\bf p},t)\/\phi_{0}({\bf r},{\bf p},t)}\,, \label{Hextuext}
\een
which linearly couples to the phase-space density $N({\bf r},{\bf p},t)$ (see Eq.(\ref{microscopicphasespacedensity}) below.)

The general definitions given above encompass a large class of systems, including for instance (i) the dynamics of a fluid at equilibrium when $H_{ext}\equiv 0$ and $F_{0}=f_{eq}$ is an equilibrium (Gibbs) distribution function, (ii) the relaxation dynamics to equilibrium when $F_{0}\neq f_{eq}$ and $H_{ext}\equiv 0$, (iii) the out-of-equilibrium dynamics of a fluid in the presence of time-dependent fields when $H_{ext}\neq 0$.

\subsection{Closure problem, correlation and response functions} \label{sectionIIB}

\subsubsection{Fundamental field variable and correlation functions} 

A dynamical variable is a quantity $A$ that depends parametrically on the particles' trajectory $x(t)$, and therefore on the initial condition $x_0$.
As a consequence of the statistical description, $A$ no longer has a definite value but is instead characterized by its average over all possible initial conditions weighted by $F_0$,
\ben
\langle A\rangle(y)=\int{dx_{0}F_0(x_{0})A(x_0;y)}\,; \label{classicalaverage}
\een
here $y$ denotes the possible dependence on other parameters.

A fundamental dynamical variable in classical kinetic theory is the microscopic single-particle phase-space density
\ben
N({\bf r},{\bf p},t)&=&\sum_{j=1}^{N}{\delta\left({\bf r}-{\bf r}_{j}(t)\right)\delta\left({\bf p}-{\bf p}_{j}(t)\right)}\,, \label{microscopicphasespacedensity}
\een
where $({\bf r},{\bf p})\in V\times\mathbb{R}^{d}$.
The density $N({\bf r},{\bf p},t)$ plays a role similar to a field in field theory or of a random variable in the theory of stochastic processes; here $N({\bf r},{\bf p},t)$ is random in that its value varies due to its sensitivity to initial conditions.
The ensemble average of $N({\bf r},{\bf p},t)$ defines the single-particle distribution function
\ben
f({\bf r},{\bf p},t)=\langle N({\bf r},{\bf p},t)\rangle\/, \label{distributionfunctionf}
\een
and has the meaning of the probability distribution of particles at position ${\bf r}$ with momentum ${\bf p}$ at time $t$.
For an unconfined fluid ($v_0=0$) in thermal equilibrium at temperature $T$, $f$ is independent of space and time, and reduces to the Maxwell-Boltzmann distribution $f({\bf p})=ne^{-{\bf p}^{2}/2mk_{B}T}/(2\pi mk_{B}T)^{3/2}$ where $n=N/V$ is the particle density.
In non-equilibrium fluids, $f$ generally depends on both ${\bf r}$ and $t$.

As is typical in a statistical theory of a random variable $N(1)$, an important role in the theory is played by the fluctuations around its averaged value,
\ben
\delta N(1)=N(1)-\langle N(1)\rangle=N(1)-f(1)\,, \label{Nfluctuations}
\een
and by the associated  $n$-point correlation functions (or cumulants)
\ben
C^{(n)}(1,\dots,n)=\langle \delta N(1)\dots \delta N(n)\rangle\,. \label{correlationfunctions}
\een
For the two-point correlation function, we write $C(1,2)\equiv C^{(2)}(1,2)$.
Equal-time correlation functions, i.e. $t_1=\dots=t_n$, will be denoted by a bar as $\bar C^{(n)}(x_1,\dots,x_n;t)$.

As was emphasized by Rose \cite{Rose1979}, a statistical theory based on a discrete (i.e., a sum of delta functions) random variable such as $N(1)$ presents a serious technical difficulty in that it is strongly non-Gaussian, even for non-interacting particles, since the equal-time cumulants $\bar C^{(n)}$ are all non-vanishing and singular.
Indeed, in terms of the conventional $n$-particle correlation $g_n$ introduced in kinetic theory from the cluster expansion of the reduced distribution functions (see appendix \ref{appendixBBGKY}), we have
\ben
\bar C^{(2)}(X,X',t)&=&\delta(X-X')f(X,t)+g_2(X,X';t) \label{C2g2}
\een
and
\ben
\bar C^{(3)}(X,X',X'';t)&=&\delta(X-X')\delta(X'-X'')f(X,t)\nn\\
&+&\,\big[\delta(X-X')g_2(X',X'';t)+\text{cyclic permutations}\big]\nn\\
&+&\,g_3(X,X',X'';t) \label{C3g3}
\een
and so on, with $X=({\bf r},{\bf p})$.
Thus, even in the absence of three-particle correlations ($g_3=0$), the three-point cumulant is non-vanishing because of particle self-correlations, and this remains true at all times $t\geq t_0$.
This result precludes the applicability the Wick theorem so useful in conventional statistical field theories.

\subsubsection{Field equation and closure problem}

For convenience we introduce the shorthand notation in which a field point and the time variable are designated by a single number, i.e. $n\equiv ({\bf r}_{n},{\bf p}_{n},t_{n})$ (not to be confused with the particle labels $n$); for $n=1$, we often drop the subscript so that $1=({\bf r},{\bf p},t)$.
The delta function $\delta(1-2)$ denotes $\delta^{(3)}({\bf r}_1-{\bf r}_2)\delta^{(3)}({\bf p}_1-{\bf p}_2)\delta(t_1-t_2)$.

From the Hamilton equations (\ref{realHamiltonequations}), it is straigthforward to show that the phase-space density evolves according to
\be
\frac{\partial}{\partial t_1}N(1)=L_{1}N(1)+\int{d2 L_{12}N(1)N(2)}\/. \label{accurateKlimontovich}
\ee
where the summation is defined as
\be
\int{d1\,\dots}=\int_{V}{d{\bf r}_{1}\int_{\mathbb{R}^d}{d{\bf p}_{1}\int_{t_{0}}^{\infty}{dt_{1}\dots}}}
\ee
Here,
\ben
L_{1}\bullet &=&\frac{\partial h(1)}{\partial {\bf r}_{1}}\cdot\frac{\partial \bullet}{\partial {\bf p}_{1}}-\frac{\partial h(1)}{\partial {\bf p}_{1}}\cdot\frac{\partial \bullet}{\partial {\bf r}_{1}}=\big\{h,\bullet \big\}(1)\label{L1}
\een
is the single particle part of the Liouville operator with the total single-particle Hamiltonian $h=h_{0}+\phi_{0}$.
In the second line of Eq.(\ref{L1}), it is expressed in terms of the single-particle Poisson bracket
\ben
\big\{a,b\big\}({\bf r},{\bf p})=\frac{\partial a}{\partial {\bf r}}\cdot\frac{\partial b}{\partial {\bf p}}-\frac{\partial a}{\partial {\bf p}}\cdot\frac{\partial b}{\partial {\bf r}}\,,
\een
for any function $a({\bf r},{\bf p})$ and $b({\bf r},{\bf p})$ in $R^{2d}$ (note the curly brackets notation to avoid confusion with the $N$-particle Poisson bracket $[\cdot,\cdot]_{PB}$ defined earlier.)
Finally,
\ben
L_{12}\,\bullet=\frac{\partial}{\partial {\bf r}_{1}}\tilde{v}_{12}\cdot\left(\frac{\partial\,\bullet}{\partial {\bf p}_{1}}-\frac{\partial\,\bullet}{\partial {\bf p}_{2}}\right) \label{Ljk}
\een
is the interaction part of the Liouville operator with $\tilde{v}_{12}=v({\bf r}_{1}-{\bf r}_{2})\delta(t_{1}-t_{2})$.
Those notations are in widespread use in the literature on classical kinetic theory.
In the theory presented below, however, it will be more convenient to rewrite Eq.(\ref{accurateKlimontovich}) as
\ben
\left(\frac{\partial}{\partial t_1}-L_{1}\right)N(1)&=&\frac{1}{2}\int{d2\int{d3 \gamma_{3}(1,2,3)N(2)N(3)}}\nn\\
&=&\frac{1}{2}\gamma_{3}(1,2,3)N(2)N(3)\label{accurateKlimontovich2}
\een
in terms of the ``bare interaction vertex'',
\ben
\gamma_{3}(1,2,3)&=&\frac{\partial}{\partial {\bf p}_{1}}\cdot\big({\bf w}_{12}\/\delta(1-3)+{\bf w}_{13}\/\delta(1-2)\big)\label{hatgamma3}\\
{\bf w}_{12}&=&\frac{\partial}{\partial {\bf r}_1}\tilde{v}_{12}\nn
\een
which is symmetric in its last two arguments.
The second line of Eq.(\ref{accurateKlimontovich2}) illustrates the summation convention over repeated indices of dummy variables that we shall use throughout the paper.

Averaging Eq.(\ref{accurateKlimontovich2}) over the initial conditions, we obtain the equation of evolution of the phase-space distribution $f(1)$,
\ben
\left[\frac{\partial}{\partial t_1}-L_{1}\right]f(1)-\{u^{mf}(1),f(1)\}&=&\frac{1}{2}\gamma_{3}(1,2,3)C(2,3)\,, \label{equationf}
\een
where $u^{mf}$ is the mean-field (a.k.a. Vlasov) potential
\ben
u^{mf}({\bf r},t)&=&\int{d2 f(2) \tilde{v}(1-2)}\,. 
\label{umf}
\een
Equation (\ref{equationf}) couples $f(1)$ to the (equal-time) two-point correlation function $C(1,2)$, and corresponds to the first equation of the ordinary BBGKY hierarchy; the right-hand side (rhs) is usually referred to as the collision integral.
Similarly, as a consequence of the quadratic non-linearity of the field equation (\ref{accurateKlimontovich2}), the evolution of $C^{(n)}$ at each order $n$ involves the next higher-order correlation $C^{(n+1)}$; those equations can be straightforwardly derived using the equation for the density fluctuations $\delta N$ obtained by subtracting Eqs.(\ref{accurateKlimontovich2}) and (\ref{equationf}).
Thus, for the two-point correlation function, we find
\be
\left[\frac{\partial}{\partial t_1}-L_{1}\right]C(1,1')-\Sigma^{mf}(1,2)C(2,1')=\frac{1}{2}\gamma_{3}(1,2,3)C^{(3)}(2,3,1')
\ee
where
\ben
\Sigma^{mf}(1,1')&=&\gamma_3(1,2,3)f(2)\delta(1'-3)\label{sigmamf}
\een

The hierarchy of evolution equations for the $C^{(n)}$ is different from the ordinary BBGKY hierarchy, which instead involves the correlation functions $g_{n}$ (a quick summary of BBGKY is given in appendix \ref{appendixBBGKY}.)
The BBGKY can be recovered by considering the equal-time limit of the equations for the $C^{(n)}$'s.
The ``simplest'' closures of the BBGKY hierarchy lead to the most famous kinetic equations, in particular the Vlasov equation, the Landau equation \cite{Landau1936} for so-called weakly-interacting gases, the Boltzmann equation \cite{Boltzmann1872} for dilutes gases, and the Lenard-Balescu-Guernsey equation \cite{LenardBalescuGuernsey} for weakly-coupled plasmas.
Those closures rely, among additional hypothesis, on a systematic ordering of the correlations $g_n$ in terms of an adequately chosen small parameter \cite{Balescu1997}.
As summarized in table \ref{tableclosure}, each closure amounts to keeping only certain contributions in Eqs.(\ref{C2g2})-(\ref{C3g3}).
In particular, these popular closures systematically neglect the $n$-particle correlation functions $g_n$ with $n\geq 3$, which is inappropriate whenever correlations are strong.
A major challenge of kinetic theory is to incorporate the effects of those neglected terms in a self-consistent manner.
In Sec.\ref{sectionVII}, we shall see how the present theory addresses the problem.

\begin{center}
\begin{table}[t]
\setlength{\extrarowheight}{.1cm}
\begin{tabular}{m{2.6cm}||c|c}
{\bf Closure} & $\bar{C}^{(2)}$ & $\bar{C}^{(3)}$ \\\hline
Vlasov  & $0$ & $0$\\\hline
Landau  & $\delta(x-x')f(x,t)$ & $0$\\\hline
Boltzmann & $\delta(x-x')f(x,t)$ & $\left[\delta(x-x')g_2(x',x'';t)+c.p.\right]$\\
& & $\hspace{.2cm}+\delta(x-x')\delta(x'-x'')f(x,t)$\\\hline
LBG & $g_2(x,x';t)+\delta(x-x')f(x,t)$ & $0$\\\hline
Book-Frieman \cite{BookFrieman} & $g_2(x,x';t)+\delta(x-x')f(x,t)$ & $\left[\delta(x-x')g_2(x',x'';t)+c.p.\right]$\\
& & $\hspace{.2cm}+\delta(x-y)\delta(y-z)f(x,t)$
\end{tabular}
\caption{The most popular closures of classical statistical physics amount, among additional hypothesis, to replace Eqs.(\ref{C2g2})-(\ref{C3g3}) for $\bar{C}^{\/(2),(3)}$ by the terms displayed in the table.
Those approximations systematically neglect the $n$-particle correlation functions $g_n$ with $n\geq 3$, which is inappropriate whenever correlations are strong.
One of the challenge of kinetic theory is to incorporate the effects of those neglected terms in a self-consistent manner.
\newline
For comparison, the term $g_2$ kept in the Lenard-Balescu-Guernsey (LBG) equation but discarded in the Boltzman equation allows the renormalization of the bare interaction into a dynamically screened two-particle interaction between the particles of the plasma.
On the contrary, the contribution to $\bar{C}^{(3)}$ kept in the Boltzmann approximation but discarded in the Lenard-Balescu-Guernsey equation, is responsible for the bare two-particle interactions (large-angle scattering) describing the two-body collision in a dilute gas.
}
\label{tableclosure}
\end{table}
\end{center}

\subsubsection{Linear response functions and fluctuation-dissipation theorem}

We now consider the linear response of the system defined above to weak, external perturbations $\phi$ that couples linearly to the single-particle density $N(X,t)$ as
\be
H_{pert}({\bf r},{\bf p},t)=\int{d{\bf r}d{\bf p}\,N({\bf r},{\bf p},t)\delta\phi({\bf r},{\bf p},t)}\,.
\ee
By linearizing with respect to $\delta\phi$ the Liouville equation (\ref{Liouvilleequation}) with $H_{tot}$ replaced by $H_{tot}+H_{pert}$, we find that to first order in the external perturbation the phase-space distribution $f({\bf r},{\bf p},t)$ is modified from its value in the absence of perturbation by the amount,
\be
\delta f(1)=\int{d1'\chi^R(1;1')\delta\phi(1')}\,,
\ee
in term of the retarded response function,
\ben
\chi^{R}(1,1')&=&\frac{\delta f(1)}{\delta \phi(1')}=\theta(t-t')\theta(t'-t_0)\Big\langle\big[N(1),N(1')\big]_{PB}\Big\rangle\/,\label{retartedBAresponse}
\een
for all $t,t'\geq t_{0}$.
The retarded response function is not symmetrical in its arguments but instead $\chi^R(1',1)=\chi^{A}(1,1')$, where $\chi^A$ is the {\it advanced} response functions,
\ben
\chi^{A}(1;1')&=&-\theta(t'-t)\theta(t-t_0)\Big\langle\big[N(1),N(1')\big]_{PB}\Big\rangle\,. \label{chiAdefinition}
\een
In the Poisson bracket, $N(1)$ is evaluated along the unperturbed trajectory governed by $H_{tot}$ (interaction representation).
Other useful properties in the derivation of the present theory are given in appendix \ref{appendixpropertiesPoissonbracket}.

The two-point correlation $C$ and response functions $\chi^{R,A}$ play a fundamental role in the present theory.
The correlation function gives information about the likelihood of fluctuations in the fluid and determines the effect of particle collisions in the collision integral of Eq.(\ref{equationf})
The response functions carry the dynamical information on how fluctuations are propagated in time.
In classical mechanics, those functions are related to averages of dynamical variables of different nature, namely the averaged {\it product} of field variables for $C$ and their averaged {\it Poisson bracket} for $\chi^{R,A}$.
This result is to be contrasted with the equivalent result in quantum mechanics where both correlation and response functions are related to the average of products of the same dynamical variables, on which more is said later.

Only for the special case of a system in thermal equilibrium ($F_{0}=f_{eq}$ and $H_{ext}\equiv 0$), the response and correlation function are simply related according to
\be
\chi^{R}(1,1')&=&\beta\theta(t-t')\theta(t'-t_0)\frac{d}{dt'}C(1,1')\,,
\ee
with $\beta=1/k_B T$.
This result is known as the fluctuation-dissipation theorem \cite{NoteFluctuationDissipation}.

\section{Renormalization on the closed-time contour} \label{sectionIII}

The present theory is based on the definition of a generating functional $\tilde{\Omega}[\phi]$ that contains all the information about the dynamics under investigation.
Here $\phi$ represents some external time-dependent potential, which perturbs the system and, just like $\phi_{0}$ in Eq.(\ref{Hextuext}), couples linearly to the microscopic phase-space density in the total Hamiltonian.
The functional is designed such that its derivatives with respect to $\phi$ generate the phase-space distribution $f$, the correlation and response functions $C$ and $\chi^{R/A}$ of the system.

The functional $\tilde{\Omega}[\phi]$ plays a role similar to the grand-potential $\Omega[v_0]$ in the density functional theory of classical fluids used to study their static properties \cite{HansenMcDonald3rd}; in this case, the first derivative with respect to the confining potential $v_0$ gives the particle density $\delta\Omega[v_0]/\delta v_0({\bf r})=\rho({\bf r})$ and the second derivative equals the density correlation function $\delta\Omega[v_0]/\delta v_0({\bf r})\delta v_0({\bf r'})=\langle\delta\rho({\bf r})\delta\rho({\bf r}')\rangle_{eq}$ \cite{HansenMcDonald3rd}.
The existence of such a generating functional for the calculation of dynamical properties is far from obvious.
For instance, if one imagines a straightforward extension of the static theory in which $\tilde{\Omega}[\phi_0(1)]$ is a functional of time-dependent external potentials $\phi_0({\bf r},{\bf p},t)$ that generates the phase-space density $\delta \tilde{\Omega}/\delta \phi_0(1)=f(1)$, one is immediately led to a contradiction.
Indeed, the symmetry of the second-order derivative requires
\be
\frac{\delta}{\delta\phi_0(2)}\frac{\delta\tilde{\Omega}}{\delta\phi_0(1)}=\frac{\delta}{\delta\phi_0(1)}\frac{\delta\tilde{\Omega}}{\delta\phi_0(2)}\,,
\ee
while causality requires
\be
\frac{\delta}{\delta\phi_0(2)}\frac{\delta\tilde{\Omega}}{\delta\phi_0(1)}=\frac{\delta f(1)}{\delta\phi_0(2)}=\chi^{R}(1,2)\neq \chi^{R}(2,1) 	
\ee
since, say when $t_1>t_2$, $\chi^{R}(1,2)=\left\langle\left[N(1),N(2)\right]_{PB}\right\rangle$ and $\chi^{R}(2,1)=0$ (see Eq.(\ref{retartedBAresponse})).
Both requirements could never be satisfied together and therefore the search for such a functional is doomed to fail.

In order to cope with the problem, we shall define $\tilde{\Omega}$ over a wider set of dynamics using the idea of closed-time contour originally introduced by Schwinger in the early 1960's to deal with quantum systems \cite{Schwinger1961}.
Because quantum mechanics propagates probability amplitudes instead of probability densities, the mere notion of closed-time contour arises more manifestly in quantum than in classical mechanics.
Thus, if at time $t\geq t_{0}$ a quantum system is in the state $|\Psi(t)\rangle$, the expectation value of an observable $\hat{O}$ at that time can be expressed as
\ben
\langle O\rangle(t)&=&\langle\Psi(t)|\hat{O}|\Psi(t)\rangle\label{QlaOra0}\\
&=&\langle\Psi(t_{0})|U(t_{0},t)\hat{O}U(t,t_{0})|\Psi(t_{0})\rangle\label{QlaOra}
\een
in terms of the propagator $\hat{U}(t,t')$ from time $t'$ to $t$, and the initial state $|\Psi(t_{0})\rangle$ at time $t_{0}$.
If we read the time arguments of the propagators in Eq.(\ref{QlaOra}) from right to left we may say that the system evolves from $t_0$ to time $t$ after which the operator $\hat{O}$ acts; then the system evolves back along the real axis from time $t$ to $t_{0}$.
Schwinger first noticed that one could imagine that the forward and backward time evolutions are governed by different dynamics (e.g., if a different external potential acts on each branch), and showed how this can be turned into an ingenious device to probe (generate) the properties of the physical system under investigation simply by suitably perturbing its dynamics along the closed-time contour that goes from $t_{0}$ to $t$ back to $t_{0}$.
The closed-time contour approach was further developed by Keldysh \cite{Keldysh1964} and many others to obtain an elegant and powerful treatment of non-equilibrium quantum systems in terms of non-equilibrium Green's functions \cite{RammerSmith1986,CalzettaHu2008,KrempSchlangesKraeft}.
In contrast, in classical mechanics, the expectation value (\ref{QlaOra0}) becomes
\be
\langle O\rangle(t)=\int{dx_{0} F(x_{0})O(x(t))}\,,
\ee
and the concept of closed-time contour is much less apparent.
Nevertheless, we shall show how a similar idea can be fruitfully developed to the classical case.

First we shall define the notions of closed-time contour and extended dynamics.
Then we shall use these concepts to define the action functional $\tilde{\Omega}[\phi]$ and investigate its generating properties.
Finally we shall show how $\tilde{\Omega}[\phi]$ provides a formal solution to the closure problem.

\subsection{Closed time contour}

\begin{figure}[t]
\begin{center}
\includegraphics[scale=.6]{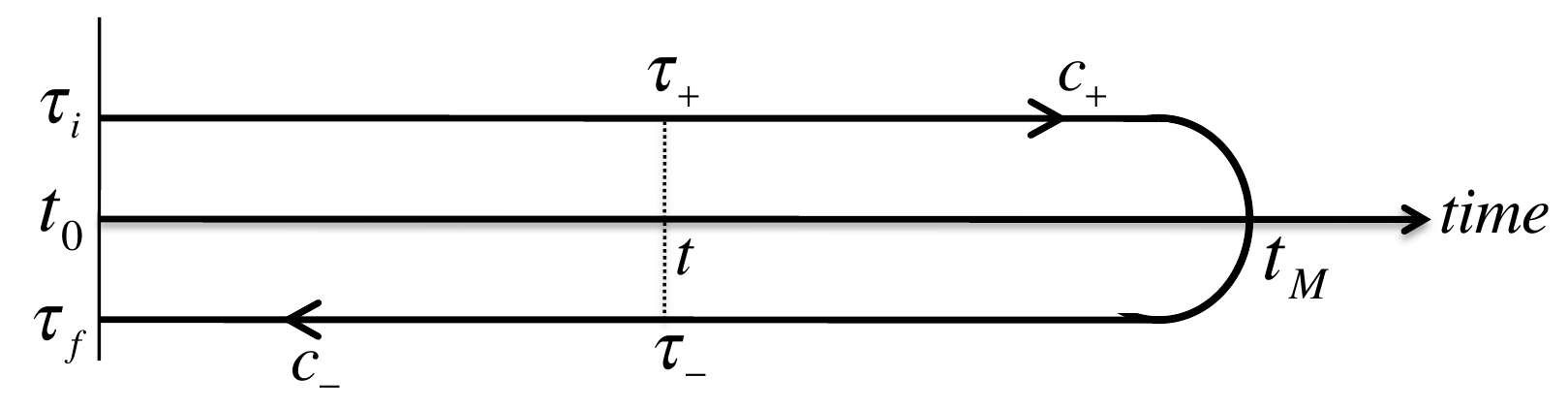}
\end{center}
\caption{Schematic illustration of the closed-time contour. For any time $t\in [t_{0},t_{M}]$, $\tau_{+}$ is the time on the forward branch $c_{+}$ and $\tau_{-}$ is the time on the backward branch $c_{-}$ of the closed-time contour such that $t=t(\tau_{+})=t(\tau_{-})$.} \label{figure1}
\begin{center}
\includegraphics[scale=.5]{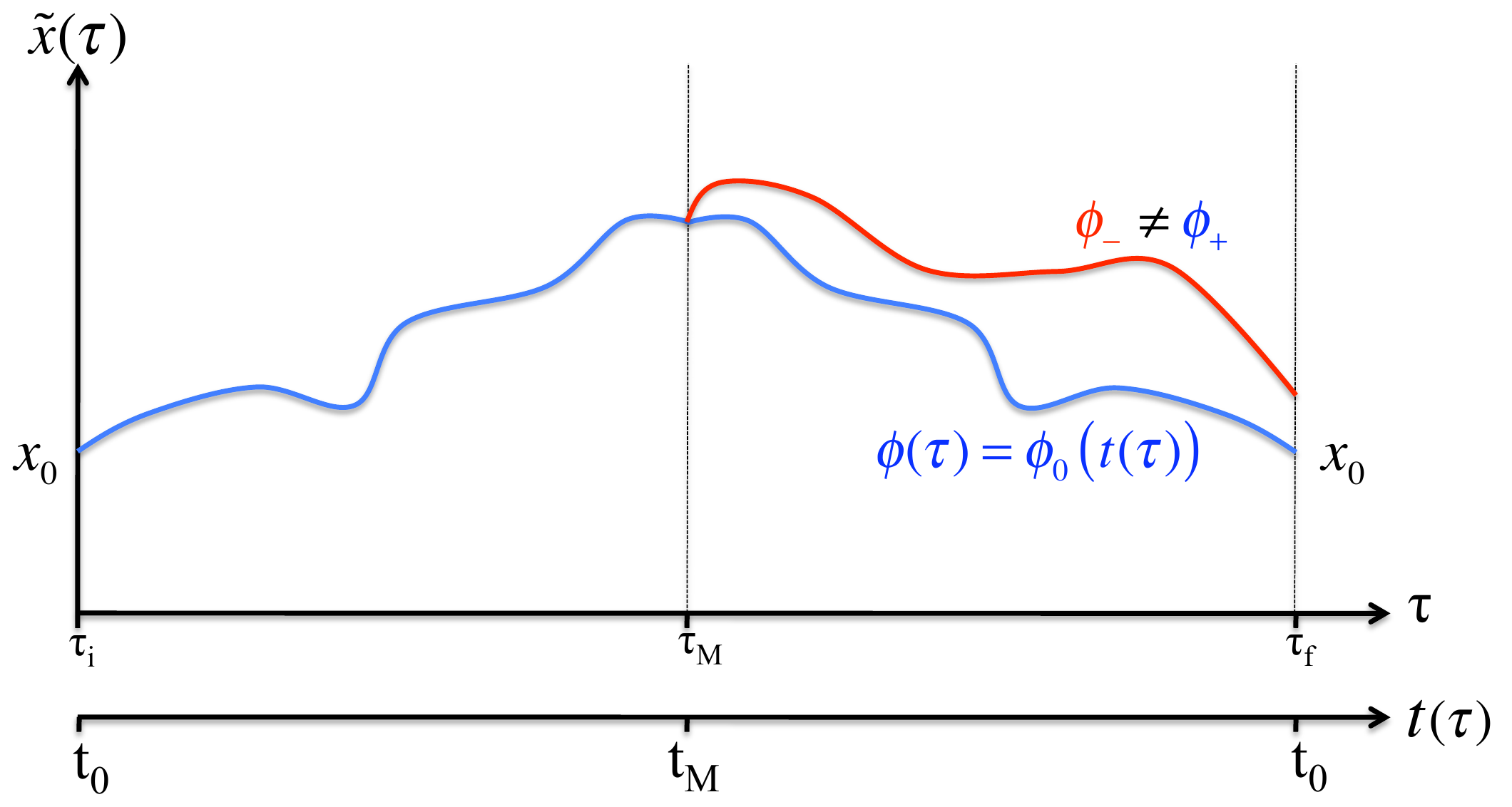}
\end{center}
\caption{Illustration of the extended dynamics $\tilde{x}(\tau)$ on the closed-time contour.
For a physical potential, the dynamics retraces the forward trajectory on the return branch.}
\label{figure2}
\end{figure}
We shall extend the dynamics under investigation along a closed path in time that, as illustrated in Fig.(\ref{figure1}), goes from the initial time $t_{0}$ to some time $t_{M}$ in the future and back to the initial time $t_{0}$.
The value of $t_{M}$ can be chosen arbitrarily as long as it is larger than the largest time of interest; for definiteness, we shall set $t_{M}=+\infty$ in the following.
The closed-time contour can be conveniently defined by parameterizing the physical time $t=t(\tau)$ in terms of a time variable $\tau$ in such way that if $\tau$ runs from an initial $\tau_{i}$ to a final $\tau_{f}$ then $t(\tau)$ monotonically increases from the initial time $t_{0}$ to the maximum time $t_{M}$ at $\tau_{M}$ and then monotonically decreases from $t_{M}$ back to $t_{0}$ \cite{PossibleParametrization}.
In fact, the chosen parametrization is inconsequential since the final results are independent of it.

In the following, for any physical time $t\in [t_{0},t_{M}]$, we will denote by $\tau_{+}$ the unique time on the forward branch and $\tau_{-}$ the unique time on the backward branch of the closed-time contour such that $t=t(\tau_{+})=t(\tau_{-})$.
We will also distinguish certain quantities defined on the closed-time contour variable $\tau$ by a tilde.
Thus we write $\tilde{1}=({\bf r}_1,{\bf p}_1,\tau_1)=(X_1,\tau_1)$ or simply $\tilde{1}=(X,\tau)$ when no confusion is possible.
In these notations, $\delta(\tilde 1-\tilde 2)=\delta^{(3)}({\bf r}_1-{\bf r}_2)\delta^{(3)}({\bf p}_1-{\bf p}_2)\delta(\tau_1-\tau_2)$.

\subsection{Dynamics on the closed time contour} \label{sectionIIIB}

Let us first imagine that the dynamics $x(t)=\big\{{\bf r}_{j}(t),{\bf p}_{j}(t)\big\}_{j=1,N}$ under investigation over the physical time interval $t_{0}\leq t\leq t_{M}$, or equivalently $\tau_{i}\leq \tau\leq\tau_{M}$, extends to the backward branch $\tau_{M}\leq\tau\leq\tau_{f}$ of the contour in such way that, as illustrated in Fig.(\ref{figure2}), the extended dynamics retraces backward the same phase-space trajectory along $[\tau_{M},\tau_{f}]$.
In other words, if $\tilde{\bf r}_j(\tau)$ and $\tilde{\bf p}_j(\tau)$ denote the positions and momenta of this extended dynamics, we require that for all $\tau\in[\tau_{i},\tau_{f}]$,
\ben
\D\tilde{\bf r}_j(\tau)={\bf r}_j\big(t(\tau)\big)\quad,\quad\D\tilde{\bf p}_j(\tau)={\bf p}_j\big(t(\tau)\big)\,. \label{tildeqpequalqp}
\een
By differentiation of Eq.(\ref{tildeqpequalqp}) with respect to $\tau$, we easily find that the extended dynamics is Newtonian and governed by the Hamiltonian,
\ben
\hspace{-.2cm}{\cal{H}}_{\phi_0}(\tilde x,\tau)&=&t'(\tau)H_{tot}(\tilde x,t(\tau))=t'(\tau)\left(H_{N}(\tilde x)+\sum_{j=1}^{N}{\phi_{0}({\tilde r}_{j},{\tilde p}_{j},t(\tau))}\right) \label{physicalcase} 
\een
with $\tilde{x}=(\tilde{\bf r}_{1},\dots,\tilde{\bf r}_{N};\tilde{\bf p}_{1},\dots,\tilde{\bf p}_{N})$.
This extended dynamics then can be regarded as a special case of a larger set of dynamics on the closed-time contour and characterized by the Hamiltonians
\ben
{\cal{H}}_{\phi}(\tilde x,\tau)&=&t'(\tau)\left(H_{N}(\tilde x)+\sum_{j=1}^{N}{\phi({\tilde{\bf r}}_{j},{\tilde{\bf p}}_{j},\tau)}\right)\/. \label{calHphi}\\
&=&t'(\tau)\left(H_{N}(\tilde x)+\oint{dxd\tau{\cal{N}}(\tilde x,\tau)\phi(\tilde x,\tau)}\right)\nn
\een
where $\phi$ is any external, time-dependent potential defined on the closed-time contour, and
\be
\D{\cal{N}}({\bf r},{\bf p},\tau)&=&\D\sum_{j=1}^{N}{\delta\left({\bf r}-\tilde{\bf r}_{j}(\tau)\right)\delta\left({\bf p}-\tilde{\bf p}_{j}(\tau)\right)}
\ee
is the single-particle phase-space density.
Just like in calculus it is often very useful to consider a real function of the real variable as a complex function of the complex variable, considering the physical dynamics under consideration as an element of the set of extended dynamics will allow us to derive properties of the physical dynamics that are difficult to obtain otherwise.

We can distinguish between two types of external potentials $\phi$, namely physical and non-physical potentials; we shall denote physical potentials with a subscript $p$, e.g. $\phi_{p}$.
A {\em physical potential} is identical on both branches of the contour, i.e. $\phi_p(\tau_+)=\phi_p(\tau_-)$, while a non-physical potential is not.
Any potential with real, physical origin such as $\phi_{0}$ in Eq.(\ref{Hextuext}), defines a physical potential through $\phi_{p}(\tau)=\phi_{0}(t(\tau))$.
By construction, under the influence of a physical potential, the extended dynamics retraces its forward trajectory backwards on the return branch of the closed-time contour, while for a non-physical potential, forward and backward trajectories are generally different.
Both cases are illustrated in Fig. \ref{figure2}.

\subsection{Generating action functional}

The extended dynamics governed by ${\cal{H}}_{\phi}$ can also be described by the Lagrangian
\begin{eqnarray}
{\cal{L}}_{\phi}(\tilde r,{\tilde v},\tau)&=&\tilde p\cdot{\tilde v}-{\cal{H}}_{\phi}(\tilde r,\tilde p,\tau)\,,
\end{eqnarray}
with the velocity $\tilde{v}=d\tilde{r}/d\tau$ and the short-hand notation $\tilde r=(\tilde{\bf r}_1,\dots,\tilde{\bf r}_N)$.
Given an initial condition $x_{0}$ and the corresponding trajectory $\D\left(\tilde{r}(\tau),\tilde{v}(\tau)\right)$, we define the total action over the closed-time contour as
\ben
{\cal{S}}[\phi;x_{0}]&=&\int_{\tau_{i}}^{\tau_{f}}{d\tau\,{\cal{L}}_{\phi}\left(\tilde{r}(\tau),\tilde{v}(\tau),\tau\right)}\,-{\cal{S}}_{b}[\phi,x_{0}]\,,\nn\\
&=&\int_{-\infty}^{+\infty}{d\tau\,\Pi(\tau){\cal{L}}_{\phi}\left(\tilde{r}(\tau),\tilde{v}(\tau),\tau\right)}\,-{\cal{S}}_{b}[\phi,x_{0}]\,, \label{classicalclosedtimeaction}
\een
with the boundary term,
\be
\D{\cal{S}}_{b}[\phi,x_{0}]=\frac{1}{2}\left(\tilde p(\tau_i)+\tilde p(\tau_f)\right)\left(\tilde r(\tau_i)-\tilde r(\tau_f)\right)\/.
\ee
The role of ${\cal{S}}_b$ is to cancel boundary terms that arise when performing the variations of the first term with respect to $\phi$.
In the second line, we include the integration range $[\tau_i,\tau_f]$ in the integrand through the ``window'' function
\ben
\Pi(\tau)=\theta(\tau-\tau_i)\theta(\tau_f-\tau)=\theta\left(t(\tau)-t_0\right) \label{windowfunction}
\een
where $\theta$ is the (Heaviside) step function \cite{Heavisidefunction} so that $\Pi(\tau)=1$ when $\tau_i<\tau<\tau_f$ and $\Pi(\tau)=0$ when $\tau<\tau_i$ or $\tau>\tau_f$.
As we shall see, this apparently insignificant rewriting in fact allows us to self-consitently include the contribution of initial conditions within the theory.

We then define the effective action functional $\tilde{\Omega}[\phi]$ such as
\ben
\D e^{-\frac{i}{s}\tilde{\Omega}[\phi]}=\lel e^{\frac{i}{s}{\cal{S}}[\phi;x_{0}]}\rir={\cal{Z}}[\phi]\,. \label{actionfunctional}
\een
Here $s$ is a fixed parameter with the dimension of an action, or energy times time, to make the exponent dimensionless.
Its value may be set to the quantum of action $\hbar$ if one regards $e^{\frac{i}{\hbar}{\cal{S}}[\phi;x_{0}]}$ as the classical contribution to the total quantum amplitude to go from points $\tilde{x}(\tau_{i})=x_{0}$ to $\tilde{x}(\tau_{f})$ \cite{FeynmanHibbs}.
However the identification $s=\hbar$ is not required since Eq.(\ref{actionfunctional}) could also be defined without reference to its quantum origin.
Accordingly, we find that the actual value of $s$ is inconsequential since all the general results derived from Eq.(\ref{actionfunctional}) are independent of it.
(The situation here is reminiscent of the problem encountered in the early days of classical statistical mechanics \cite{Fermi}, where an arbitrarily chosen size $h_0$ of the unit cell in phase-space was introduced in order to count the accessible classical states; while the physical laws were not affected by the value of $h_0$, the arbitrariness was later removed using the principles of quantum mechanics, leading to $h_0=\hbar$.) It may for instance be useful to chose $s$ to be purely imaginary, $s\to is$ , and write
\be
\tilde{\Omega}[\phi]=-s\ln{\cal{Z}}[\phi]=-s\ln\lel e^{\frac{1}{s}{\cal{S}}[\phi;x_{0}]}\rir \,.
\ee
in order to circumvent difficulties related to complex logarithms, to ensure nice properties of the effective action (e.g. convexity), etc.; we shall not delve into these technical points here.

We remark that the action functional vanishes at any physical potential $\phi_p$,
\be
\tilde{\Omega}[\phi_p]=0\quad,\quad\forall\phi_p\,.
\ee
Indeed, the total action ${\cal{S}}$ vanishes
\be
{\cal{S}}[\phi_p;x_{0}]=0\quad,\quad \forall x_{0}\,,
\ee
since the contributions over both branches of the closed-time contour have opposite sign by construction,
\be
\int_{\tau_{i}}^{\tau_{M}}{d\tau\,{\cal{L}}_{\phi}(\tau)}&=&-\int_{\tau_{M}}^{\tau_{f}}{d\tau\,{\cal{L}}_{\phi}(\tau)}\,,
\ee
and ${\cal{S}}_{b}=0$ since $\tilde r(\tau_i)=\tilde r(\tau_f)$.
As a consequence, ${\cal{Z}}[\phi_p]=\int{dx_{0}F(x_{0})}=1$, and therefore $\tilde{\Omega}[\phi_p]=0$.
However, since $\tilde{\Omega}[\phi]$ generally takes on nonzero values outside the subset of physical potentials, its functional derivatives at a $\phi_p$ can take finite values \cite{examplederivatives}.
As mentioned earlier, those functional derivatives are indeed quite remarkable since they are simply related to key physical quantities.

\subsection{Functional derivatives of $\tilde\Omega[\phi]$}

\subsubsection{Quick comparison with quantum action functionals}

Action functionals of the form (\ref{actionfunctional}), i.e. the exponential of an action, play a fundamental role in field theory, which often deals with generating field integrals (i.e. traces) of the form \cite{Browntextbook},
\ben
\D e^{-i\tilde{\Omega}_{Q}[\phi]/\hbar}&=&\left\langle\int{{\cal{D}}\Psi\,e^{\frac{i}{\hbar}({\cal{S}}_{Q}[\Psi]+\int{\phi\Psi})}}\right\rangle \label{quantumfieldtheoryactionfunctional}\\
&=&\iint{d\psi_{1}d\psi_{2}\langle\psi_{1}|\hat\rho_{0}|\psi_{2}\rangle\int_{\psi_1}^{\psi_2}{{\cal{D}}\Psi\,e^{\frac{i}{\hbar}({\cal{S}}_{Q}[\Psi]+\int{\phi\Psi})}}}\nn
\een
where $\Psi$ is the field, ${\cal{S}}_{Q}[\Psi]$ is the action, $\hat\rho_{0}$ is the initial density matrix, and $\phi$ is an external source term that couples linearly to $\Psi$.
In that case, the whole exponent in Eq.(\ref{quantumfieldtheoryactionfunctional}) is linear in $\phi$, and the successive functional derivatives $\delta^{(n)}\tilde{\Omega}_{Q}[\phi]/\delta\phi(1)\dots\delta\phi(n)$ simply insert the field $\Psi$ in the field integral and generate the averaged (time-ordered) products $\langle\delta\phi(1)\dots\delta\phi(n)\rangle$ of the field, alternatively referred to as the time-ordered correlation functions, propagators or Green's functions of the theory.
In contrast, the classical action functional (\ref{actionfunctional}) is highly non-linear on the external potential $\phi$ since the phase-space trajectory $\tilde{x}(\tau)$ in ${\cal{S}}[\phi;x_{0}]$ implicitly depends on the external potential $\phi$ as well.
If, as already mentionned in the previous paragraph, one regards $\tilde{S}[\phi;x_0]$ as the classical contribution to the total quantum amplitude to go from $x_{0}$ to $\tilde{x}(\tau_{f})$, we see that the reduction of the quantum field integral from all possible field configurations to the classical path only is responsible for the non-linearity in the external perturbation of the classical action (\ref{actionfunctional}).
As a consequence, two types of quantities arise when differentiating the classical action functional $\tilde{\Omega}[\phi]$ instead of only one, namely the time-ordered correlation functions, in quantum field theory.
The explicit linear dependence of ${\cal{S}}$ on $\phi$ generates the correlations of the phase-space density as in field theory, while the implicit non-linearities generate terms involving the Poisson bracket $[\cdot,\cdot]_{PB}$ such as the response functions $\chi^{R,A}$, Eq.(\ref{retartedBAresponse}) \cite{anotherdistinction}.

\subsubsection{First and second derivatives.}

In the neighborhood of the external potential $\phi_0$, $\tilde{\Omega}[\phi]$ can be expanded in the series,
\ben
\tilde{\Omega}[\phi_0+\delta\phi]&=&\sum_{n=0}^{\infty}{\frac{1}{n!}\oint{d\tilde 1\dots d\tilde n \Omega^{(n)}(\tilde 1,\dots,\tilde n)\delta\phi(\tilde 1)\dots\delta\phi(\tilde n)}}\,, \label{functionalderivativesexpansion}
\een
where we introduce the short-hand notation,
\be
\oint{d\tilde 1\dots}=\int_{-\infty}^{+\infty}{d\tau_1 t'(\tau_1)\int_{V}{d{\bf q}_1\int{d{\bf p}_1\dots}}}
\ee
Here $\tilde\Omega^{(0)}=\tilde{\Omega}[\phi_0]$ and, for $n\geq 1$, $\tilde\Omega^{(n)}$ is the $n$-th functional derivative at potential $\phi_{0}$,
\be
\Omega^{(n)}(\tilde 1,\dots,\tilde n)\equiv\frac{\delta^{(n)}\tilde\Omega[\phi]}{\delta\phi(\tilde 1)\dots\delta\phi(\tilde n)}\Big|_{\phi=\phi_0}\,.
\ee
The latter can be obtained by a systematic perturbation expansion in powers of the variations $\delta\phi$ around $\phi_0$.
We are particularly interested in this work in the first two derivatives $n=1,2$.
For clarity, we just report here the main results and give their proofs in appendix \ref{sectionproof}; in addition, the differentiability properties of $\tilde{\Omega}[\phi]$ are discussed more thoroughly in appendix \ref{differentiabilityproperites}.

The first functional derivative generates the phase-space distribution function $f$ of the dynamics under investigation,
\ben
\frac{\delta\tilde\Omega}{\delta\phi(\tilde 1)}\Big|_{\phi=\phi_0}=\Pi(\tau)\big\langle{\cal{N}}({\bf r},{\bf p},\tau)\big\rangle=\Pi(\tau)f({\bf r},{\bf p},t(\tau))\equiv \tilde{f}(\tilde 1)\,, \label{domegadphi}
\een
where we recall that $\Pi(\tau)$ is the window function defined by Eq.(\ref{windowfunction}).
Accordingly, its second functional derivative,
\ben
\frac{\delta^{(2)}\tilde\Omega[\phi]}{\delta\phi(\tilde 1)\delta\phi(\tilde 2)}\Big|_{\phi=\phi_0}&=&\frac{\delta\/\tilde f(\tilde 1)}{\delta\phi(\tilde 2)}\Big|_{\phi=\phi_0}\equiv\chi(\tilde{1},\tilde{2})\,. \label{secondderivative}
\een
can be regarded as the linear response function {\it along the closed-time contour} of the system's dynamics.
The direct calculation of $\chi$ developed in appendix \ref{sectionproof} leads to 
\ben
\chi(\tilde{1},\tilde{2})=\Pi(\tau_1)\Pi(\tau_2)\Big({\cal{R}}(\tilde{1},\tilde{2})-\frac{i}{s}C(\tilde{1},\tilde{2})\Big)\,, \label{d2omegadphi2}
\een
and consist in the linear combination of two terms.
The last term is the two-body correlation function
\ben
C(\tilde 1,\tilde 2)=\Big\langle \delta{\cal{N}}(\tilde{1})\delta{\cal{N}}(\tilde{2})\Big\rangle\/,
\een
and, as discussed earlier, comes from the explicit linearity of the action ${\cal{S}}$ in the external potential.
The first term in (\ref{d2omegadphi2}) originates from the non-linerity of the action and involves the many-body Poisson bracket $[.,.]_{PB}$,
\ben
\D {\cal{R}}(\tilde{1},\tilde{2})&=&\frac{1}{2}\Big\langle {\cal{T}}_{c}\left[{\cal{N}}(\tilde{1}),{\cal{N}}(\tilde{2})\right]_{PB}\Big\rangle\,. \label{calR}
\een
In Eq.(\ref{calR}), we have introduced the chronologically ordered Poisson bracket of two dynamical variables $A(\tau)$ and $B(\tau)$ as
\be
{\cal{T}}_{c}\left[A(\tau_1),B(\tau_2)\right]_{PB}=\theta(\tau_1-\tau_2)\left[A(\tau_1),B(\tau_2)\right]_{PB}+\theta(\tau_2-\tau_1)\left[B(\tau_2),A(\tau_1)\right]_{PB}\,.
\ee
Accordingly, $\chi$ can also be expressed as \cite{notesymmetryofchi},
\ben
\chi(\tilde{1},\tilde{2})=\theta(\tau_1-\tau_2)\chi^{>}(\tilde{1},\tilde{2})+\theta(\tau_2-\tau_1)\chi^{<}(\tilde{1},\tilde{2})\,, \label{chichi>chi<}
\een
where for $\tau_1>\tau_2$ (dropping the window functions for simplicity),
\ben
\chi^{>}(\tilde{1},\tilde{2})=\frac{1}{2}\Big\langle\left[{\cal{N}}(\tilde{1}),{\cal{N}}(\tilde{2})\right]_{PB}\Big\rangle-\frac{i}{s}C(\tilde{1},\tilde{2})\,, \label{chi>}
\een
and for $\tau_2>\tau_1$,
\ben
\chi^{<}(\tilde{1},\tilde{2})\!=\!-\frac{1}{2}\Big\langle\left[{\cal{N}}(\tilde{1}),{\cal{N}}(\tilde{2})\right]_{PB} \Big\rangle-\frac{i}{s}C(\tilde{1},\tilde{2})\,. \label{chi<}
\een

\subsection{Connection to physical response and correlation functions.}

Remarkably, the function $\chi$ contains information about both the physical correlations and response properties of the physical system, which can easily be extracted as follows.

The physical correlation function $C(1,2)$ defined in Sec.\ref{sectionII} is easily extracted from $\chi$ by taking the sum of Eqs.(\ref{chi>}) and (\ref{chi<}) \cite{notenotationfortimevariable},
\ben
C(1,2)=\frac{is}{2}\big(\chi^{>}(1,2)+\chi^{<}(1,2)\big)\quad,\quad \forall t_1,t_2\geq t_0\,. \label{physicalCfromchi>chi<}
\een
The connection to the physical response can be seen by calculating the variation $\delta f$ of the phase-space density due to a change in the external potential $\delta \phi_0$ around $\phi_0$.
We have,
\begin{eqnarray*}
\delta f({\bf x}_{1},t_{1})&=&\oint{d\tilde 2\,\chi(\tilde 1,\tilde 2)\delta \phi_0(\tilde 2)}\\
&=&\int{d{\bf x}_{2}\int_{t_{0}}^{t_1}{dt_{2}\,\chi^{>}({\bf x}_{1},t_{1};{\bf x}_{2},t_{2})\delta \phi_0({\bf x}_{2},t_{2})}}\\
&&\hspace{0.5cm}+\int{d{\bf x}_{2}\int_{t_1}^{\infty}{\chi^{<}({\bf x}_{1},t_{1};{\bf x}_{2},t_{2})\delta \phi_0({\bf x}_{2},t_{2})}}\\
&&\hspace{1cm}+\int{d{\bf x}_{2}\int_{\infty}^{t_{0}}{\chi^{<}({\bf x}_{1},t_{1};{\bf x}_{2},t_{2})\delta \phi_0({\bf x}_{2},t_{2})}}\\
&=&\int{d{\bf x}_{2}\int_{t_{0}}^{t_{1}}{dt_{2}\left[\chi^{>}({\bf x}_{1},t_{1};{\bf x}_{2},t_{2})-\chi^{<}({\bf x}_{1},t_{1};{\bf x}_{2},t_{2})\right]\delta \phi_0({\bf x}_{2},t_{2})}}\,.
\end{eqnarray*}
Using Eqs.(\ref{chi>}-\ref{chi<}), we find $\delta f(1)=\int{d{\bf x}_{2}\int_{-\infty}^{\infty}{dt_{2}\chi^{R}(1,2)\delta \phi_{0}(2)}}$, with
\be
\chi^{R}(1,2)&=&\theta(t_{1}-t_{2})\left[\chi^{>}(1,2)-\chi^{<}(1,2)\right]\\
&=&\theta(t_{1}-t_{2})\theta(t_2-t_0)\big\langle\left[N(1),N(2)\right]_{PB}\big\rangle\/,
\ee
which is just the retarded response function (\ref{retartedBAresponse}) of traditional perturbation theory recalled in Sec. \ref{sectionIIB}.

In conclusion, simple linear combinations of the second order derivatives $\chi$ yield to key dynamical properties of the physical system under investigation, namely $C$ and $\chi^{R,A}$.
The simulateneous occurence of both $C$ and $\chi^{R}$, two quantities of different nature, is remarkable and further discussed in appendix \ref{Discussionoccurence} with regard to its counterpart in quantum field theory.

\subsection{Generalized BBGKY hierarchy for the extended dynamics} \label{sectionIIIF}

We consider the statistical dynamics on the closed-time contour governed by ${\cal{H}}_{\phi}$ defined in Eq.(\ref{calHphi}).
We assume that at the initial time $\tau_{i}$, the initial positions and momenta $\tilde{x}(\tau_i)$ are distributed according to the same distribution $F_{0}(x_{0})$ as the physical system under investigation (see Sec.\ref{sectionII}.)
The notations used here for $L_1,\gamma_3,\dots$ are those of Sec.\ref{sectionIIB} with the real time variables $t_1,t_2\dots$ simply replaced by $\tau_1,\tau_2\dots$.

The microscopic phase-space density ${\cal{N}}(\tilde{1})$ evolves according to,
\ben
\left[\frac{1}{t'(\tau)}\frac{\partial}{\partial \tau}-L_{\tilde{1}}\right]{\cal{N}}(\tilde{1})&=&\frac{1}{2}\gamma_{3}(\tilde{1},\tilde{2},\tilde{3}){\cal{N}}(\tilde{2}){\cal{N}}(\tilde{3})\,.\label{dcalNphidt1}
\een
By averaging over the initial state, we obtain the evolution equation for the phase-space density $\tilde{f}$ defined by Eq.(\ref{domegadphi}),
\ben
\left[\frac{1}{t'(\tau)}\frac{\partial}{\partial \tau}-L_{\tilde 1}\right]\tilde f(\tilde 1)-\{u^{mf}(\tilde 1),\tilde f(\tilde 1)\}=\frac{1}{2}\gamma_{3}(\tilde 1,\tilde 2,\tilde 3)C(\tilde 2,\tilde 3)+\Delta(\tau)f(1)\,,\nn\\ \label{dfphidt10}
\een
in terms of the two-point correlation function $C$ and of
\ben
\Delta(\tau)=\frac{1}{t'(\tau)}\frac{d\Pi(\tau)}{d\tau}&=&\frac{1}{t'(\tau)}\Big[\delta(\tau-\tau_i)-\delta(\tau-\tau_f)\Big]=\delta(t(\tau)-t_0)\,.\label{Deltatau}
\een
Following the traditional BBGKY approach, the evolution equation (\ref{dfphidt10}) would be regarded as the first equation of the hierarchy of evolution equations between the successive equal-time cumulants $C^{(n)}$.
Our approach allows us to instead regard Eq.(\ref{dfphidt10}) as the first equation of a ``larger'' hierarchy in terms of the successive functional derivatives $\Omega^{(n)}(1,\dots,n)$, which combine information on both correlation and response functions.
The advantage of this extended hierarchy is that it can be formally closed at second order $n=2$ using the procedure discussed below.

To obtain this extended hierarchy, we note the equal-time property
\be
\chi(X_1,\tau;X_2,\tau)=-\frac{i}{s}C(X_1,\tau;X_2,\tau)\,,
\ee
as can simply be seen by setting $\tau_1=\tau_2=\tau$ in Eq.(\ref{chichi>chi<}).
The collision integral in Eq.(\ref{dfphidt1}) can then be rewritten in terms of $\chi$ and yields the evolution equation
\ben
\left[\frac{1}{t'(\tau)}\frac{\partial}{\partial \tau}-L_{\tilde 1}\right]\tilde f(\tilde 1)-\{u^{mf}(\tilde 1),\tilde f(\tilde 1)\}=\frac{1}{2}\gamma_{3}(\tilde 1,\tilde 2,\tilde 3)\chi(\tilde 2,\tilde 3)+\Delta(\tau)\tilde f(1)\,, \nn\\\label{dfphidt1}
\een
which relates the first derivative $f=\delta\tilde\Omega/\delta\phi$ to the second derivative $\chi=\delta^{2}\tilde\Omega/\delta\phi^{2}$ of the generating functional.
Successive functional differentiations of Eq.(\ref{dfphidt1}) with respect to $\phi$ generate the extended hierarchy between the $\Omega^{(n)}$'s.
The first functional derivative yields the evolution equation for $\chi$ (see footnote \cite{Notesuccessivedifferentiation} for details),
\ben
\lefteqn{\D\left[\frac{1}{t'(\tau)}\frac{\partial}{\partial \tau}-L_{1}\right]\chi(\tilde 1,\tilde 1')-\Sigma^{mf}(\tilde 1,\tilde 2)\cdot\chi(\tilde 2,\tilde 1')}&\label{dchi11pdtau}\\
&\D=s_{2}(\tilde 1,\tilde 1')+\frac{is}{2}\gamma_{3}(\tilde 1,\tilde 2,\tilde 3)\tilde\Omega^{(3)}(\tilde 2,\tilde 3,\tilde 1')+\Delta(\tau)\Pi(\tau')C(\tilde 1,\tilde 1')/is\,,\nn
\een
where we define \cite{Notesuccessivedifferentiation},
\be
s_{2}(\tilde 1,\tilde 1')\equiv\left\{\delta(\tilde 1-\tilde 1'),\tilde f(\tilde 1)\right\}(\tilde 1)\,.
\ee
The mean-field kernel is defined as in Eq.(\ref{sigmamf}),
\ben
\Sigma^{mf}(\tilde 1,\tilde 1')&=&\gamma_3(\tilde 1,\tilde 2,\tilde 3)\tilde f(\tilde 2)\delta(\tilde 1'-\tilde 3)\,.\label{sigmamfcc}
\een
In the rhs of Eq.(\ref{dchi11pdtau}), the second term comes from the direct differentiation of the collision integral in Eq.(\ref{dfphidt1}), and the last term includes the boundary conditions.

Similarly, functional differentiation of Eq.(\ref{dchi11pdtau}) yields the evolution equation for $\Omega^{(3)}$,
\ben
\lefteqn{\D\left[\frac{1}{t'(\tau)}\frac{\partial}{\partial \tau'}-L_{1}\right]\Omega^{(3)}(\tilde 2,\tilde 3,\tilde 1)-\Sigma^{mf}(\tilde 1,\tilde 2')\Omega^{(3)}(\tilde 2',\tilde 2,\tilde 3)}&&\label{dchi111pdtauGmf}\\
\D&=&s_{3}(\tilde 2\tilde 3\tilde 1)+\frac{is}{2}\gamma_{3}(\tilde 1\tilde 2'\tilde 3')\left(\Omega^{(4)}(\tilde 2',\tilde 3',\tilde 2,\tilde 3)+\frac{2}{is}\chi(\tilde 2',\tilde 2)\chi(\tilde 3',\tilde 3)\right)\nn\\
\D&&+\Delta(\tau')\Omega^{(3)}(\tilde 2,\tilde 3,\tilde 1)\,,\nn
\een
where
\ben
s_{3}(\tilde 2\tilde 3\tilde 1)=\left\{\delta(\tilde 1-\tilde 2),\chi(\tilde 1,\tilde 3)\right\}(\tilde 1)+\left\{\delta(\tilde 1-\tilde 3),\chi(\tilde 1,\tilde 2)\right\}(\tilde 1)\,, \label{s3}
\een
and so on and so forth.
The hierarchy thus obtained for the $\Omega^{(n)}$'s suffers from the closure problem whereby the equation for $\tilde\Omega^{(n)}$ depends on the next order response function $\tilde\Omega^{(n+1)}$.
In the following we show that this hierarchy can be closed at the $n=2$ level.

\subsection{Formal closure of the hierarchy \cite{Noteonrenormalizationapproach}} \label{sectionIIIG}

\subsubsection{Effective potential and vertex functions}

The method uses a Legendre transformation of $\tilde\Omega[\phi]$ \cite{Browntextbook,ZinnJustin}, a functional of the external potentials $\phi$, to obtain a function $\tilde{\Gamma}[f_{\phi}]$ of the phase-space densities $f_{\phi}$ given by
\begin{eqnarray}
\tilde{\Gamma}[f_{\phi}]=-\tilde{\Omega}[\phi]+\oint{d\tilde 1f_{\phi}(\tilde 1)\,\phi(\tilde 1)}\,. \label{legendretransformation}
\end{eqnarray}
We shall refer to $\tilde{\Gamma}$ as the {\em effective potential}; the latter can be regarded as the generalization of the Helmholtz free energy used in density functional theory  \cite{HansenMcDonald3rd}.
The transformation (\ref{legendretransformation}) shifts attention from the potential that perturbs the system to the phase-space density that describes its effect.

The functional derivatives of the Legendre transform $\Gamma[f_{\phi}]$ at the phase-space distribution $f_\phi=f$ under investigation,
\begin{eqnarray*}
\Gamma^{(n)}(\tilde 1,\dots,\tilde n)=\frac{\delta^{n}\tilde{\Gamma}[f_{\phi}]}{\delta f_{\phi}(\tilde 1)\dots\delta f_{\phi}(\tilde n)}\Bigg|_{f_\phi=f}\,,
\end{eqnarray*}
are intimately related to those of $\tilde{\Omega}[\phi]$.
In statistical field theory, the functions $\Gamma^{(n)}$ are usually referred to as {\it vertex functions}.
For $n=2$, we shall write $\Gamma\equiv\Gamma^{(2)}$.		

The first derivative simply equals the external potential,
\ben
\frac{\delta\,\tilde{\Gamma}}{\delta\, f_{\phi}(\tilde 1)}\Bigg|_{f_\phi=\tilde f}&=&\phi_0(\tilde 1)\,. \label{dFdrho1}
\een
In the special case when $\phi_{0}=0$ as in equilibrium or relaxation problems, the previous relation shows that the physical density $f$ under investigation is an extremum of the effective potential $\tilde\Gamma$.
As such, $\tilde\Gamma$ plays a role similar to a thermodynamic potential in equilibrium statistical mechanics.

The higher-order derivatives can be obtained by differentiation Eq.(\ref{dFdrho1}) with respect to $\phi$ and using the chain rule of differentiation.
For instance, for the second-order derivative, we find
\ben
\hspace{-.5cm}\oint{d\tilde 2\,\frac{\delta^{2}\Gamma[\phi]}{\delta f_{\phi}(\tilde 1)\delta f_{\phi}(\tilde 2)}\frac{\delta{f_{\phi}(\tilde 2)}}{\delta \phi(\tilde 1')}}\!\!=\!\!\oint{d\tilde 2\,\frac{\delta^{2}\Gamma[\phi]}{\delta f_{\phi}(\tilde 1)\delta f_{\phi}(\tilde 2)}\frac{\delta^{2}\tilde{\Omega}[\phi]}{\delta\phi(\tilde 2)\delta\phi(\tilde 1')}}=\delta_{c}(\tilde 1-\tilde 1')\label{d2gammad2omega}
\een
where the $\delta_c$-function $\delta_{c}(\tilde 1-\tilde 1')=\frac{1}{t'(\tau)}\delta(\tilde 1-\tilde 1')$ satisfies
\ben
\oint{d\tilde 2 \delta_{c}(\tilde 1-\tilde 2)\tilde f(\tilde 2)}=\tilde f(\tilde 1)\,.\label{deltacDirac}
\een
With short-hand notations, Eq.(\ref{d2gammad2omega}) writes as
\ben
\Gamma(\tilde 1,\tilde 2)\cdot\chi(\tilde 2,\tilde 1')=\chi(\tilde 1,\tilde 2)\cdot\Gamma(\tilde 2,\tilde 1')=\delta_{c}(\tilde 1-\tilde 1') \label{chigammainverse}\,,
\een
and shows that $\Gamma$ can be regarded as the inverse of the response function $\chi$.
Taking the derivative of (\ref{chigammainverse}) with respect to $\phi$ yields a relation between $\Gamma^{(3)}$ and $\Omega^{(3)}$,
\ben
\Omega^{(3)}(\tilde 1,\tilde 2,\tilde 3)&=&-\chi(\tilde 1,\tilde 1')\chi(\tilde 2,\tilde 2')\chi(\tilde 3,\tilde 3')\Gamma^{(3)}(\tilde 1',\tilde 2',\tilde 3')\,,\label{chi3gamma3}
\een
and so on and so forth \cite{ZinnJustin}.

\subsubsection{Closure of the hierarchy}

The relation (\ref{chi3gamma3}) provides the starting point for closing the extended BBGKY hierarchy linking the $\Omega^{(n)}$.
Firstly, we can now express the collision integral (\ref{dchi11pdtau}) as
\be
\frac{is}{2}\gamma_{3}(\tilde 1\tilde 2\tilde 3)\Omega^{(3)}(\tilde 2,\tilde 3,\tilde 1')
&=&\frac{is}{2}\gamma_{3}(\tilde 1\tilde 2\tilde 3)\left[-\chi(\tilde 2,\bar{2})\chi(\tilde 3,\bar{3})\chi(\tilde 1',\bar{1}')\Gamma^{(3)}(\bar{2},\bar{3},\bar{1}')\right]\\
&\equiv&\Sigma(\tilde 1,\tilde 2)\cdot\chi(\tilde 2,\tilde 1')\,,
\ee
in terms of the memory function kernel $\Sigma$,
\ben
\Sigma(\tilde 1,\tilde 1')&=&-\frac{is}{2}\gamma_{3}(\tilde 1\tilde 2\tilde 3)\chi(\tilde 2,\bar{2})\chi(\tilde 3,\bar{3})\Gamma^{(3)}(\bar{2},\bar{3},\tilde 1') \label{Sigmadefinitiondetailed}\\
&=&-\frac{is}{2}\gamma_{3}(\tilde 1\tilde 2\tilde 3)\frac{\delta\chi(\tilde 2,\tilde 3)}{\delta f_\phi(\tilde 1')}\label{Sigmadefinition}
\een
The equation of evolution (\ref{dchi11pdtau}) becomes
\ben
\Big[G(\tilde 1,\tilde 2)-\Sigma^{mf}(\tilde 1,\tilde 2)-\Sigma(\tilde 1,\tilde 2)\Big]\cdot\chi(\tilde 2,\tilde 1')=s_{2}(\tilde 1,\tilde 1')+\Delta(\tau)\Pi(\tau')C(\tilde 1,\tilde 1')/is \nn\\\label{dchidtausigma}
\een
where for convenience we have defined
\be
G(\tilde 1,\tilde 2)=\left[\frac{1}{t'(\tau)}\frac{\partial}{\partial \tau}-L_{1}\right]\delta(\tilde 1-\tilde 2)\,.
\ee
Equation (\ref{dchidtausigma}) for the closed-time contour response function $\chi$ will play here a role similar to the Dyson equation in quantum field theory \cite{ZinnJustin,KrempSchlangesKraeft}.
In the latter case, the equation describes the evolution of the time-dependent Green's functions, from which all other dynamical properties including correlation and response functions can be calculated.
Here, in constrast, Eq.(\ref{dchidtausigma}) describes the evolution of the quantity $\chi$, defined in Eq.(\ref{d2omegadphi2}), which combines in one single quantity, two properties of different nature in classical mechanics, namely the correlation and response functions (see also discussion in appendix \ref{Discussionoccurence}).

Secondly, Eq.(\ref{dchidtausigma}) allows us to derive another relation that expresses $\Gamma^{(3)}$ in terms of $\Sigma$.
To this end, we multiply Eq.(\ref{dchidtausigma}) on the right by the inverse $\chi^{-1}=\Gamma$, which leads to a relation between $\Sigma$ and $\Gamma$,
\be
G(\tilde 1,\tilde 1')-\Sigma^{mf}(\tilde 1,\tilde 1')&-&\Sigma(\tilde 1,\tilde 1')\\
&=&\left\{\Gamma(\tilde 1,\tilde 1'),f(\tilde 1)\right\}(\tilde 1)+\Delta(\tau)\Pi(\tau')C(\tilde 1,\tilde 2)\Gamma(\tilde 2,\tilde 1')/is\,.
\ee
By taking the functional derivative of the previous equation with respect to $f$ and combining the result with Eq.(\ref{Sigmadefinitiondetailed}), we obtain the following integro-differential equation for the memory function kernel,
\bs \label{Sigmaf}
\ben
\lefteqn{\Big\{\Sigma(\tilde 1,\tilde 1'),\tilde f(\tilde 1')\Big\}(\tilde 1')}&&\nn\\
&=&\frac{is}{2}\gamma_{3}(\tilde 123)s_3(23\tilde 1')\label{sigmadeltaf}\\
&+&\frac{is}{2}\gamma_{3}(\tilde 123)\chi(2,\bar{2})\chi(3,\bar{3})\left(\gamma_{3}(\tilde 1'2'3')+\frac{\delta\Sigma(\tilde 1',\bar{2})}{\delta f_\phi(\bar{3})}\Big|_{f_\phi=f}\right) \label{sigmaregf}\\
&+&\Delta(\tau')\frac{1}{2}\gamma_{3}(\tilde 123)\chi(2,\bar{2})\chi(3,\bar{3})\frac{\delta}{\delta f(\bar 3)}\left[C(\tilde 1',\bar 2')\Gamma(\bar 2',\bar 2)\right]\label{sigma0f}
\een
\es
The set of equations consisting of the first two equations (\ref{dfphidt1}) and (\ref{dchidtausigma}) of the hierarchy, together with (\ref{Sigmadefinitiondetailed}) and (\ref{Sigmaf}) is closed and is fully equivalent to the hierarchy of evolution equations for the $\Omega^{(n)}$.
Equation (\ref{Sigmaf}) is a functional equation for the memory function that can be used to generate self-consistent approximations in terms of $\gamma_3$ and $\chi$ (see also appendix \ref{closedunclosedsigma}).

\subsection{The memory function kernel}

Equation (\ref{Sigmaf}) implies that the memory function $\Sigma$ can be splitted into three parts as
\ben
\Sigma(1,1')&\equiv&\Sigma^{(0)}(X_1,X_1^\prime;\tau)\Delta(\tau')+\Sigma^{\delta}(X_1,X_1^\prime;\tau)\delta(\tau-\tau')+\Sigma^{reg}(1,1')\,.\nn\\
\label{splittingSigma}
\een
The physical interpretation of the different components of $\Sigma$ is discussed in detail in the following sections.
We just give here a short description of their origin.

The first term
\be
\big\{\Sigma^{(0)}(X_1,X_1^\prime;\tau),f(\tilde 1')\big\}(\tilde 1')=\frac{1}{2}\gamma_{3}(\tilde 1\tilde 2\tilde 3)\chi(\tilde 2,\bar{2})\chi(\tilde 3,\bar{3})\frac{\delta}{\delta f(\bar 3)}\left[C(\tilde 1',\bar 2')\chi^{-1}(\bar 2',\bar 2)\right]
\ee
describes the effect of initial correlations in the initial state at time $t_0$ on the dynamics of the system at $t>t_0$.
This term can be shown to vanish for Gaussian initial conditions but, as was recalled in Sec.\ref{sectionIIB}, this simplification is always invalid here since the phase-space density ${\cal{N}}$ is always strongly non-Gaussian \cite{Rose1979}.
The neglect of this term has deep consequences related to so-called {\it Bogolyubov's condition} of weakening (suppression) of the initial correlations and to the transition from reversible to irreversible dynamical equations \cite{Bogoliubov1962}.

The second term, proportional to the delta function $\delta(\tau-\tau')$, is singular in time and satifies,
\ben
\big\{\Sigma^{(\delta)}(X_1,X_1^\prime;\tau),f(\tilde 1')\big\}(\tilde 1')&=&\frac{is}{2}\gamma_{3}(\tilde 1\tilde 2\tilde 3)s_3(\tilde 2\tilde 3\tilde 1')\nn\\
&=&\gamma_{3}(\tilde 1\tilde 2\tilde 3)\big\{\delta(\tilde 1'-\tilde 2),C(\tilde 1',\tilde 3)\big\}(\tilde 1')\label{Sigmasingularpart}
\een
This term is related the instantaneous effect of correlations on the effective interactions between particles (at equilibrium, those correlations are the so-called 'static' correlations \cite{HansenMcDonald3rd}.)
This term naturally combines to the mean-field contribution $\Sigma^{mf}$, Eq.(\ref{sigmamfcc}), such that
\be
\Sigma_{tot}^{(\delta)}\equiv\Sigma^{mf}+\Sigma^{(\delta)}\,,
\ee
renormalizes the instantaneous mean-field effect with instantaneous correlations effects beyond the mean-field.
In the absence of particle correlations (Vlasov approximation), $\Sigma^{(\delta)}$ vanishes.

Finally, the last term describes memory, non-markovian correlation effects in the dynamics and satisfies
\be
\lefteqn{\D\big\{\Sigma^{(reg)}(\tilde 1,\tilde 1'),\tilde f(\tilde 1')\big\}(\tilde 1')}&\nn\\
&=\D\frac{is}{2}\gamma_{3}(\tilde 1\tilde 2\tilde 3)\chi(\tilde 2,\bar{2})\chi(\tilde 3,\bar{3})\left(\gamma_{3}(\tilde 1'\bar 2\bar 3)+\frac{\delta\Sigma(\tilde 1',\bar{2})}{\delta f_\phi(\bar{3})}\Big|_{f_\phi=f}\right)\,.
\ee

\section{Real-time formalism (I)} \label{sectionIV}

The previous approach allowed us to derive a closed, self-consistent set of equations to describe the evolution of the single-particle phase-space distribution function $f$ of the system.
However, being expressed in terms of quantities defined along the closed-time contour, this formulation lacks physical transparency and does not appeal to intuition.
In the next two sections, we shall re-express these exact results in terms of quantities defined on the real-time axis, namely the correlation $C$ and response functions $\chi^{R,A}$ and a set of memory functions.
In the following section, Sec.\ref{sectionV}, we will recast the overall approach into a form that leads to the same results {\it directly} in terms of physical quantities defined on the real-time axis.
In the present section, we focus on the evolution equations, discuss their physical content, and make contacts with previous works.
The careful derivation of the evolution equations for $C$, $\chi^{R,A}$ done in the next subsection is somewhat lenghty; the reader uninterested in those details may safely skip ahead to Eqs.(\ref{geenralCchiRchiA}) in Sec.\ref{sectionIV2}, where the results are given and then discussed and compared with previous works.

\subsection{From closed-time to real-time representations} \label{SectionIV1}

Throughout the section, we will use the (Pauli) matrices defined as
\be
\sigma_{x}=\left(\begin{array}{cc}
0&1\\
1&0
\end{array}
\right)
\,,\,\sigma_{y}=\left(\begin{array}{cc}
0&-i\\
i&0
\end{array}
\right)
\,,\,\sigma_{z}=\left(\begin{array}{cc}
1&0\\
0&-1
\end{array}
\right)\,.
\ee

\subsubsection{Real-time (matrix) representation of $\chi$}

The most straightforward transcription of the closed-time contour formalism is obtained by mapping the function $\chi(1,1')$ onto the $2\times 2$-matrix \cite{RammerSmith1986},
\ben
\Bar{\Bar{\chi}}(1,1^{\prime})&=&\left(
\begin{array}{l}
\chi_{++}(1,1^{\prime})\quad\chi_{+-}(1,1^{\prime})\\
\chi_{-+}(1,1^{\prime})\quad\chi_{--}(1,1^{\prime})
\end{array}\,,
\right)\label{barbarchi}
\een
defined such that  the $jk$-component, $i,j=\pm$, is $\chi(1,1^{\prime})$ with $1$ lying on $c_{i}$ and $1^{\prime}$ lying on $c_{j}$ where $c_{+}$ ($c_{-}$) is the upper (lower) part of the closed-time contour (see Fig.\ref{figure1}).
Explicitly, using Eq.(\ref{d2omegadphi2}), the matrix components are
\be
\chi_{++}(1,1^{\prime})&=&\frac{1}{2}\llangle{\cal{T}}_{c}\left[N(1),N(1^{\prime})\right]_{PB}\rrangle-\frac{i}{s}C(1,1')\\
\chi_{--}(1,1^{\prime})&=&\frac{1}{2}\llangle{\cal{T}}_{a}\left[N(1),N(1^{\prime})\right]_{PB}\rrangle-\frac{i}{s}C(1,1')\\
\chi_{+-}(1,1^{\prime})&=&-\frac{1}{2}\llangle\left[N(1),N(1')\right]_{PB}\rrangle-\frac{i}{s}C(1,1')\\
\chi_{-+}(1,1^{\prime})&=&\frac{1}{2}\llangle\left[N(1),N(1')\right]_{PB}\rrangle-\frac{i}{s}C(1,1')
\ee
where ${\cal{T}}_{c}$ and ${\cal{T}}_{a}$ are the chronological and anti-chronological time-ordering operators on the real-time axis $[t_{0},+\infty]$.
This matrix representation has, however, the disadvantage that it does not explicitly take account of the linear dependence of its components; for instance, 
\be
\chi_{++}+\chi_{--}=\chi_{+-}+\chi_{-+}\/.
\ee
It is possible to chose another representation that not only removes this redundancy of information but also involves the physically relevant quantities.
Indeed we note that simple linear combinations of the $\chi_{jk}$'s yield the correlation and response functions of interest,
\be
\chi^{R}&=&\chi_{++}-\chi_{+-}=\chi_{-+}-\chi_{--}\\
\chi^{A}&=&\chi_{++}-\chi_{-+}=\chi_{+-}-\chi_{--}\\
\frac{2}{is}C&=&\chi_{+-}+\chi_{-+}=\chi_{++}+\chi_{--}\,.
\ee
Using these relations, it is straightforward to show that the orthogonal transformation \cite{Keldyshtransformation} defined as $\hat{\chi}=Q\Bar{\Bar{\chi}}Q^{\dagger}$ with the orthogonal matrix $Q=(1-i\sigma_y)/\sqrt{2}$ defines the equivalent real-time representation,
\ben
\D\hat{\chi}=Q\Bar{\Bar{\chi}}Q^{\dagger}=\left(
\begin{array}{ccc}
\D0&&\chi^{A}\\\\
\D\chi^{R}&&\frac{2}{is}{\cal{C}}
\end{array}
\right)\,,
\label{hatchi}
\een
which has the remarkable property that it depends on the correlation function $C$ and response functions $\chi^{R,A}$ only.

\subsubsection{From closed-time to real-time integrals}

The theory derived in Sec.\ref{sectionIII} involves integrals over the closed-time contour of the form $\oint{d2\Sigma(1,2)\chi(2,1')}$.
Here we provide useful identities to express such quantities in terms of integrals along the real axis (see e.g. \cite{KrempSchlangesKraeft}.)

We remark that the quantities $\chi$ and $\Sigma$ defined before belong to the class of functions of two closed-time variables of the form
\ben
f(\tau,\tau')&=&f^{0}(\tau')\big[\delta_{c}(\tau'-\tau_i)-\delta_{c}(\tau'-\tau_f)\big]\label{Keldyshfunction}\\
&+&f^{\delta}(\tau)\delta_{c}(\tau-\tau')+\theta(\tau-\tau')f^{>}(\tau,\tau')+\theta(\tau'-\tau)f^{<}(\tau,\tau')\,,\nn
\een
where for simplicity we do not write the possible dependence on phase-space variables.
In matrix form, $f$ can be represented by $\Bar{\Bar{f}}$ as in Eq.(\ref{barbarchi}) or by $\hat{f}=Q\Bar{\Bar{f}}Q^\dagger$ as in Eq.(\ref{hatchi}), namely
\be
\D\hat{f}=\left(
\begin{array}{ccc}
\D0&&f^{A}\\\\
\D f^{R}&&\frac{2}{is}{f^{C}}
\end{array}
\right)
\ee
where the direct calculation of the retarded (R), advanced (A) and correlation (C) components in physical time gives the following expressions,
\bs\label{fRAc}
\ben
f^{R}(t,t')&=&f^{\delta}(t)\delta(t-t')+\theta(t-t')\left[f^{>}(t,t')-f^{<}(t,t')\right]\label{fR}\\
f^{A}(t,t')&=&f^{\delta}(t)\delta(t-t')-\theta(t'-t)\left[f^{>}(t,t')-f^{<}(t,t')\right]\label{fA}\\
\frac{2}{is}f^{C}(t,t')&=&2f^{(0)}(t')\delta(t'-t_0)+\left(f^{>}(t,t')+f^{<}(t,t')\right)\label{fC}\,,
\een
\es
In the special case $f=\chi$, equations (\ref{fRAc}) give back $f^{R,A}=\chi^{R,A}$ and $f^{C}=C$.
With $f=\Sigma$, the retarded and advanced terms $\Sigma^{A,R}$ contain the singular part $\Sigma^{\delta}$ that arises from static correlations, while $\Sigma^{C}$ contains the part $\Sigma^{(0)}$ due to the initial conditions (see Eqs.(\ref{SigmaRAtot}-\ref{SigmaC}) below).
Finally, for the delta function $\delta_c$ defined by Eq.(\ref{deltacDirac}), we obtain the matrix representation $\hat\delta_c(\tilde 1,\tilde 1')=\frac{1}{t'(\tau)}\delta(1-1')\sigma_{z}$.

We are interested in integrals of the kind
\begin{eqnarray}
h(\tau,\tau')=\oint{d\bar{\tau}f(\tau,\bar\tau)g(\bar\tau,\tau')}\,,
\end{eqnarray}
where $f$ and $g$ are of the form (\ref{Keldyshfunction}).
It is easy to show that $h(\tau,\tau')$ also belongs to the class (\ref{Keldyshfunction}).
By splitting the closed-time contour integral into integrals over the forward and backward branches, namely $\oint_{c}{d\tau}=\int_{t_{0}}^{\infty}{dt}-\int_{t_{0}}^{\infty}{dt}$, it is straigthforward to derive the following relations,
\ben
h^{R}(t,t')&=&f^{R}\cdot g^{R}(t,t')=\int_{t_{0}}^{\infty}{dt''f^{R}(t,t'')g^{R}(t'',t')}\nn\\
h^{A}(t,t')&=&f^{A}\cdot g^{A}(t,t')=\int_{t_{0}}^{\infty}{dt''f^{A}(t,t'')g^{A}(t'',t')}\nn\\
h^{C}(t,t')&=&f^{C}\cdot g^{R}(t,t')+f^{R}\cdot g^{C}(t,t')\,. \label{cK}
\een
In matrix form, those properties can be summarized as
\be
\Bar{\Bar{h}}(1,1')=\Bar{\Bar{f}}(1,2)\sigma_{z}\Bar{\Bar{g}}(2,1')\quad,\quad\hat{h}(1,1')=\hat{f}(1,2)\sigma_{x}\hat{g}(2,1')\,,
\ee
with
\be
\hat{h}=\left(
\begin{array}{ccc}
0 & & f^A\cdot g^A\\
f^R\cdot g^R & & \frac{2}{is}\left(f^R\cdot g^C+f^C\cdot g^A\right)
\end{array}
\right)\,.
\ee 

\subsection{Equations of motion} \label{sectionIV2}

With the help of the previous identities, we can easily re-express the evolution equation (\ref{dchidtausigma}) of $\chi(\tilde 1,\tilde 1')$ into equations for the physically relevant quantities $C(1,1')$ and $\chi^{R,A}(1,1;)$ in real time.
In matrix form, Eq.(\ref{dchidtausigma}) writes
\ben
\left(\hat{G}-\hat{\Sigma}^{mf}-\hat\Sigma\right)\sigma_{z}\hat{\chi}(1,1')&=&\left(
\begin{array}{cc}
G-\Sigma^{mf}-\Sigma^{A}&0\\
-\frac{2}{is}\Sigma^{C}&G-\Sigma^{mf}-\Sigma^{R}
\end{array}
\right)\cdot\hat{\chi}(1,1')\nn\\
&=&\{\delta(1-1'),f(1)\}\,\sigma_{x} \label{matrixformrealtimeequations}
\een
where we recall that the phase-space density $f(1)$ evolves according to Eq.(\ref{equationf}).
In Eq.(\ref{matrixformrealtimeequations}), we introduced the short-hand notations,
\be
G(1,2)&=&-\frac{\partial}{\partial t_{2}}\delta(1-2)+\left\{h_{0}(2)+\phi_0(2),\delta(1-2)\right\}(2)\\
\Sigma^{mf}(1,1')&=&\gamma_3(1,2,3)f(2)\delta(1'-3) 
\ee
Explicitly in terms of the components, we obtain the coupled evolution equations for the correlation and response functions \cite{Noteofadjointness},
\bs \label{geenralCchiRchiA}
\ben
\left[\frac{\partial}{\partial t}-L_1\right]C(1,1')&-&\int_{t_{0}}^{\infty}{dt_2\int{dX_2\,\Sigma_{tot}^{R}(1,2)C(2,1')}}\label{geenralCchiRchiAC}\\
&=&\int_{t_{0}}^{\infty}{dt_2\int{dX_2\,\Sigma^{C}(1,2)\chi^{A}(2,1')}}+\delta(t-t_0)C(1,1')\nn\\
\left[\frac{\partial}{\partial t}-L_1\right]\chi^R(1,1')&-&\int_{t_0}^{\infty}{dt_2\int{dX_2\,\Sigma_{tot}^{R}(1,2)\chi^R(2,1')}}\label{geenralCchiRchiAR}\\
&=&\left\{\delta(1-1'),f(1)\right\}(1)\nn\\
\left[\frac{\partial}{\partial t}-L_1\right]\chi^A(1,1')&-&\int_{t_0}^{\infty}{dt_2\int{dX_2\,\Sigma_{tot}^{A}(1,2)\chi^A(2,1')}}\label{geenralCchiRchiAA}\\
&=&\left\{\delta(1-1'),f(1)\right\}(1)\nn
\een
\es
Here we have combined the mean-field contribution $\Sigma^{mf}$ to $\Sigma^{R,A}$, 
\ben
\Sigma_{tot}^{R,A}(1,1')&=&\Sigma^{mf}(1,1')+\Sigma^{R,A}(1,1')\nn\\
&=&\Sigma_{tot}^{\delta}\delta(t-t')\pm\Theta\big(\mp(t-t')\big)\big(\Sigma^{>}(1,1')-\Sigma^{<}(1,1')\big)\nn\\
&\equiv&\Sigma_{tot}^{\delta}\delta(t-t')\pm\Theta\big(\mp(t-t')\big)\Sigma^{\triangle}(1,1')\label{SigmaRAtot}
\een
and, according to Eq.(\ref{fC}), $\Sigma^{C}$ is given by
\ben
\Sigma^{C}(1,1')&=&is\Sigma^{(0)}(X_1,X_1^\prime;t)\delta(t'-t_0)+\frac{is}{2}\big(\Sigma^{>}(t,t')+\Sigma^{<}(t,t')\big)\nn\\
&\equiv&\Sigma^{C,0}(X_1,X_1^\prime;t)\delta(t'-t_0)+\Sigma^{C,reg}(t,t')\label{SigmaC}\,,
\een
where in the last expression we introduce notations where $is$ does not appear explicitely since indeed $is\Sigma^{(0)}$ and $is(\Sigma^>-\Sigma^<)$ are independent of $s$ (see e.g. Eq.(\ref{physicalCfromchi>chi<}) for a similar property.)
With Eqs.(\ref{SigmaRAtot}-\ref{SigmaC}), the evolution equations (\ref{geenralCchiRchiA}) can be further details as follows:
for all $t,t'\geq t_{0}$,
\bs\label{setCchiRchiA}
\ben
\lefteqn{\left[\frac{\partial}{\partial t}-L_1\right]C(1,1')}&&\label{setCchiRchiA1}\\
\D&-&\int{dX_2\,\Sigma_{tot}^{\delta}(X_1,X_2;t_1)C(X_2,t_1,1')}-\int_{t_{0}}^{t}{dt_2\int{dX_2\,\Sigma^{\triangle}(1,2)C(2,1')}}\nn\\
\D&=&\int{dX_2\,\Sigma^{C,0}(X_1,X_2;t_1)\chi^{A}(X_1,t_0;1')}+\int_{t_{0}}^{t'}{dt_2\int{dX_2\,\Sigma^{C,reg}(1,2)\chi^{A}(2,1')}}\nn\\
&&\hspace{1cm}+\delta(t-t_0)C(1,1')\nn
\een
and for all $t_{0}\leq t'\leq t$,
\ben
\lefteqn{\left[\frac{\partial}{\partial t}-L_1\right]\chi^R(1,1')}&&\label{setCchiRchiA2}\\
\D&-&\int{dX_2\,\Sigma_{tot}^{\delta}(X_1,X_2;t_1)\chi^R(X_2,t_1,1')}-\int_{t'}^{t}{dt_2\int{dX_2\,\Sigma^{\triangle}(1,2)\chi^R(2,1')}}\nn\\
\D&=&\big\{\delta(1-1'),f(1)\big\}(1)\nn
\een
and for all $t_{0}\leq t\leq t'$,
\ben
\lefteqn{\left[\frac{\partial}{\partial t}-L_1\right]\chi^A(1,1')}&&\label{setCchiRchiA3}\\
\D&-&\int{dX_2\,\Sigma_{tot}^{\delta}(X_1,X_2;t_1)\chi^A(X_2,t_1,1')}+\int_{t'}^{t}{dt_2\int{dX_2\,\Sigma^{\triangle}(1,2)\chi^A(2,1')}}\nn\\
&=&\big\{\delta(1-1'),f(1)\big\}(1)\nn
\een
\es
In the remaining of this section, we discuss general properties of Eq.(\ref{geenralCchiRchiA}) and make contact with previous approaches.
The full closure of the set of equations for $f$,$\chi^{R,A}$ and $C$ in real-time is completed in Sec.\ref{sectionV}, where we transcribe the closure relation (\ref{Sigmaf}) in terms of $\Sigma^{R,A,C}$ and $f,\chi^{R,A},C$.

\subsection{Discussion}

\paragraph{Initial conditions.}
Equations (\ref{equationf}) and (\ref{setCchiRchiA1}-\ref{setCchiRchiA3}) specify an initial-value problem for which $t=t'=t_0$ is the initial time.
They require initial conditions for $f(X,t_0)$, $\chi^{R}(X,t_0;X',t_0)=-\chi^{A}(X,t_0;X',t_0)$ and $C(X,t_0;X't_0)$.
Since $\chi^{R,A}$ must comply with Eq.(\ref{EqualtimechiRA}), i.e. 
\be
\chi^{R/A}(X,t_0;X',t_0)=\mp\left\{\delta(X-X'),f(X,t_0)\right\}(X)\,,
\ee
their initial value is fixed when $f(X,t_0)$ is given.
Therefore, the initial conditions are completely specified by the given of $f(X,t_0)$ and $C(X,t_0;X't_0)$.

\paragraph{Reversibility.} The equations (\ref{setCchiRchiA1}-\ref{setCchiRchiA3}) form a set of causal equations with time history integrals.
The presence of those memory integrals is a property of the exact statistical dynamics defined in Sec.\ref{sectionII} and, despite their formal resemblance with (generalized) Langevin equations, they should be distinguished from phenomenological non-equilibrium equations.
Thus the complete information on time reversibility and other symmetries of the microscopic Hamilton dynamics resides in the time dependent, non-local memory kernels: the equations themselves do not single out a direction of time.
The transition to irreversible dynamics can be induced by approximation, e.g., as alluded earlier, when neglecting the dependence on the initial conditions (Bolgolyubov's condition).

\paragraph{Many-body effects.} The original non-linearity of the microscopic equations (\ref{accurateKlimontovich}) has become separated into several effects \cite{followrosediscussion}.
Given a certain level of fluctuations, $C(1,1')$, the term $\Sigma_{tot}^{R}$ can cause them to grow or decay, and when viewed as a matrix in its momentum indices, transfer fluctuations from of component of $C$ to another.
The singular part $\Sigma^{mf}+\Sigma^{\delta}$ is related to the instantaneous effect of the mean-field and of the correlations, respectively, while the regular part $\Sigma^{\triangle}$ describes non-Markovian (i.e. delayed) correlation effects.
Another effect contained $\Sigma_{tot}^{R}$ is the renormalization of the spectrum of fluctuations by the many-body interactions.

The term on the left-hand side of Eq.(\ref{setCchiRchiA1}) can be interpreted as follows (a more complete discssion is given in Sec.\ref{formalsolutionC}.)
Given a fluctuation $\delta N$ of the density around its average $f$, let $\delta\phi(1)$ be the potential defined such as
\be
\delta N(1)=\chi^{R}(1,2)\delta\phi(2)\,.
\ee
In other words, we imagine that $\delta N$ can be created by disturbing the system by the external potential $\delta\phi$.
From Eq.(\ref{geenralCchiRchiA}), we have
\be
\left(G-\Sigma^{mf}-\Sigma^{R}\right)(1,2)\cdot \delta N(2)=\{\phi(1),f(1)\}\,.
\ee
Multiplying the previous equation by $\delta N(1')=\chi^{R}(1',3)\delta\phi(3)=\chi^{A}(3,1')\delta\phi(3)$ and ensemble averaging yields
\be
\left(G-\Sigma^{mf}-\Sigma^{R}\right)(1,2)\cdot C(1,1')=\{\langle \phi(1)\phi(3)\rangle,f(1)\}\cdot\chi^{A}(3,1')\,.
\ee
Comparing this with Eq.(\ref{geenralCchiRchiA}), we may interpret $\Sigma^{C}$ as an effective random source of potential fluctuations.
Since, as we shall find later, $\Sigma^{C}$ includes terms which are quadratic in $C$, this term can be interpreted as a source of non-linear noise which drives the fluctuations $C$, and couples modes of different wave numbers.

More on those properties will be discussed in Sec.\ref{formalsolutionC}.
In the next two subsections, we compare the results derived so far with previous works, starting with fluids at equilibrium and then with non-equilibrium kinetic theories.
The comparison with those previous works is not thorough and would certainly deserve a more detailed analysis.

\subsection{Equilibrium limit, fluctuation-dissipation theorem and detailed balance} \label{sectionVI}

When considering a system at equilibrium, we generally implicitly set the initial time $t_0=-\infty$.
As a consequence, statistical averages like the correlation and response functions are invariant under time translation, e.g. \cite{equilibriumtheory}
\be
C(t_1,t_1^\prime)=C(t_1+s,t_1^\prime+s)\quad,\quad\forall s\in\mathbb{R}\,,
\ee
and are functions of the time difference $t=t_1-t_1^\prime$, i.e.
\be
C(1,1')\equiv C_{eq}(X_1,X_1^\prime;t_1-t_1^\prime)\quad,\quad \forall t_1,t_1^\prime\,.
\ee
As recalled in Sec. \ref{sectionII}, a fundamental property of equlibrium fluids is the fluctuation-dissipation theorem, which simply relates $C_{eq}$ and $\chi_{eq}^{R,A}$ as
\ben
\chi_{eq}^{R}(X,X';t)-\chi_{eq}^{A}(X,X';t)=-\beta\frac{\partial}{\partial t}C_{eq}(X,X';t)\,.\label{SigmaCRA}
\een
with the inverse temperature $\beta=1/K_BT$.
By rewriting the evolution equations (\ref{setCchiRchiA}) in terms of $t=t_1-t_1^\prime$, it is straigthforward to show that these equations are consistent with the fluctuation-dissipation theorem if and only if the equilibrium memory function kernels satisfy a relation similar to Eq.(\ref{SigmaCRA}) \cite{sigmaC0cancels},
\ben
\Sigma_{eq}^{R}(X,X';t)-\Sigma_{eq}^{A}(X,X';t)&=&-\beta\frac{\partial}{\partial t}\Sigma_{eq}^{C,reg}(X,X';t)\,.\label{SigmaCRA}
\een
This relation describes detailed balance of the collision mechanisms that control the occurrence of spontaneous fluctuations and their damping.

It turns out that the evolution equations obtained by setting $t_0=-\infty$ and $t=t_1=t_1^\prime$ in Eqs.(\ref{setCchiRchiA}) are rather cumbersome to manipulate.
In the study of the dynamics of equilibrium correlation functions, it is more customary to rather work with the following {\it explicit} expression of the correlation function,
\ben
C_{M}(X,X';t)=C(X,t;X',0)=\llangle\delta N(X,t)\delta N(X',0)\rrangle\,.\label{CMdefinition}
\een
This expression is one possible representation of $C_{eq}(X,X',t)$ (corresponding to $s=0$ in $C(t+s,s)=C_{eq}(t)$.); contrary to $C_{eq}$, $C_{M}$ favors an initial time, $t_0=t'=0$.
Previous works on the equilibrium kinetic theory of fluids and particular Mazenko's theory \cite{Mazenko1974} discussed in the introduction are theories for $C_{M}(t)$.
In our approach, the evolution equation for $C_{M}$ can simply be obtained by setting both the initial time $t_0$ and the time $t'$ to zero in Eq.(\ref{setCchiRchiA1}).
We immediately obtain $\forall t>0$,
\be
\left[\frac{\partial}{\partial t}-L_{1}(X)\right]C_M(X,X';t)&-&\int{dX_2\,\Sigma_{tot}^{\delta}(X;X_2;t)C_M(X_2,X',t)}\\
&&\hspace{-1cm}-\int_{0}^{t}{d\bar{t}\int{dX_2\,\Sigma^{\triangle}(X,t;X_2,\bar{t})C_M(X_2,X';\bar{t})}}=0\,.
\ee
This evolution equation has the well-known form of memory function equations in the equilibrium theories \cite{equilibriumtheory}.
In particular, it corresponds to Mazenko's memory function $\phi$ defined in \cite{Mazenko1974} with the identification,
\be
\Sigma_{tot}^{\delta}(X,\bar{X};t)&=&-\phi^{s}(X,\bar{X})\\
\Sigma^{\triangle}(X,t;X',t')&=&-\phi^{c}(X,X';t-t')\,.
\ee

\subsection{Contact with other non-equilibrium kinetic theories}  \label{sectionVII}

In this section, we explore the correspondence with previous non-equilibrium kinetic theories.
First we make contact with some of the most popular closures of the BBGKY hierarchy.
We shall see that those closures systematically neglect the effects of the memory function kernels $\Sigma^{\triangle, C}$.
Then we compare our results to the theory of Rose \cite{Rose1979} mentioned in the introduction.

\subsubsection{Reduction to popular BBGKY closures} \label{sectionpopularclosures}

\begin{table}[t]
\setlength{\extrarowheight}{.1cm}
\begin{center}
\begin{tabular}{m{2.6cm}||c|c||c|c||c|c|c}
{\bf Closure} & $A$ & $B$ & $C$ & $D$ & $\Sigma^{\delta}$ & $\Sigma^{\triangle}$ & $\Sigma^{C}$ \\\hline
Landau  & $\times$ & $0$ & $0$ & $0$ & 0 & 0 & 0 \\\hline
LBG & $\times$ & $\times$ & $0$ & $0$ & 0 & 0 & 0 \\\hline
Boltzmann & $\times$ & $0$ & $\times$ & $0$ & Eq.(\ref{approximationsigmadeltaboltzmann}) & 0 & 0 \\\hline
Book-Frieman \cite{BookFrieman} & $\times$ & $\times$ & $\times$ & 0 & Eq.(\ref{approximationsigmadeltaboltzmann}) & 0 & 0
\end{tabular}
\caption{Left side: terms kept ($\times$) and dropped ($0$) in the second BBGKY equation (\ref{secondBBGKY}) for the Landau, Boltzmann, Lenard-Balescu-Guernsey and Book-Frieman closures. The latter, which was proposed to improve the close collisions in the Lenard-Balescu-Guernsey equation, simply amounts to setting the three-body correlation $g_3$ equal to zero. The same closures are expressed in terms of the cumulants $C^{(n)}$ in table \ref{tableclosure}.
Right side: in the approach developed in this paper, the approximations for $C$ and $D$ correspond to the approximations for the memory functions $\Sigma^{\delta,\triangle,C}$ given in the last three columns.
}
\label{tableclosure2}
\end{center}
\end{table}

The BBGKY hierarchy is a hierarchy for the equal-time correlation functions $g_n$ (for convenience, we give a short reminder of the BBGKY hierarchy in appendix \ref{appendixBBGKY}.)
Traditional closures of the BBGKY hierarchy are performed at the level of the evolution equation for $g_2$, which reads
\ben
\lefteqn{\hspace{-.8cm}\left[\frac{\partial}{\partial t}-\left(L_{1}+L_{1'}\right)\right]g_{2}(1,1')=\underbrace{L_{11'}f(1)f(1')}_{A}}&&\label{secondBBGKY}\\
&&\hspace{-.8cm}+\underbrace{\!\int{\!\!d2 L_{12} f(1)g_{2}(1',2)+L_{1'2} f(1')g_{2}(1,2)+\left[L_{12}+L_{1'2}\right]f(2)g_{2}(1,1')}}_{B}\nn\\
&&\hspace{-.8cm}+\,\underbrace{L_{11'}g_2(1,1')}_{C}+\underbrace{\int{d2 \left(L_{12}+L_{1'2}\right)g_{3}(1,1',2)}}_{D}\,,\nn
\een
where, throughout this sub-section, we write $1=({\bf r},{\bf p})$ and all quantities are evaluated at the same time $t=t_1=\dots=t_n$.
They rely, among other hypothesis, on an assumed ordering of the correlations $g_n$ in terms of an adequately chosen small parameter $\lambda$ \cite{Balescu1997},
\be
g_{n}=O(\lambda^{n})\,.
\ee
The Landau, Boltzmann and Lenard-Balescu-Guernsey closures give evolution equations for $f$ that are second order in $\lambda$.
The Landau closure is valid for so-called weakly coupled systems and $\lambda$ is the dimensionless strength of the potential ($v(r)=\lambda \bar v(r)$ in Eq.(\ref{HN}).)
The perturbation parameter in the Boltzmann equation is the density parameter $\lambda=n l_{C}^{3}$ where $n$ is the particle density and $l_{C}$ is the correlation length.
The Lenard-Balescu-Guernsey equation is an equation for weakly-coupled plasmas and $\lambda$ is the so-called plasma parameter $\lambda=q^2/a k_{B}T$ ($q$ is the particles charge, $a=n^{-1/3}$ is the interparticle spacing.)
In the Lenard-Balescu-Guernsey equation, the plasma is represented as a weakly coupled system in which the collisions are due to the interactions via an effective, dynamically screened potential.
For those popular closures, table \ref{tableclosure2} shows the terms that are kept ($\times$) and dropped ($0$) in the evolution equation (\ref{secondBBGKY}).

In our approach, the equation for $g_2(1,1')=C(1,1')-\delta(1-1')f(1)$ can be easily obtained from Eq.(\ref{setCchiRchiA1}) as
\bs\label{g2formtheory}
\ben
\lefteqn{\hspace{-.8cm}\left[\frac{\partial}{\partial t}-\left(L_{1}+L_{1'}\right)\right]g_{2}(1,1')=L_{11'}f(1)f(1')}&&\label{g2formtheorya}\\
&&\hspace{-.8cm}+\!\int{\!\!d2 L_{12} f(1)g_{2}(1',2)+L_{1'2} f(1')g_{2}(1,2)+\left[L_{12}+L_{1'2}\right]f(2)g_{2}(1,1')}\label{g2formtheoryb}\\
&&\hspace{-.8cm}+\int{d2 \Sigma^{\delta}(12)g_{2}(2,1')+\Sigma^{\delta}(1'2)g_{2}(2,1)}\label{g2formtheoryc}\\
&&\hspace{-.8cm}+\int{d2 \Sigma^{\triangle}(12)g_{2}(2,1')+g_{2}(1,2)\Sigma^{\triangle}(21')}\label{g2formtheoryd}\\
&&\hspace{-.8cm}+\int{d2 \Sigma^{C}(12)\chi^A(2,1')+\chi^R(1,2)\Sigma^{C}(21')}\,,\label{g2formtheorye}
\een
\es
where $t=t'=t_2$ is implied.
We note that in the rhs, the terms on lines (\ref{g2formtheorya}) and (\ref{g2formtheoryb}), which are just the $A$ and $B$ terms of Eq.(\ref{secondBBGKY}), come from the mean-field term $\Sigma^{mf}$, while all the other terms come from memory functions $\Sigma^{R,A,C}$.
The term $C+D$ of Eq.(\ref{secondBBGKY}) are replaced by the last three lines in Eq.(\ref{g2formtheory}).

In order to appreciate better the physics contained in those memory function terms, we translate the tradiational closures discussed before in terms of approximations of Eq.(\ref{g2formtheory}); the results are summarized in table \ref{tableclosure2}.

Both the Landau and Lenard-Balescu-Guernsey closures discard all the terms involving the memory functions, which amounts to setting $\Sigma^{\delta,\triangle,C}=0$ (Landau retains Eq.(\ref{g2formtheorya}) only and Landau-Balescu-Guernsey discards the terms (\ref{g2formtheoryc}) through (\ref{g2formtheorye}).)
This simply tells us that memory functions will account for effects not accounted in those closures, for instance by adding the effects of static ($\Sigma^{\delta}$) and dynamic ($\Sigma^{\triangle}$) correlations on the effective, dynamically screened interparticle potential through which particles mutually interact.

The Boltzmann and Book-Frieman equations are recovered by (i) discarding Eq.(\ref{g2formtheoryb}), (ii) setting $\Sigma^{\triangle}=\Sigma^{C}=0$, and (iii) using the following approximation for $\Sigma^{\delta}$,
\ben
\Sigma^{\delta}(11')&=&\gamma_{3}(123)\delta(1'-2)\delta(1'-3)=\frac{\partial \tilde v(1-1')}{\partial {\bf r}}\cdot \frac{\partial \delta(1-1')}{\partial {\bf p}}\,,\label{approximationsigmadeltaboltzmann}
\een
which, together with Eq.(\ref{g2formtheoryc}) yields the term $C$ of the second BBGKY equation (\ref{secondBBGKY}) that describes bare two-particle interactions (uncorrelated binary collisions).
The approximation (\ref{approximationsigmadeltaboltzmann}) can be obtained using Eq.(\ref{Sigmasingularpart}) together with the approximation $C(11')\approx \delta(1-1')f(1)$ (see table \ref{tableclosure}.)
Indeed, we obtain
\be
\{\Sigma^{\delta}(1,1'),f(1')\}(1')&=&\gamma_{3}(123)\{\delta(1'-3),\delta(1'-2)f(1')\}(1')\\
&=&\gamma_{3}(123)\{\delta(1'-3)\delta(1'-2),f(1')\}(1')\,,
\ee
which implies Eq.(\ref{approximationsigmadeltaboltzmann}).

In summary, this quick comparison reveals that inclusion of any coherent approximation for the memory functions $\Sigma^{R,A,C}$ will bring in physics information that is systematically neglected in traditional closures.

\subsubsection{Connection with MSR and Rose's theories} \label{contactwithMSRRose}

\paragraph{General remarks.}
The present theory retains two main imprints of quantum mechanics, namely the classical amplitude $e^{\frac{i}{s}{\cal{S}}}$ with the action parameter $s$ and the closed-time contour idea of Schwinger.
In contrast, MSR \cite{MartinSiggiaRose} is based on the concepts of canonical quantization using field doubling via conjugate response field and, in Rose \cite{Rose1979}, on the second quantization using the occupation number representation in phase-space.
Thus, MSR and Rose's theories recast the classical problem in a form identical to quantum mechanics, while our approach should be seen as a classical limit of the quantum theory.

Another difference is in the nature of the external coupling .
In MSR and Rose, as usually is the case in quantum field theory \cite{Browntextbook}, the coupling corresponds to a source term in the evolution equation of the field, while here the coupling is an additional term in the Hamiltonian.
The response function in MSR and Rose thus describes the response to an infinitesimal source of particles while in the present theory it describes the response to a disturbance produced by an externally applied force field.
Since all three theories use similar renormalization techniques based on a generating functional and a Legendre transform, the equations obtained look alike.
As we shall see, our results encompass the results of Rose's renormalized kinetic theory \cite{Rose1979} for any initial conditions.
The ingenious formal devices introduced by MSR and Rose naturally arise in the present formalism.

\paragraph{Technical comparison.}
Briefly speaking, Rose's theory is a theory for the correlation function $C$ and the response function $R$ that measures the linear response of the phase-space density to an infinitesimal source term $\eta$ in the field equation (\ref{equationf}), i.e.
\ben
\left[\frac{\partial}{\partial t_1}-L_{1}\right]f(1)-\{u^{mf}(1),f(1)\}-\frac{1}{2}\gamma_{3}(1,2,3)C(2,3)=\eta(1)\label{equationfRose}
\een
and
\be
R(1,1')=\frac{\delta f(1)}{\delta\eta(1')}\big|_{\eta=0}\,,
\ee
The function $R$ can also be regarded as the Green's function of the linearized version of Eq.(\ref{equationf}).

On the other hand, in our theory $\chi^R$ describes the linear response to a variation $\phi_{0}+\delta \phi_{0}$ of the external potential.
In the presence of $\delta\phi_0$, the evolution equation (\ref{equationf}) can be regarded as Eq.(\ref{equationfRose}) with the source term
\be
\eta(1)=\left\{\delta \phi_{0}(1),f(1)\right\}\,.
\ee
which has the particularity of conserving the number of particles (since the momentum integral vanishes.)
In the linear regime, we can write
\be
\delta f(1)&=&\chi^{R}(1,2)\delta \phi_{0}(2)\\
&&=R(1,2)\eta(2)=-\big\{R(1,2),f(2)\big\}(2)\delta \phi_0(2)
\ee
and therefore,
\ben
\chi^{R}(1,1')=-\big\{R(1,1'),f(1')\big\}(1')\,. \label{chiRR}
\een
When introduced into the evolution equations (\ref{geenralCchiRchiA}) for $\chi^{R,A}$ and $C$, we find
\ben
\left[\frac{\partial}{\partial t}-L_1(1)\right]R(1,1')-\int{d2\,\Sigma_{tot}^{R}(1,2)R(2,1')}&=&\delta(1-1')\label{equationR}
\een
and
\ben
\left[\frac{\partial}{\partial t}-L_1(1)\right]C(1,1')&-&\int{d2\,\Sigma_{tot}^{R}(1,2)C(2,1')}\label{equationCR}\\
&=&\int{d2\,\Big\{\Sigma^{C}(1,2),f(2)\Big\}(2)R(1',2)}+\delta(t-t_0)C(1,1')\nn
\een
These equations are equivalent to the (Dyson) equations (63) and (64) of \cite{Rose1979} derived by H. Rose, with the correspondence
\ben
\Sigma_{-+}(1,2)&=&\Sigma_{tot}^{R}(1,2)\nn\\
\Sigma_{--}(1,2)&=&\Big\{\Sigma^{C,reg}(1,2),f(2)\Big\}(2) \label{sigmatrose--}\\
0&=&\Sigma^{C,0}\nn\,.
\een
The last equation is the consequence of the fact that Rose assumes Gaussian initial conditions.
Although as noted by Rose in his appendix A, formalisms developed by Deker \cite{Deker} and more recently by Andersen \cite{Andersen2000} exist to extend their validity to general initial conditions, we have not performed the extension ourself yet.
The closure relations, i.e. the functional equations for the memory functions, in Rose and our formalism (see Sec.\ref{sectionV}) look quite different (simply because they involve different starting points and ingredients) and are thus difficult to compare.
Moreover, Rose points out that the closure relations (52)-(62) of his paper are ``exact if and only if the initial conditions satisfy a Wick-type theorem'', i.e. are Gaussian.
Nevertheless, assuming Gaussian initial correlations in our formalism, the lowest order approximations in both approaches are the same (see Sec.\ref{sectionV}.)
At this point, it is not clear which approach is more handy when it comes to practical calculations.

\subsection{Formal solution for $C(1,1')$. Dependence on initial conditions.} \label{formalsolutionC}

In this paragraph, we show that it is possible to solve at least formally the evolution equation Eq.(\ref{geenralCchiRchiAC}) for the correlation function $C(1,1')$ in terms of the response functions $\chi^{R,A}$ and Green's function $R$, respectively.
When introduced in the collision integral Eq.(\ref{equationf}), this solution may be used to build a kinetic equation for the distribution function $f$.
In its simplest approximation, this approach corresponds to the quasi-linear theory used for instance in plasma physics to derive the Lenard-Balescu equation \cite{LandauLifshitz10}.
The solution given below is exact and takes care of the initial conditions.

\paragraph{General expression.} Using the definitions of Sec.\ref{SectionIV1}, it is straightforward to show that the inverse $\Gamma$ of the closed-time response function $\chi$ defined by Eq.(\ref{chigammainverse}) can be represented as \cite{matrixforgamma},
\be
\hat\Gamma=\left(
\begin{array}{cc}
0&\Gamma^{A}\\
\Gamma^{R}&\frac{2}{is}\Gamma^{C}
\end{array}
\right)\,,
\ee
where the components $\Gamma^{R}$ and $\Gamma^{A}$ merely are the inverse of the retarded (advanced) response function $\chi^{R}$ amd $\chi^{A}$, respectively, i.e.
\ben
\begin{array}{l}
\Gamma^{R}(1,2)\cdot\chi^{R}(2,1')=\chi^{R}(1,2)\cdot\Gamma^{R}(2,1')=\delta(1-1')\\
\Gamma^{A}(1,2)\cdot\chi^{A}(2,1')=\chi^{A}(1,2)\cdot\Gamma^{A}(2,1')=\delta(1-1')
\end{array}
\,,\label{chiRAgammaRA}
\een
and $\Gamma^{C}$ is related to the correlation function as,
\ben
\Gamma^{C}(1,1')&=&-\Gamma^{R}(1,2) C(2,3)\Gamma^{A}(3,1') \label{gammaCC}\/.
\een
Equation (\ref{gammaCC}) implies the following formal solution of its evolution equation (\ref{geenralCchiRchiAC})
\ben
\hspace{-.5cm}C(1,1')&=&-\chi^{R}(1,2) \Gamma^{C}(2,3)\chi^{A}(3,1')=-\chi^{R}(1,2)\chi^{R}(1',3)\Gamma^{C}(2,3)\label{gammaCCinverse}
\een
in terms of $\chi^{R,A}$ and of the quantity $\Gamma^{C}$.
Another expression can be obtained using the relation (\ref{chiRR}) between $\chi^{R}$ and the Green's function $R$ in Eq.(\ref{gammaCCinverse}) and yields
\ben
C(1,1')=R(1,2)R(1',3){\cal{S}}(2,3)\,, \label{spectralbalance}
\een
where we introduce the symmetric kernel,
\ben
{\cal{S}}(1,1')=-\Big\{\big\{\Gamma^{C}(1,1'),f(1)\big\},f(1')\Big\}\,. \label{calS}
\een
Equation (\ref{gammaCC}) can easily be verified by substitution into the evolution equation (\ref{geenralCchiRchiAC}), while Eq.(\ref{spectralbalance}) is solution of
\ben
\left[\frac{\partial}{\partial t}-L_1\right]C(1,1')&-&\int_{t_{0}}^{\infty}{dt_2\int{dX_2\,\Sigma_{tot}^{R}(1,2)C(2,1')}}\label{geenralCchiRchiACwithS}\\
&=&\int_{t_{0}}^{\infty}{dt_2\int{dX_2\,{\cal{S}}(1,2)R(1',2)}}+\delta(t-t_0)C(1,1')\nn\,,
\een
obtained by rewritting the rhs of (\ref{geenralCchiRchiAC}) with Eq.(\ref{chiRR}) \cite{calSsigmamm}.
They may be considered as a generalized form of the fluctuation-dissipation theorem in the sense that they express the overall level of density flucutations $C$ as balance between forcing (through $\Gamma^{C}$ or ${\cal{S}}$) and dissipation (encapsulated in $\chi^{R,A}$ or $R$) \cite{Krommes2002}.
In the next paragraph we show how Eqs.(\ref{gammaCC}-\ref{spectralbalance}) can be further split into more fundamental components.

\paragraph{Detailed expression.} 
Introducing Eq.(\ref{gammaCC}) into Eq.(\ref{geenralCchiRchiAC}) leads to the following relation between $\Gamma^{C}$ and $\Sigma^{C}$,
\ben
\big\{\Gamma^{C}(1,1'),f(1)\big\}(1')=\Sigma^{C}(1,1')+\delta(t-t_0)C(1,2)\Gamma^{A}(2,1')\,, \label{gammaCf}
\een
which, as proved in details in appendix \ref{appendixonGammaC}, imply the following structure for $\Gamma^{C}(1,1')$,
\ben
\Gamma^{C}(1,1')&=&\Gamma^{reg}(1,1')+\delta(t'-t_0)\Gamma^{\delta}(X_1,X_{1}^\prime;t)+\delta(t-t_0)\Gamma^{\delta}(X_1^\prime,X_1;t')\nn\\
&&\hspace{3cm}+\delta(t-t_0)\delta(t'-t_0)\Gamma^{(0)}(X_1,X_1^\prime)\,, \label{GammaCstructure}
\een
where $\Gamma^{reg,\delta,0}$ are given by
\bs\label{Gammareggammadelta}
\ben
\big\{\Gamma^{reg}(1,1'),f(1)\big\}(1')&=&\frac{is}{2}\Big(\Sigma^{>}(1,1')+\Sigma^{<}(1,1')\Big)\\
\big\{\Gamma^{\delta}(X_1,X_1^\prime;t),f(1)\big\}(1')&=&is\Sigma^{(0)}(X_1,X_1^\prime;t)
\een
\es
and
\be
\big\{\big\{\Gamma^{(0)}(X_1,X_1^\prime),f(X_1,t_0)\big\},f(X_1^\prime,t_0)\big\}&=&C_{0}(X_1,X_1^\prime)\,, \label{Gamma0}
\ee
where $C_{0}(X_1,X_1^\prime)=C(X_1,t_0;X_1^\prime,t_0)$ is the iniital value of the correlation function.
Similarly, equation (\ref{calS}) implies
\be
{\cal{S}}(1,1')&=&{\cal{S}}^{reg}(1,1')+\delta(t'-t_0){\cal{S}}^{\delta}(X_1,X_{1}^\prime;t)+\delta(t-t_0){\cal{S}}^{\delta}(X_1^\prime,X_1;t')\nn\\
&&\hspace{3cm}+\delta(t-t_0)\delta(t'-t_0){\cal{S}}^{0}(X_1,X_1^\prime)  \label{calSstructure}
\ee
with
\be
{\cal{S}}^{reg}(1,1')&=&\frac{is}{2}\Big\{\Sigma^{>}(1,1')+\Sigma^{<}(1,1'),f(1')\Big\}(1')\\
{\cal{S}}^{\delta}(X_1,X_1^\prime;t)&=&is\Big\{\Sigma^{(0)}(X_1,X_1^\prime;t),f(1')\Big\}(1')\\
{\cal{S}}^{0}(X_1,X_1^\prime)&=&C_{0}(X_1,X_1^\prime)
\ee
Substituting those expressions into Eqs.(\ref{gammaCC}-\ref{spectralbalance}), we obtain expressions that describe the temporal evolution of the correlation function from its initial value $C(X_1,t_0;X_1^\prime,t_0)$ to $C(X_1,t;X_1^\prime,t')$ at times $t,t'\geq t_0$,
\bs \label{solutionforC}
\ben
\hspace{-.5cm}C(1,1')&=&\!\!\iint{ dX_2dX_3 \chi^{R}(1;X_2,t_0)\Gamma^{(0)}(X_2,X_3)\chi^{A}(X_3,t_0;1')}\label{solutionforCwithchigamma}\\
&+&\int_{t_0}^{t}{\!\!dt_2\!\!\iint{ dX_2dX_3 \chi^{R}(1,2)\Gamma^{\delta}(X_2,X_3;t_2)\chi^{A}(X_3,t_0;1')}}\nn\\
&+&\int_{t_0}^{t'}{\!\!dt_3\!\!\iint{ dX_2dX_3 \chi^{R}(1;X_2,t_0)\Gamma^{\delta}(X_2,X_3;t_3)\chi^{A}(3;1')}}\nn\\
&+&\int_{t_0}^{t}{\!\!dt_2\!\!\int_{t_0}^{t'}{dt_3\iint{dX_2dX_3\chi^{R}(1,2)\Gamma^{reg}(23)\chi^{A}(3,1')}}}\nn\\
&=&\!\!\iint{ dX_2dX_3 R(1;X_2,t_0)R(1';X_3,t_0)C_{0}(X_2,X_3)}\label{solutionforCwithRcalS}\\
&+&\int_{t_0}^{t}{\!\!dt_2\!\!\iint{ dX_2dX_3 R(1,2)R(1';X_3,t_0){\cal{S}}^{\delta}(X_2,X_3;t_2)}}\nn\\
&+&\int_{t_0}^{t'}{\!\!dt_3\!\!\iint{ dX_2dX_3 R(1;X_2,t_0)R(1',3){\cal{S}}^{\delta}(X_2,X_3;t_3)}}\nn\\
&+&\int_{t_0}^{t}{\!\!dt_2\!\!\int_{t_0}^{t'}{dt_3\iint{dX_2dX_3 R(1,2)R(1',3){\cal{S}}^{reg}(23)}}}\,.\nn
\een
\es
We can shed light on the different components of Eq.(\ref{solutionforC}) by considering the dynamics of the density fluctuations $\Delta N(1)$.
An equation of motion for $\Delta N(1)=N(1)-f(1)$ is easily obtained by substracting Eq.(\ref{equationf}) from Eq.(\ref{accurateKlimontovich2}).
By substracting the quantity $\Sigma^{R}(1,2)\delta N(2)$ on both sides of the resulting equation, we find
\be
\lefteqn{\left(\frac{\partial}{\partial t_1}-L_{1}\right)\delta N(1)-\Sigma^{mf}(1,2)\delta N(2)-\Sigma^{R}(1,2)\delta N(2)}&&\\
&&=\frac{1}{2}\gamma_{3}(1,2,3)\big(N(2)N(3)-C(2,3)\big)-\Sigma^{R}(1,2)\delta N(2)\equiv\delta{\cal{F}}(1)\,,
\ee
which can formaly be integrated in terms of the Green's function $R(1,1')$ (recall Eq.(\ref{equationR})) as
\ben
\delta N(1)=\int{dX_2 R(1;X_2,t_0)\delta N(X_2,t_0)}+\int_{t_0}^{t}{dt_2\int{dX_2 R(1,2)\delta{\cal{F}}(2)}} \label{equationdeltaN}
\een
Thus the microscopic phase-space density fluctuation is written as the sum of two terms.
The first term represents the propagation in the fluid of the density fluctuation that would be caused by a small initial perturbation $\delta N(X_2,t_0)$ in $f$.
Roughly speaking, it describes how a microscopic flucutation propagates in the fluid {\it on average and neglecting non-linear effects}.
The second term in Eq.(\ref{equationdeltaN}) describes the corrections to this interpretation that manifest  in the equation of motion in the source term $\delta{\cal{F}}$ and in the solution (\ref{equationdeltaN}) through non-local effects in both space and time.
Using Eq.(\ref{equationdeltaN}) to build the correlation function $C(1,1')=\langle\delta N(1)\delta N(1')\rangle$ lead to the four contributions in Eq.(\ref{solutionforCwithRcalS}) with
\ben
{\cal{S}}^{reg}(1,1')&=&\big\langle \delta{\cal{F}}(1)\delta{\cal{F}}(1') \big\rangle\nn\\
{\cal{S}}^{\delta}(X_1,X_{1}^\prime;t)&=&\big\langle \delta{\cal{F}}(1)\delta N(X_1^\prime,t_0)\big\rangle  \label{calSdelta}
\een
Thus, the first term in Eq.(\ref{solutionforCwithRcalS}) describes the contribution obtained when propagating the initial fluctuations as if they were small, independent perturbations on the fluid.
The three other terms represent the corrections to this picture, which allows us to express the components of the spectral function ${\cal{S}}$ in terms of correlation functions between the noise term $\delta{\cal{F}}$ and the initial fluctuations $\delta N$: ${\cal{S}}^{reg}(1,1')$ is the autocorrelation function of the noise term $\delta{\cal{F}}$ while ${\cal{S}}^{reg}$ describes the influence the initial conditions. 

Equation (\ref{solutionforCwithRcalS}) exhibits the influence of initial correlations in the initial state; the latter can be important when considering ultrafast relaxation processes when $t,t'$ approach $t_0$.
The influence of initial correlations is often neglected in kinetic theory by invoking the condition of complete suppression of initial correlations (a.k.a Bogolyubov's condition \cite{Bogoliubov1962}) according to which all initial corraletions existing at $t_0$ are damped at a sufficiently large time $t-t_0>>t_{cor}$.
Inspection of Eq.(\ref{solutionforCwithRcalS}) suggests that $t_{cor}$ is determined by the decorrelation time of ${\cal{S}}^{\delta}(X_1,X_{1}^\prime;t)$ in Eq.(\ref{calSdelta}) and by the properties of $R(1,1')$.
Then for times $t-t_0>>t_{cor}$, we set ${\cal{S}}^{\delta}\equiv 0$ and
\be
C_{0}(X_1,X_1^\prime)=C_{0}^{ideal}(X_1,X_1^\prime)=f_0({\bf r},{\bf p})\delta({\bf r}-{\bf r}^\prime)\delta({\bf p}-{\bf p}^\prime)
\ee
where $f_0({\bf r},{\bf p})=\big\langle \sum_{j=1}^{N}{\delta\left({\bf r}-{\bf r}_{j})\right)\delta\left({\bf p}-{\bf p}_{j}\right)} \big\rangle$ is the initial single-particle distribution function such that
\be
\hspace{-.5cm}C(1,1')&=&\!\!\iint{ dX_2dX_3 R(1;X_2,t_0)R(1';X_3,t_0)C_{0}^{ideal}(X_2,X_3)}\\
&+&\int_{t_0}^{t}{\!\!dt_2\!\!\int_{t_0}^{t'}{dt_3\iint{dX_2dX_3 R(1,2)R(1',3){\cal{S}}^{reg}(23)}}}\,.\nn
\ee
When dynamical correlations are alltogether discarded ($\Sigma^{R,A,C}$=0), then
\be
\hspace{-.5cm}C(1,1')&=&\!\!\iint{ dX_2dX_3 R^{mf}(1;X_2,t_0)R^{mf}(1';X_3,t_0)C_{0}^{ideal}(X_2,X_3)}
\ee
with $\left[\frac{\partial}{\partial t}-L_1(1)\right]R(1,1')-\Sigma^{mf}(1,2)R(2,1')=\delta(1-1')$ corresponding to the linearized Vlasov equation, which in plasma physics is known as the quasi-linear theory and leads to the Lenard-Balescu-Guernsey equation.

\section{Real-time formalism (II)} \label{sectionV}

The previous section focused on the derivation of the coupled evolution equations for the correlation and response functions in terms of the $\Sigma^{R,A,C}$
Here we concentrate on translating to real time the closure relations (\ref{Sigmadefinitiondetailed}) and (\ref{Sigmaf}) to obtain the equivalent relations for the real-time components $\Sigma^{R,A,C}$.
Although it is possible to directly translate Eq.(\ref{Sigmadefinition}) in terms of the matrix components $\chi^{R,A,C}$, $\Sigma^{R,A,C}$ and of the components of the tensor obtained from the different components of the three-point vertex $\Gamma^{(3)}$ \cite{Daligaultunpublished,GuangZhaoZhou}, the manipulations are not trivial and we prefer the approach described below.
A considerable advantage of this alternative approach is that it allows us to introduce an equivalent formulation of the closed-time contour approach, which directly generates the correlation $C$ and response functions $\chi^{R,A}$.

\subsection{Real-time representation of the generating functional}

Given a potential $\phi(\tau)$ on the closed-time contour, we define the physical component $\phi_p(t)$ and non-physical $\phi_{\Delta}(t)$ on the real-time axis as
\ben
\left\{
\begin{array}{l}
\D\phi_{p}(X,t)=\frac{1}{2}\left[\phi(X,\tau_+)+\phi(X,\tau_-)\right]\\\\
\D\phi_{\Delta}(X,t)=\phi(X,\tau_+)-\phi(X,\tau_-)
\end{array}
\right.
\label{phitophipphid}
\een
where $t=t(\tau_+)=t(\tau_-)$.
In this representation, physical potentials are simply characterized by $\phi_\Delta \equiv 0$.
Similarly, we define the 'physical' and 'non-physical' components of the phase-space density as \cite{classicalquantumcomponents},
\ben
\left\{
\begin{array}{l}
\D {\cal{N}}_{p}(X,t)=\frac{1}{2}\left[{\cal{N}}(X,\tau_+)+{\cal{N}}(X,\tau_-)\right]\\\\
\D {\cal{N}}_{\Delta}(X,t)={\cal{N}}(X,\tau_+)-{\cal{N}}(X,\tau_-)
\end{array}
\right.
\,. \label{NtoNpNd}
\een
For a physical potential, the phase-space density is equal on both sides of the contour and therefore ${\cal{N}}_{\Delta}(X,t)=0$ and ${\cal{N}}_{p}(X,t)$ equals the phase-space density of the system under investigation in the external potential $\phi_p$.
With these definitions, the coupling term in the action (\ref{classicalclosedtimeaction}) writes
\ben
\oint{d\tilde{1}{\cal{N}}(\tilde 1)\phi(\tilde 1)}=\int_{t_0}^{\infty}{\int{dX_1 \phi_p(1){\cal{N}}_{\Delta}(1)+\phi_\Delta(1){\cal{N}}_{p}(1)}}\,,\label{phipNdeltaphideltaNp}
\een
which shows that $\phi_p$ linearly couples ${\cal{N}}_{\Delta}$ while $\phi_\Delta$ couples to ${\cal{N}}_{\Delta}$; accordingly, variations with respect to $\phi_\Delta$ will generate the physical phase-space distribution under investigation.

Equation (\ref{phitophipphid}) defines a one-to-one mapping between the potentials $\phi$ and the vector of potentials $(\phi_p,\phi_{\Delta})$.
As a consequence, the generating functional $\tilde\Omega[\phi]$ can be replaced by the action functional $\Omega[\phi_p,\phi_{\Delta}]$ defined as
\ben
\Omega[\phi_p,\phi_{\Delta}]\equiv \tilde{\Omega}[\phi]\,. \label{newgeneratingfunctional}
\een
In terms of the new variables, the Taylor expansion (\ref{functionalderivativesexpansion}) around the external potential $\phi_0$ becomes,
\ben
\lefteqn{\Omega[\phi_0+\delta\phi_p,\delta\phi_\Delta]=\sum_{j,k=0}^{\infty}{\frac{1}{j!k!}\int{d1\dots dj\int{d1'\dots dk'}}}}&&\label{functional2derivativesexpansion}
\\
&&\hspace{1cm}\Omega_{\small\underbrace{1\dots 1}_{j\text{ times}}\underbrace{\/2\dots 2}_{k\text{ times}}}\big(1\dots n;1'\dots n'\big)\,\delta\phi_p(1)\dots\delta\phi_p(j)\,\phi_\Delta(1')\dots\delta\phi_\Delta(k')\nn
\een
where
\be
\Omega_{\small\underbrace{1\dots 1}_{j\text{ times}}\underbrace{\/2\dots 2}_{k\text{ times}}}\big(1\dots j;1'\dots k'\big)=\frac{\delta^{j+k}\Omega}{\delta\phi_{p}(1)\dots\delta\phi_{p}(j)\delta\phi_{\Delta}(1')\dots\delta\phi_{\Delta}(k')}\Bigg|_{\phi_p=\phi_0,\phi_\Delta=0}\,.
\ee
Explicit expressions for these derivatives can be obtained either from the $\tilde{\Omega}^{(n)}[\phi]$ or from a systematic perturbation expansion as shown in appendix \ref{sectionproof} \cite{noteOmega1dots12dots2}.

The first derivatives generate the phase-space density $f({\bf r},{\bf p},t)$ under investigation,
\be
\frac{\delta \Omega}{\delta \phi_{p}(1)}\Bigg|_{\phi_p=\phi_0,\phi_\Delta=0}=0\quad,\quad\frac{\delta \Omega}{\delta \phi_{\Delta}(1)}\Bigg|_{\phi_p=\phi_0,\phi_\Delta=0}=f(1)\/.
\ee
Most remarkably, the second (partial) derivatives directly generate the retarded and advanced response functions and correlation functions,
\ben
\left\{
\begin{array}{ll}
\D\Omega_{11}(1,2)=0\\
\D\Omega_{12}(1,2)=\chi^{A}(1,2)\\
\D\Omega_{21}(1,2)=\chi^{R}(1,2)\\
\D\Omega_{22}(1,2)=\frac{1}{is}C(1,2)
\end{array}
\right.
\label{omegaij}
\een
We therefore obtain the noteworthy result that a generating functional defined in terms of real-time quantities exists that generates the phase-space distribution, the correlation function $C$ and response functions $\chi^{R,A}$.

We now continue the closure procedure by adapting the steps of Sec.\ref{sectionIII} to the new action functional (\ref{newgeneratingfunctional}).
We define the effective potential as the Legendre transform,
\ben
\Gamma[f_{p},f_{\Delta}]=-\Omega[\phi_{p},\phi_{\Delta}]+\int{d1\,f_{p}(1)\phi_{\Delta}(1)+f_{\Delta}(1)\phi_{p}(1)}\,. \label{deltaGammarealdphi}
\een
Its deriatives at $f_p=f,f_\Delta=0$ define the vertex functions
\be
\Gamma_{\underbrace{1\dots 1}_{j\text{ times}}\underbrace{\/2\dots 2}_{k\text{ times}}}(1\dots j;1'\dots k')=\frac{\delta^{j+k}\Gamma}{\delta f_{p}(1)\dots\delta f_{p}(j)\delta f_{\Delta}(1')\dots\delta f_{\Delta}(k')}\Bigg|_{f_p=f,f_\Delta=0}\,.
\ee
The first derivative generates the external potential $\phi_0$,
\be
\frac{\delta \Gamma}{\delta f_{p}(1)}\Bigg|_{f_p=f,f_\Delta=0}=0\quad,\quad\frac{\delta \Gamma}{\delta f_{\Delta}(1)}\Bigg|_{f_p=f,f_\Delta=0}&=&\phi_{0}(1)\,.
\ee
while, as shown in the appendix \ref{appendix1A}), the second-order derivatives directly yield the $\Gamma^{R,A,C}$, i.e. \cite{Noteonmatrixrepresentation}
\ben
\left\{
\begin{array}{ll}
\D\Gamma_{11}(1,1')=0\\
\D\Gamma_{12}(1,1')=\Gamma^A(1,1')\\
\D\Gamma_{21}(1,1')=\Gamma^R(1,1')\\
\D\Gamma_{22}(1,1')=\frac{1}{is}\Gamma^{C}(1,1')
\end{array}
\right.
\,. \label{gammaij}
\een
Higher order vertex functions can be obtained by taking the successive derivatives with respect to the potentials with the chain rule of differentiation.
The direct calculations become rapidly cumbersome with the growing number of indices and variables and, in appendix \ref{appendix1}, we present a graphical method that allows their calculation in a much more economical manner.
We just give here the results for those derivatives of importance in the present work, namely
\bs \label{omegaijk}
\ben
-\Omega_{221}(123)&=&\Omega_{22}(1\bar{1})\Omega_{21}(2\bar{2})\Omega_{12}(3\bar{3})\Gamma_{121}(\bar{1}\bar{2}\bar{3})\nn\\
&+&\Omega_{21}(1\bar{1})\Omega_{22}(2\bar{2})\Omega_{12}(3\bar{3})\Gamma_{211}(\bar{1}\bar{2}\bar{3})\nn\\
&+&\Omega_{21}(1\bar{1})\Omega_{21}(2\bar{2})\Omega_{12}(3\bar{3})\Gamma_{221}(\bar{1}\bar{2}\bar{3})\label{omega221}\\
-\Omega_{222}(123)&=&\Big[\Omega_{21}(1\bar 1)\Omega_{22}(2\bar 2)\Omega_{22}(3\bar 3)\Gamma_{211}(\bar 1\bar 2\bar 3)+\text{c.p}\Big]\nn\\
&+&\Big[\Omega_{21}(1\bar 1)\Omega_{21}(2\bar 2)\Omega_{22}(3\bar 3)\Gamma_{221}(\bar 1\bar 2\bar 3)+\text{c.p}\Big]\nn\\
&+&\Omega_{21}(1\bar 1)\Omega_{21}(2\bar 2)\Omega_{21}(3\bar 3)\Gamma_{222}(\bar 1\bar 2\bar 3)\label{omega222}
\een
\es
where c.p. stands for cyclic permutation of the triplet $(1\bar 1,2\bar 2,3\bar 3)$.
These equations are the transcription in terms of real-time quantities of the equation (\ref{chi3gamma3}) derived before.

\subsection{Real-time representation of the closure}

The same steps described in Secs.\ref{sectionIIIF}-\ref{sectionIIIG} to derive the closure can be followed with the evolution equations for ${\cal{N}}_{p,\Delta}$ as a starting point.
The main steps are given in appendix \ref{keystepsrecastformalism} and can be summarized as follows.
From the evolution equation for ${\cal{N}}_{p,\Delta}$, we first obtain the evolution equation for $f_{p,\Delta}$ by averaging and for $\chi^{R,A}$ and $C$ by functional differentiation using Eqs.(\ref{omegaij}).
The collision operators in those equations involve the third order derivatives $\Omega_{ijk}$ and are rexpressed in in terms of memory function $\Sigma^{R,A,C}$ using Eqs.(\ref{omegaijk}).
The overall procedure directly leads to the evolution equations (\ref{geenralCchiRchiA}) for $\chi^{R,A}$ and $C$ together with the following closure relations
\bs \label{sigmaRAC}
\ben
\D\Sigma^{R}(1,1')&=&\D\frac{1}{2}\gamma_3(123)\frac{\delta C(2,3)}{\delta f_p(1')}\Big|_{f_p;f_\Delta=0}\label{sigmaR}\\
\D\Sigma^{A}(1,1')&=&\D is\gamma_3(123)\frac{\delta \chi^R(2,3)}{\delta f_\Delta(1')}\Big|_{f_p;f_\Delta=0}\label{sigmaA}\\
\D\Sigma^{C}(1,1')&=&\D\frac{is}{2}\gamma_3(123)\frac{\delta C(2,3)}{\delta f_\Delta(1')}\Big|_{f_p;f_\Delta=0}\label{SigmaCC}
\een
\es
or, after performing the functional differentiations, 
\bs \label{sigmaRACdetail}
\ben
\D\Sigma^{R}(1,1')&=&\D-\frac{is}{2}\gamma_{3}(123)\Big[\Omega_{21}(2\bar{2})\Omega_{21}(3\bar{3})\Gamma_{221}(\bar 2,\bar 3,1')\\
&&\hspace{3cm}+2\Omega_{22}(2\bar{2})\Omega_{21}(3\bar{3})\Gamma_{121}(\bar 2,\bar 3,1')\Big]\nn\\
\D\Sigma^{A}(1,1')&=&-is\gamma_{3}(123)\Big[\Omega_{12}(2\bar{2})\Omega_{21}(3\bar{3})\Gamma_{122}(\bar 2,\bar 3,1')\\
&&\hspace{3cm}+\Omega_{12}(2\bar{2})\Omega_{22}(3\bar{3})\Gamma_{112}(\bar 2,\bar 3,1')\Big]\nn\\
\D\Sigma^{C}(1,1')&=&-\frac{(is)^2}{2}\gamma_{3}(123)\Big[\Omega_{22}(2\bar{2})\Omega_{22}(3\bar{3})\Gamma_{112}(\bar 2,\bar 3,1')\\
&&\hspace{.1cm}+2\Omega_{21}(2\bar{2})\Omega_{22}(3\bar{3})\Gamma_{212}(\bar 2,\bar 3,1')+\Omega_{21}(2\bar{2})\Omega_{21}(3\bar{3})\Gamma_{222}(\bar 2,\bar 3,1')\Big]\,.\nn
\een
\es
These equations for $\Sigma^{R,A,C}$ are the transcription in real-time of the closed-time contour formulas (\ref{Sigmadefinition}) and (\ref{Sigmadefinitiondetailed}).
The set of equations consisting of the evolution equations (\ref{equationf}) and (\ref{geenralCchiRchiA}) together with Eqs.(\ref{sigmaRAC}) is formally closed.

As in Sec.\ref{sectionIIIG}, the evolution equations (\ref{geenralCchiRchiA}) can be used to derive additional relations between the memory and vertex functions to derive functional equations for the memory functions $\Sigma^{R,A,C}$.
The details of the calculations are given in appendix \ref{keystepsrecastformalism}.
The resulting closure relations can be expressed in a variety of equivalent forms in terms of the quantities $C$, $\chi^{R,A}$, $R$, $\Sigma^{R,A,C}$ and ${\cal{S}}$ encountered before, and we choose to give them in the following form (which is most directly comparable with previous works, e.g. \cite{Rose1979}):
\ben \label{SigmaRAdeltaSigma}
\D\Sigma^{R}(1,1')&=&\gamma_3(123)\gamma_3(\bar 2\bar 3 1')R(2\bar 2)C(3\bar 3)+\gamma_{3}(123)R(2\bar 2)C(3 \bar 3)\frac{\delta \Sigma^{R}(\bar 2\bar 3)}{\delta f_p(1')}\nn\\
&&\hspace{3cm}+\frac{1}{2}\gamma_{3}(123)R(2\bar 2)R(3 \bar 3)\frac{\delta {\cal{S}}(\bar 2\bar 3)}{\delta f_p(1')}
\een
and (see Sec.\ref{formalsolutionC}),
\ben
{\cal{S}}^{reg}(1,1')&=&\frac{1}{2}\gamma_3(123)\gamma_3(1'\bar 2\bar 3)C(2\bar 2)C(3\bar 3)+\frac{1}{2}\gamma_3(123)C(2\bar 2)C(3\bar 3)\frac{\delta\Sigma^R(1'\bar 3)}{\delta f_p(\bar 2)}\label{calSdetailed}\\
&+&is\gamma_3(123)\chi^{R}(2\bar 2)C(3\bar 3)\frac{\delta\Sigma^R(1'\bar 3)}{\delta f_\Delta(\bar 2)}+is\gamma_3(123)\chi^{R}(2\bar 2)\chi^{R}(3\bar 3)\frac{\delta\Sigma^C(1'\bar 3)}{\delta f_\Delta(\bar 2)}\nn
\een
and
\be
{\cal{S}}^{\delta}(X_1,X_1^\prime;t)&=&\frac{1}{2}\gamma_{3}(123)\chi^{R}(2\bar 2)\chi^{R}(3\bar 3)\Big[is\frac{\delta}{\delta f_\Delta(\bar 3)}C(X_1^\prime,t_0;4)\Gamma^{A}(4\bar 2)\Big]
\ee

In the lowest order approximation,
\be
\Sigma^{R}(1,1')&=&\gamma_3(123)\gamma_3(\bar 2\bar 3 1')R(2\bar 2)C(3\bar 3)\\
{\cal{S}}^{reg}(1,1')&=&\frac{1}{2}\gamma_3(123)\gamma_3(1'\bar 2\bar 3)C(2\bar 2)C(3\bar 3)\\
{\cal{S}}^{\delta}&=&0
\ee
we recover the famous Direct Interaction Approximation (DIA) derived by Kraichnan \cite{Kraichnan1959} in the context of fluid turbulence and later extended to the Klimontovich equations by Dubois and Espedal \cite{DuboisEspedal1979,Krommes2002}.
Further discussions on the application of those closure relations to physical problems would certainly require substantial additional work, which is beyond the scope of the present paper.

\section{Summary} \label{sectionVIA}

We have presented the foundations of a theory to derive closed, self-consistent approximations for calculating the statistical dynamics of classical Hamiltonian systems, which can describe both equilibrium and non-equilibrium states.
The theory, which unifies and encompasses previous results for classical Hamiltonian systems with any initial conditions, can be regarded (i) as the classical  mechanical counterpart of the theory of non-equilibrium Green's functions in quantum field theory and (ii) as a generalization to dynamical problems of the density functional theory of fluids in equilibrium.
The present approach avoids many of the complications inherent to previous works \cite{Andersen2000}; in particular the theory is valid for any initial state and treats equilibrium and non-equilibrium states in a unified manner.
Interestingly, several of the key ideas and tricks of the previous theories (e.g, the need for both correlation and response functions, operator doubling, imposed commutation relations and other causality constraints...) arise naturally here albeit in a different form (e.g., the operator doubling in MSR is replaced by the coupling to physical and non-physical phase-space densities ${\cal{N}}_{p,\Delta}$, which generate both response and corrrelation functions..)

The main ingredients of the present approach are the following.
Given a Hamiltonian system in an external potential $\phi_0$ and in the statistical state described by $F_0$ at initial time $t_0$, the effective action functional (\ref{newgeneratingfunctional}) defined as
\be
\Omega[\phi_p,\phi_{\Delta}]=-s\ln{\cal{Z}}[\phi_p,\phi_\Delta]\,,
\ee
where $s$ is a fixed parameter with the dimension of an action, contains all information about the dynamical properties of the system at times $t\geq t_0$.
In particular, its first derivative at $\phi_p=\phi_0,\phi_\Delta=0$ gives the single-particle phase-space distribution function,
\be
\frac{\delta \Omega}{\delta \phi_{\Delta}(1)}=f(1)\,,
\ee
with $1=({\bf r},{\bf p},t)$, and its second derivatives generate the correlation function and retared and advanced response functions,
\be
\frac{\delta^2 \Omega}{\delta \phi_{\Delta}(1)\delta \phi_{p}(2)}&=&\chi^{R}(1,2)\\
\frac{\delta^2 \Omega}{\delta \phi_{p}(1)\delta \phi_{\Delta}(2)}&=&\chi^{A}(1,2)\\
\frac{\delta^2 \Omega}{\delta \phi_{\Delta}(1)\delta \phi_{\Delta}(2)}&=&C(1,2)/is\,.
\ee
Higher-order derivatives systematically generate combinations of products and Poisson brackets of the fundamental field $N(1)$.
Using traditional renormalization techniques involving the Legendre transform of the action function $\Omega[\phi_p,\phi_\Delta]$, a closed, self-consistent set of equations of motion is derived for the single-particle phase-space distribution function $f$, the correlation function $C=\langle \delta f\delta f \rangle$, the retarded and advanced density response functions $\chi^{R,A}$ to external potentials, and the associated memory functions $\Sigma^{R,A,C}$,
\ben
\left[\frac{\partial}{\partial t_1}-L_{1}\right]f(1)-\{u^{mf}(1),f(1)\}&=&\frac{1}{2}\gamma_{3}(1,2,3)C(2,3)\,,\nn
\een
and
\be
\left[\frac{\partial}{\partial t}-L_1\right]\!\chi^{R}(1,1')-\int_{t_0}^{\infty}{\!\!\!\!\int{d2\,\Sigma_{tot}^{R}(1,2)\chi^{R}(2,1')}}&=&\big\{\delta(1-1'),f(1)\big\}(1)\\
\left[\frac{\partial}{\partial t}-L_1\right]\!\chi^A(1,1')-\int_{t_0}^{\infty}{\!\!\!\!\int{d2\,\Sigma_{tot}^{A}(1,2)\chi^A(2,1')}}&=&\big\{\delta(1-1'),f(1)\big\}(1)\\
\left[\frac{\partial}{\partial t}-L_1\right]\!C(1,1')-\!\int_{t_{0}}^{\infty}{\!\!\!\!\int{d2\,\Sigma_{tot}^{R}(1,2)C(2,1')}}&=&\int_{t_{0}}^{\infty}{\!\!\!\!\int{d2\,\Sigma^{C}(1,2)\chi^{A}(2,1')}}
\ee
with the memory function kernels satisfying
\ben
\D\Sigma^{R}(1,1')&=&\D\frac{1}{2}\gamma_3(123)\frac{\delta C(2,3)}{\delta f_p(1')}\Big|_{f_p;f_\Delta=0}\nn\\
\D\Sigma^{A}(1,1')&=&\D is\gamma_3(123)\frac{\delta \chi^R(2,3)}{\delta f_\Delta(1')}\Big|_{f_p;f_\Delta=0}\nn\\
\D\Sigma^{C}(1,1')&=&\D\frac{is}{2}\gamma_3(123)\frac{\delta C(2,3)}{\delta f_\Delta(1')}\Big|_{f_p;f_\Delta=0}\,.
\een
Moreover, the memory functions satisfy functional differential equations, e.g. (\ref{SigmaRAdeltaSigma}), that may be used systematic approximations.

In conclusion, the purpose of the present paper was to lay down the foundations of a theory for the construction of renormalized kinetic equations of classical systems of particles in and out of equilibrium.
We hope that these results will serve to (re)stimulate further research on this challenging topic.
Of course further works calls for applications of the approach to realistic problems; work along those lines is under way.

\begin{acknowledgements}
The author wishes to thank H. Rose for stimulating discussions and encouragements.
This work was performed for the U.S. Department of Energy by Los Alamos National Laboratory under contract DE-AC52-06NA25396.
\end{acknowledgements}



\appendix

\section{BBGKY hierarchy} \label{appendixBBGKY}

The BBGKY hierarchy \cite{Balescu1997} is an hierarchy of equations coupling the {\it equal-time} reduced distribution function $f_n$,
\be
\lefteqn{f_n(X_1,\dots,X_n)}\\
&&=\frac{N!}{(N-s)!}\int{dX_{n+1}\dots dX_{N} F(X_1,\dots, X_N;t)}\,.
\ee
The latter can also be expression as ensemble average of microscopic fields, e.g.
\be
f_1(X,t)&=&\left\langle\sum_{j=1}^{N}\delta(x-{\bf x}_j(t))\right\rangle\\
f_{2}(X,X';t)&=&\left\langle\sum_{j=1}^{N}\sum_{k\neq j}^{N}\delta(x-{\bf x}_j(t))\delta(x'-{\bf x}_k(t))\right\rangle
\ee
and
\be
\lefteqn{f_{3}(x,x',x'';t)}\\
&=&\left\langle\sum_{j=1}^{N}\sum_{k\neq j}^{N}\sum_{l\neq j,k}^{N}\delta(x-{\bf x}_j(t))\delta(x'-{\bf x}_k(t))\delta(x''-{\bf x}_l(t))\right\rangle\,,
\ee
and so forth.

The equations of the BBGKY hierarchy can be obtained by direct integration of the Liouville equation (\ref{Liouvilleequation}).
The $n$-particle reduced distribution $f_n(X_1,\dots,X_n)$ represents -- up to the factor $N!/(N-s)!$ -- the probability density of finding simultaneously $n$ particles in the specified phase-space positions $X_1,\dots,X_n$.
In practice, those equations are often recast in terms of the correlation functions defined from the cluster representation of the reduced distribution functions for $n\geq 2$,
\be
f_{2}(x,x';t)&=&g_{2}(x,x',t)+f_1(x,t)f_1(x',t)\\
f_{3}(x,x',x'';t)&=&g_3(x,x',x'',t)+\left[f_1(x;t)g_2(x',x'';t)+c.p.\right]\\
&&\hspace*{1cm}+f_1(x,t)f_1(x',t)f_1(x'',t)
\ee
and so on, where c.p. means cyclic permutation.
In terms of the standard notations (\ref{L1}) and (\ref{Ljk}), the first two equations are
\be
\left[\frac{\partial}{\partial t}-L_{1}\right]f_1(1)&=&\int{d2 L_{12} f_1(1)f_1(2)+L_{12}g_{2}(12)}
\ee
and
\be
\left[\frac{\partial}{\partial t}-\left(L_{1}+L_{1'}\right)\right]g_{2}(1,1')&=&L_{11'}f_1(1)f_1(1')+\int{d2 \big[L_{12} f_1(1)g_{2}(1',2)+L_{1'2} f_1(1')g_{2}(1,2)\big]}\nn\\
&+&\int{d2 \left(L_{12}+L_{1'2}\right)f_1(2)g_{2}(1,1')}\nn\\
&+&\,L_{11'}g_2(1,1')+\int{d2 \left(L_{12}+L_{1'2}\right)g_{3}(1,1',2)}\,.
\ee

\section{Useful properties} \label{appendixpropertiesPoissonbracket}

We gather here some basic properties that are often used in the remainder of the paper:
\begin{itemize}
\item Equal-time Poisson bracket:
\ben
\lefteqn{\left[N(X,t),N(X',t)\right]_{PB}}&&\label{equaltimePoissonpsdensity}\\
&&=\left(\frac{\partial}{\partial {\bf r}}\cdot\frac{\partial}{\partial {\bf p}^{\prime}}-\frac{\partial}{\partial {\bf r}^{\prime}}\cdot\frac{\partial}{\partial {\bf p}}\right)\left[\delta(X-X^{\prime}){\cal{N}}(X,t)\right]\nn\\
&&=\left\{\delta(X-X'),N(X,t)\right\}(X)\nn\,.
\een
Therefore,
\ben
\chi^{R/A}(X,t;X',t)=\mp\left\{\delta(X-X'),f(X,t)\right\}(X)\,. \label{EqualtimechiRA}
\een
\item From rotational symmetry,
\ben
\int{dX_2 L_{12}\delta(X_1-X_2)}=0\,. \label{L12delta12}
\een
\item Unlike their quantum counterparts, the $n$-point correlation functions $C^{(n)}$ are fully symmetric with respect to there arguments.
\end{itemize}

\section{First and second derivatives of $\tilde\Omega[\phi]$} \label{sectionproof}

In this section, we present the main steps leading to Eqs.(\ref{domegadphi}) and (\ref{d2omegadphi2}) \cite{additionalpropeties}.
Since $\tilde{\Omega}[\phi]$ depends on $\phi$ through $S[\phi;x_{0}]$, its functional derivatives are combinations of the total action's derivatives.
For instance, we easily find,
\ben
\frac{\delta \tilde{\Omega}}{\delta \phi(\tilde 1)}\Bigg|_{\phi_{0}}&=&-\left\langle \frac{\delta {\cal{S}}}{\delta \phi(\tilde 1)}\Bigg|_{\phi_{0}}\right\rangle\,, \label{generalfirstderivative} 
\een
and
\ben
\D\frac{\delta^{2} \tilde{\Omega}}{\delta \phi(\tilde 1)\delta \phi(\tilde 1')}\Bigg|_{\phi_{0}}&=&\D-\llangle \frac{\delta^{2} {\cal{S}}}{\delta \phi(\tilde 1)\delta \phi(\tilde 1')}\Big|_{\phi_{0}}\rrangle \label{generalsecondderivative}\\
&&-\frac{i}{s}\llangle \left(\frac{\delta {\cal{S}}}{\delta \phi(\tilde 1)}\Big|_{\phi_{0}}-\llangle \frac{\delta {\cal{S}}}{\delta \phi(\tilde 1)}\Big|_{\phi_{0}}\rrangle\right)\left(\frac{\delta {\cal{S}}}{\delta \phi(\tilde 1')}\Big|_{\phi_{0}}-\llangle \frac{\delta {\cal{S}}}{\delta \phi(\tilde 1')}\Big|_{\phi_{0}}\rrangle\right)\rrangle\,.\nn
\een
When ${\cal{S}}$ is linear in the external potential, the term in $\delta^{2}{\cal{S}}/\delta \phi(\tilde 1)\delta \phi(\tilde 1')$ vanishes and Eq.(\ref{generalsecondderivative}) equals the two-point correlation of $\delta {\cal{S}}/\delta \phi(\tilde 1)$; this is the typical situation in field theory as discussed before.
In the present theory, the action is non-linear in $\phi$, $\delta^{2}{\cal{S}}/\delta \phi(\tilde 1)\delta \phi(\tilde 1')$ is nonzero and generates, in addition to the correlation function, the term ${\cal{R}}(1,1')$ related to the response functions $\chi^R$.

The derivatives of ${\cal{S}}[\phi;x_{0}]$ at a given potential $\phi_{0}$ can be obtained by a systematic perturbation expansion in powers of the variations $\delta\phi$ around $\phi_{0}$.
To this end, we introduce the following notations.
With $\phi=\phi_{0}+\delta\phi$, we write the total Hamiltonian as
\be
{\cal{H}}_{\phi}={\cal{H}}_{\phi_{0}}+\delta{\cal{H}}\,,
\ee
with $\delta{\cal{H}}=\int{dX\,{\cal{N}}(X,\tau)\delta\phi(X,\tau)}$.
We denote by $x(\tau)=(q(\tau),p(\tau))$ and $x_0(\tau)=(q_0(\tau),p_0(\tau))$ the closed-time trajectories starting at $x_0$ and governed by the Hamiltonian ${\cal{H}}_{\phi}$ and ${\cal{H}}_{\phi_{0}}$, respectively.
Given a dynamical variable $A$, we  denote by $A=A_{0}+\delta A=A_{0}+A_{1}+A_{2}+\dots$ its expansion in power of $\delta\phi$, where $A_{n}$ is the term of order $n$ in $\delta\phi$.
For instance, $\delta{\cal{H}}=\delta{\cal{H}}_1+\delta{\cal{H}}_2+\dots$ with
\ben
\delta{\cal{H}}_{1}&=&t'(\tau)\int{dX{\cal{N}}_{0}(X,\tau)\delta\phi(X,\tau)}\label{notationHamiltonians}\\
\delta{\cal{H}}_{2}&=&t'(\tau)\int{dX{\cal{N}}_{1}(X,\tau)\delta\phi(X,\tau)}\nn\\
&=&q_{1}\cdot\partial_{q}\delta H_{1}(q_{0},p_{0},\tau)+p_{1}\cdot\partial_{p}\delta H_{1}(q_{0},p_{0},\tau)\,.\nn
\een
With these notations, we rewrite the Lagrangian as
\be
{\cal{L}}_{\phi}(q,\dot q,\tau)&=&\big[p_0\cdot\dot q_0-{\cal{H}}_{\phi}(x_0)\big]+\big[\delta p\cdot \dot q_0-\dot p_0\cdot\delta q+{\cal H}_\phi(x_0)-{\cal H}_\phi(x)\big]\\
&&+\big[p_0\cdot\delta\dot q+\dot p_0\cdot\delta q\big]+\delta p\cdot\delta\dot q
\ee
so that by time integration and integrations by parts of the last line, we obtain
\ben
{\cal{S}}[\phi]&=&{\cal{S}}_0+\oint{d\tau\left(\delta p\cdot \dot q_0-\dot p_0\cdot\delta q+{\cal H}_\phi(x_0)-{\cal H}_\phi(x)\right)}\nn\\
&+&\frac{1}{2}\oint{d\tau \delta p\cdot\delta\dot q-\delta \dot p\cdot\delta q}\nn\\
&+&p_0(\tau_f)\cdot\delta q(\tau_f)+\frac{1}{2}\delta p(\tau_f)\cdot\delta q(\tau_f)-{\cal{S}}_b(\phi)\label{totalactiontobecanceled}
\een
The boundary term is
\ben
{\cal{S}}_{b}[\phi_0+\delta\phi,x_{0}]&=&p_0(\tau_f)\cdot\delta q(\tau_f)+\frac{1}{2}\delta p(\tau_f)\cdot\delta q(\tau_f) \label{boundarytermexpanded}
\een
so that, with $\phi=\phi_0+\delta\phi$, the last line of Eq.(\ref{totalactiontobecanceled}) vanishes
\ben
\delta{\cal{S}}&=&\oint{d\tau\left(\delta p\cdot \dot q_0-\dot p_0\cdot\delta q-{\cal H}_\phi(x_0)-{\cal H}_\phi(x)\right)}+\frac{1}{2}\oint{d\tau\, \delta p\cdot\delta\dot q-\delta \dot p\cdot\delta q}\,.\nn\\
\label{Sphi0plusdeltaphi}
\een
Using the Hamilton equations for the unperturbed trajectories, the Taylor expansion of the Hamiltonian around the unperturbed trajectory gives
\ben
{\cal{H}}_{\phi_{0}}(x)-{\cal{H}}_{\phi_{0}}(x_0)=-\dot{p}_{0}\cdot\delta q+\delta p\cdot\dot{q}_{0}+{\cal{H}}_{\phi_{0}}^{(2)}(\delta x)+\dots\,,\label{s1}
\een
and where the second order term is
\ben
{\cal{H}}_{\phi_{0}}^{(2)}(\delta x)&=&\frac{\delta q^{2}}{2}\cdot\partial_{qq}{\cal{H}}_{\phi_{0}}(x_{0})+\delta q\cdot \delta p\cdot\partial_{qp}{\cal{H}}_{\phi_{0}}(x_{0})+\frac{\delta p^{2}}{2}\cdot\partial_{pp}{\cal{H}}_{\phi_{0}}(x_{0})\,. \label{Hphi02}
\een
Using Eqs.(\ref{s1}) and (\ref{Hphi02}) in (\ref{Sphi0plusdeltaphi}), the total variation of the action writes
\be
\delta{\cal{S}}&=&\oint{d\tau \left[\frac{1}{2}\left(\delta p\cdot\delta\dot q-\delta \dot p\cdot\delta q\right)-{\cal{H}}_{\phi_{0}}^{(2)}(\delta x)-\delta{\cal{H}}(x)\right]}+\dots
\ee
At this stage, we have removed all the terms that cancel irrespective of $\delta\phi$.
We now proceed with the expansion in orders of $\delta\phi$.
To third order in $\delta\phi$, the integral in the last expression is
\be
\oint{d\tau \left[\frac{1}{2}\left(p_1\cdot\dot q_1-\dot p_1\cdot q_1\right)-{\cal{H}}_{\phi_{0}}^{(2)}(x_1)-\delta{\cal{H}}_{1}-\delta{\cal{H}}_{2}\right]}+O\left(\delta\phi^{3}\right)\,.
\ee
The first order variation $(q_1(\tau),p_1(\tau))$ of the trajectory satisfies
\ben
\left\{
\begin{array}{l}
\D\dot{q}_{1}=q_{1}\cdot\partial_{qp}H_{\phi_{0}}(x_{0})+p_{1}\cdot\partial_{pp}H_{\phi_{0}}(x_{0})+\partial_{p}\delta H_{1}(x_{0})\\\\
\D\dot{p}_{1}=-q_{1}\cdot\partial_{qq}H_{\phi_{0}}(x_{0})-p_{1}\cdot\partial_{pq}H_{\phi_{0}}(x_{0})-\partial_{q}\delta H_{1}(x_{0})
\end{array}
\right.\label{dq1}
\een
and therefore
\ben
\frac{1}{2}\oint{d\tau \left[p_{1}\cdot\dot{q}_{1}-q_{1}\cdot\dot{p}_{1}\right]}=\oint{d\tau {\cal{H}}_{\phi_{0}}^{(2)}(x_1)}+\frac{1}{2}\oint{\delta{\cal{H}}_{2}} \label{s3interm}
\een
Combining Eqs.(\ref{s3interm}) and (\ref{dq1}), to second-order in $\delta\phi$, the total variation of the action is
\ben
\delta{\cal{S}}&=&-\oint{d\tau t'(\tau)\int{dX{\cal{N}}_{0}(X,\tau)\delta\phi(X,\tau)}}\nn\\
&&-\frac{1}{2}\oint{d\tau t'(\tau)\int{dX{\cal{N}}_{1}(X,\tau)\delta\phi(X,\tau)}}+O(\delta\phi^{3})\,.\label{ae1}
\een
In order to express the second term in Eq.(\ref{ae1}), we first note that, using the results of linear response theory recalled section \ref{sectionII}, the density variation ${\cal{N}}_{1}$ can be written as
\be
{\cal{N}}_{1}(\tilde 1)=\oint{d\tilde 1' R(\tilde 1,\tilde 1')\delta\phi(\tilde 1')}\,,
\ee
in terms of the response function over the closed-time contour,
\be
R(\tilde 1,\tilde 1')=\theta(\tau-\tau^{\prime}) \{{\cal{N}}_{0}(\tilde 1),{\cal{N}}_{0}(\tilde 1')\}\,.
\ee
Therefore
\ben
\oint{d\tilde 1{\cal{N}}_{1}(\tilde 1)\delta\phi(\tilde 1)}&=&\oint{d\tilde 1 d\tilde 1'{\cal{R}}(\tilde 1,\tilde 1')\delta\phi(\tilde 1)\delta\phi(\tilde 1')}\nn\\
&=&\frac{1}{2}\oint{d\tilde 1\oint{d\tilde 1' \left[{\cal{R}}(\tilde 1,\tilde 1')+{\cal{R}}(\tilde 1',\tilde 1)\right]\delta\phi(\tilde 1')\delta\phi (\tilde 1)}}\nn\\
&=&\oint{d\tilde 1\oint{d\tilde 1' \chi(\tilde 1,\tilde 1')\delta\phi(\tilde 1)\delta\phi (\tilde 1')}}\,,\label{Rdphidphi}
\een
with
\be
\chi(\tilde 1,\tilde 1')=\frac{1}{2}\langle {\cal{T}}_{c}\{{\cal{N}}_{0}(\tilde 1),{\cal{N}}_{0}(\tilde 1')\}\,.
\ee
The power expansion of the total action (\ref{ae1}) becomes
\be
{\cal{S}}[\phi_{0}+\delta\phi]={\cal{S}}[\phi_{0}]-\oint{\!\!d\tilde 1{\cal{N}}_{0}(\tilde 1)\delta\phi(\tilde 1)}-\frac{1}{2}\oint{\!\!\oint{\!d\tilde 1d\tilde 1' \chi(\tilde 1,\tilde 1')\delta\phi(\tilde 1)\delta\phi(\tilde 1)}}+O(\delta\phi^{3})\,.
\ee
Therefore the first two functional derivatives of the action are
\be
\frac{\delta {\cal{S}}}{\delta\phi(\tilde 1)}\Bigg|_{\phi=\phi_{0}}=-{\cal{N}}_{0}(\tilde 1)\,,
\ee
and
\be
\frac{\delta^{2} {\cal{S}}}{\delta\phi(\tilde 1)\delta\phi(\tilde 1')}\Bigg|_{\phi=\phi_{0}}=-\frac{1}{2}{\cal{T}}_{c}\{{\cal{N}}_0(\tilde 1),{\cal{N}}_0(\tilde 1')\}\,,
\ee
which, combined with Eqs.(\ref{generalfirstderivative}) and (\ref{generalsecondderivative}), yield the desired results.

\section{On the differentiability of the action functional.} \label{differentiabilityproperites}

Although it is not at all the purpose of this paper to be concerned with mathematical details, we feel that a word of caution regarding the differentiability of $\tilde\Omega[\phi]$ and the derivatives $\Omega^{(n)}$ is in order.
Strictly speaking, the effective action functional $\tilde{\Omega}[\phi]$ is not (Fr{\'e}chet) differentiable everywhere.
It is however infinitely differentiable at any physical potentials $\phi_p$.
In that case, the term ${\cal{S}}_{b}$ in (\ref{classicalclosedtimeaction}) cancels the boundary terms resulting from the variations of $\tilde{x}(\tau_f)$ in the first term $\int_{\tau_{i}}^{\tau_{f}}{d\tau\,{\cal{L}}_{\phi}(\tilde{q}(\tau),\tilde{v}(\tau),\tau)}$, which allows for the definition of the functional derivatives.
For a non-physical potential, the cancellation does not necessarily occur and the definition of the functional derivatives is not always possible.
As a consequence, the quantities $\Omega^{(n)}$ can not be regarded as functional of $\phi$ and expressions like $\Omega^{(n+1)}\neq \delta\Omega^{(n)}/\delta\phi$ are not correct, and therefore using it could give wrong results.
This is unfortunate since such recurrence relation is very useful to derive by simple differentiation the hierarchy of properties satisfied by all the successive derivatives from a single starting relation.

To regain the benefits of such recurrence properties, we define the following formal device.
Given the ensemble average $\langle A(X,t)\rangle$ of a dynamical variable $A$, we define the generalized ensemble average as
\ben
\langle\langle A(\tilde 1) \rangle\rangle_{\phi}=\frac{1}{{\cal{Z}}[\phi]}\lel e^{\frac{i}{s}{\cal{S}}[\phi]}\,A(\tilde 1)\rir\,, \label{generalizedaverage}
\een
for any potential $\phi$.
At $\phi=\phi_0$, the generalized average reduces to the physical ensemble average (\ref{classicalaverage}),
\be
\langle\langle A(X,\tau) \rangle\rangle_{\phi=\phi_0}=\langle A(X,t(\tau)) \rangle\/.
\ee
The generalized average is a functional of $\phi$, which, like ${\cal{S}}$, is differentiable at any physical potential $\phi_p$.
Using results of the previous sections, we find its functional derivative at $\phi_0$,
\ben
\frac{\delta}{\delta\phi(1')}\langle\langle A(1) \rangle\rangle\Bigg|_{\phi_0}=-\frac{i}{s}\llangle \delta A(1) \delta{\cal{N}}(1')\rrangle+\llangle \frac{\delta A(1)}{\delta\phi(1')}\Bigg|_{\phi_0}\rrangle\,, \label{deltaAdeltaphigeneral}
\een
in terms of the fluctuations $\delta A=A-\langle A\rangle$.
Since, using the results of linear response theory, 
\ben
\frac{\delta A(1)}{\delta\phi(1')}=\frac{1}{2}{\cal{T}}_{c}\left\{A(1),{\cal{N}}(1')\right\} \label{deltaAdeltaphi}
\een
we find 
\ben
\frac{\delta\langle\langle A(1) \rangle\rangle_\phi}{\delta\phi(1')}\Bigg|_{\phi_0}=\frac{1}{2}\llangle {\cal{T}}_{c}\left\{A(1),{\cal{N}}(1')\right\}\rrangle-\frac{i}{s}\llangle \delta A(1) \delta{\cal{N}}(1')\rrangle\,.\nn\\
\label{deltallangleArrangledeltaphi1}
\een
For instance, with $A={\cal{N}}$, we shall write
\be
f_\phi(\tilde 1)\equiv \langle\langle\,{\cal{N}}(\tilde 1)\,\rangle\rangle_{\phi}\,.
\ee
The phase-space distribution is
\be
f(X,t(\tau))=f_{\phi=\phi_0}(X,\tau)
\ee
and from Eqs.(\ref{deltallangleArrangledeltaphi1}) and (\ref{d2omegadphi2}).
\be
\frac{f_\phi(\tilde 1)}{\delta\phi(\tilde 1')}\Bigg|_{\phi_0}&=&\frac{1}{2}\llangle {\cal{T}}_{c}\left\{{\cal{N}}(\tilde 1),{\cal{N}}(\tilde 1')\right\}\rrangle-\frac{i}{s}\llangle \delta {\cal{N}}(\tilde 1) \delta{\cal{N}}(\tilde 1')\rrangle\\
&=&\Omega^{(2)}(\tilde 1,\tilde 1')\,.
\ee
Therefore we can obtain the successive derivatives of $\tilde\Omega[\phi]$ by writing the $n$-th derivative as
\ben
\Omega^{(n)}=\langle \omega^{(n)}\rangle \label{step1}
\een
in terms of the dynamical variable $\omega^{(n)}$, and obtain the $(n+1)$-th derivative by functional derivative of its generalized average at $\phi=\phi_0$,
\ben
\Omega^{(n+1)}(\tilde 1,\dots,\widetilde{n+1})=\frac{\delta\langle\langle \omega^{(n)}(\tilde 1,\dots,\tilde n) \rangle\rangle_\phi}{\delta\phi(\tilde 1')}\Bigg|_{\phi_0}\,. \label{step2}
\een
This scheme can be used to systematically generate explicit expressions of the derivatives of the effective action \cite{Daligaultunpublished} as combinations of products and Poisson brackets of the fundamental field.
The steps (\ref{step1}) and (\ref{step2}) can be schematically written as
\be
\Omega^{(n+1)}(\tilde 1,\dots,\widetilde{n+1})=\frac{\delta}{\delta\phi(\widetilde{n+1})}\Omega^{n}(\tilde 1,\dots,\tilde {n})\,,
\ee
with the understanding of the caveats discussed before.
In this work, successive functional derivatives are obtained according to the scheme just defined.

\section{On the occurence of the linear combination of correlation and response functions} \label{Discussionoccurence}

As shown in appendix \ref{differentiabilityproperites}, for any dynamical variable $A(X,t)$ we have
\ben
\frac{\delta\langle A(1) \rangle}{\delta\phi(1')}\Bigg|_{\phi_0}=\frac{1}{2}\llangle {\cal{T}}_{c}\left\{A(1),{\cal{N}}(1')\right\}\rrangle-\frac{i}{s}\llangle \delta A(1) \delta{\cal{N}}(1')\rrangle\,.\label{deltallangleArrangledeltaphi}
\een
From the expression emerges a fundamental operation between dynamical variables, which we find interesting to highlight.
For any dynamical variables $A$ and $B$ defined on the closed-time contour, we define with square brackets $[\cdot,\cdot]$ the symmetric operation,
\ben
\left[A(\tau),B(\tau')\right]&:=&\frac{1}{2}\llangle{\cal{T}}_{c}\left\{A(\tau),B(\tau')\right\}\rrangle-\frac{i}{s}\llangle \delta A(\tau)\,\delta B(\tau')\rrangle\nn\\
&=&\left[B(\tau'),A(\tau)\right]\/.\label{newbracket}
\een
In term of this operation, equation (\ref{deltallangleArrangledeltaphi}) reads
\begin{eqnarray}
\frac{\delta}{\delta\phi(1')}\langle\langle A(1) \rangle\rangle\Bigg|_{\phi_0}=\left[A(1),{\cal{N}}(1')\right]\,. \label{differentialandnewbracket}
\end{eqnarray}
and
\be
\chi(1,1')=\left[{\cal{N}}(1),{\cal{N}}(1')\right]\,.
\ee
The operation (\ref{newbracket}) combines two fundamental quantities of classical statistical mechanics, namely the Poisson bracket $\left[\cdot,\cdot\right]_{PB}$ and the product of dynamical variables, which are intimately related to the response and correlation functions.
It obviously respects the dimensions thanks to the action parameter $s$ (the Poisson bracket $[A,B]_{PB}$ has the dimension of the product $AB$ divided by the product of a length times a momentum, i.e. by an action.)
The Poisson bracket carries the dynamical information on how fluctuations are propagated in time, while the correlation function gives information about the likelihood of fluctuations (occupation number.)
At equilibrium, both quantities are simply related by the fluctuation-dissipation theorem and only one is useful (one usually works with the phase-space density correlation functions.)
The emergence of the linear combination provides a justification of the fact recognized by MSR and others \cite{MartinSiggiaRose} that a time-dependent theory of classical dynamics should be expanded to involve not only correlation functions but also response functions in order to be amenable to a systematic renormalization.

We can shed some light on the occurrence of that particular combination by regarding the action function as the classical limit discussed above (see discussion after Eq.(\ref{actionfunctional}).)
In a quantum system, both fluctuation statistics (correlation) and response effects can be characterized by products of the same quantum field operator; the intensity of fluctuations involve the anti-commutator $[\cdot,\cdot]_+$ and the response is related to the commutator $[\cdot,\cdot]_-$.
In practice, both properties are contained in the time-ordered Green's functions, say $G(\tau,\tau')=\langle {\cal{T}}\hat{\Psi}(\tau)\hat{\Psi}(\tau')\rangle$, where $\hat{\Psi}(\tau)$ is the field operator (for simplicity, we do not write the other degrees of freedom).
We can re-express the Green's function in terms of the averaged commutator and anti-commutator of the field,
\be
G(\tau,\tau')&=&\langle \hat\Psi(\tau)\hat\Psi(\tau')\rangle\theta(\tau-\tau')+\langle \hat\Psi(\tau')\hat\Psi(\tau)\rangle\theta(\tau'-\tau)\\
&=&\frac{1}{2}\left\langle \left[\hat\Psi(\tau),\hat\Psi(\tau')\right]_-\right\rangle\,\text{sign}(\tau-\tau')+\frac{1}{2}\left\langle \left[\hat\Psi(\tau),\hat\Psi(\tau')\right]_{+}\right\rangle\,.
\ee
From the correspondence relations, in the classical limit, the anti-commutator $[\cdot,\cdot]_{+}$ becomes a product, the commutator becomes the Poisson bracket $i\hbar\{\Psi(\tau),\Psi(\tau')\}$ and therefore
\be
G(\tau,\tau')&\rightarrow& \frac{is}{2}\left\langle{\cal{T}}\left[\Psi(\tau),\Psi(\tau')\right]_{PB}\right\rangle+\left\langle \Psi(\tau)\Psi(\tau')\right\rangle\\
&=&is\left(\frac{1}{2}\left\langle{\cal{T}}\left[\Psi(\tau),\Psi(\tau')\right]_{PB}\right\rangle-\frac{i}{s}\left\langle \Psi(\tau)\Psi(\tau')\right\rangle\right)\,,
\ee
where we recognize the bracket (\ref{newbracket}) introduced earlier.
\begin{figure}[t]
\begin{center}
\includegraphics[scale=.8]{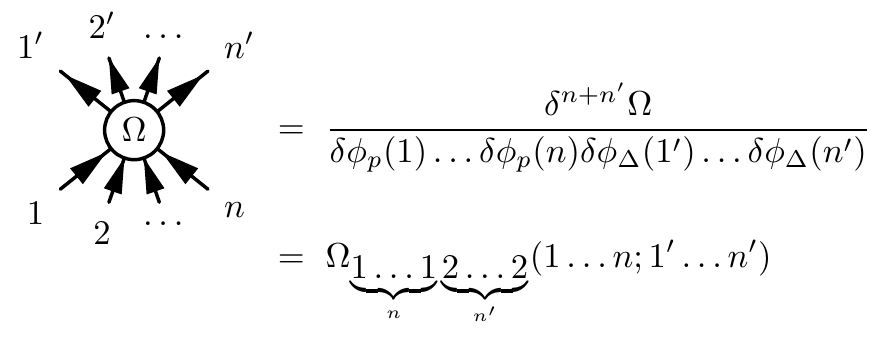}\\
\includegraphics[scale=.8]{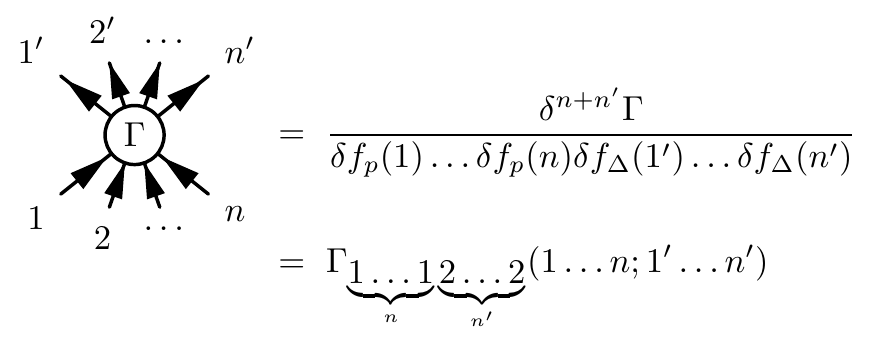}
\end{center}
\caption{Representation of derivatives of the effective effective action $\Omega$ and effective potential $\Gamma$.
An inward (outward) arrow indicates derivative with respect to the physical (non-physical) component.
} \label{figure3}
\end{figure}

\section{Closed vs unclosed expressions for $\Sigma(\tilde 1,\tilde 1')$} \label{closedunclosedsigma}

The previous closed expression for $\Big\{\Sigma(\tilde 1,\tilde 1'),f(\tilde 1')\Big\}$, Eq.(\ref{Sigmaf}), should be compared to its ``unclosed'' version, which can straightforwardly obtained by using the equations of evolution (\ref{dchi11pdtau}) and (\ref{dchi111pdtauGmf}) for $\chi$ and $\Omega^{(3)}$ on both sides of the equation (\ref{Sigmadefinitiondetailed}) for $\Sigma$.
We obtain,
\bs \label{Sigmafunclosed}
\ben
\lefteqn{\Big\{\Sigma(\tilde 1,\tilde 1'),f(\tilde 1')\Big\}(\tilde 1')}&&\nn\\
&=&\frac{is}{2}\gamma_{3}(\tilde 123)s_3(23\tilde 1')\label{sigmadeltafunclosed}\\
&+&\left(\frac{is}{2}\right)^2\gamma_{3}(\tilde 123)\gamma_{3}(\tilde 1'2'3')\big[\chi^{(4)}(232'3')-\frac{2}{is}\chi(22')\chi(33')-\chi^{(3)}(234)\chi^{-1}(45)\chi^{(3)}(52'3')\big]\nn\\\label{sigmaregfunclosed}\\
&+&\Delta(\tau')\frac{1}{2}\gamma_{3}(\tilde 123)\chi(2,\bar{2})\chi(3,\bar{3})\frac{\delta}{\delta f(\bar 3)}\left[C(\tilde 1',\bar 2')\chi^{-1}(\bar 2',\bar 2)\right]\label{sigma0funclosed}
\een
\es
thus establishing the correspondance between Eqs.(\ref{sigmaregf}) and (\ref{sigmaregfunclosed})

In his ``fully renormalized'' kinetic theory of equilibrium liquids, Mazenko \cite{Mazenko1974} uses as a starting point an expression for the memory function equivalent to the equilibrium limit of Eq.(\ref{Sigmafunclosed}).

\section{Derivation of Eq.(\ref{GammaCstructure}).} \label{appendixonGammaC}

\begin{figure}[t]
\includegraphics[scale=.75]{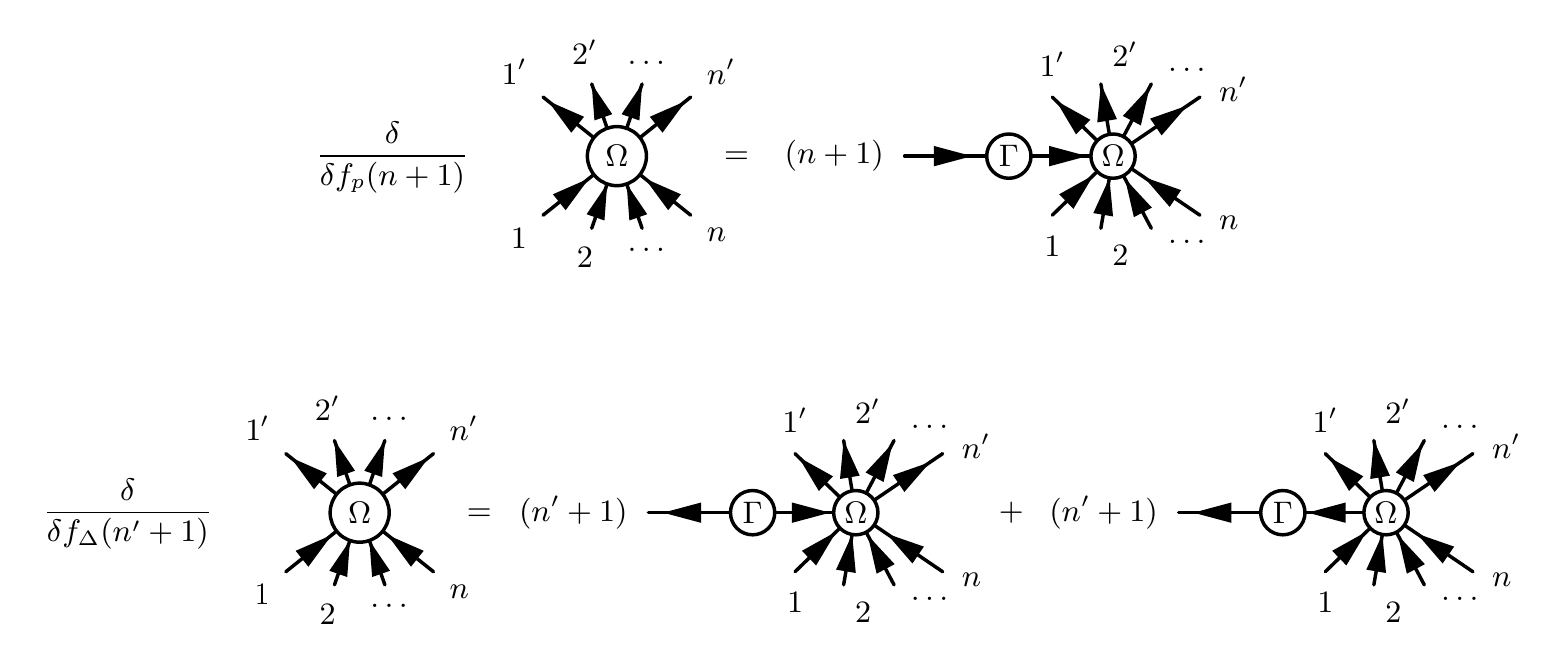}
\caption{Graphical representation of chain rule.} \label{figure4}
\end{figure}

Equation (\ref{gammaCf}), namely
\be
\big\{\Gamma^{C}(1,1'),f(1)\big\}(1')&=&\frac{is}{2}\big(\Sigma^{>}(t,t')+\Sigma^{<}(t,t')\big)+is\Sigma^{(0)}(X_1,X_1;t)\delta(t'-t_0)\\
&&+\delta(t-t_0)C(1,2)\Gamma^{A}(2,1')
\ee
implies that $\Gamma^{C}$ is of the form
\ben
\hspace{-0.5cm}\Gamma^{C}(1,1')=\Gamma^{reg}(1,1')+\delta(t'-t_0)\Gamma^{\delta}(X_1,X_{1}^\prime;t)+\delta(t-t_0)\Gamma^{(1)}(X_1,X_1^\prime;t') \label{GammaCappendix}
\een
where $\Gamma^{reg}$ and $\Gamma^{\delta}$ are given by Eqs.(\ref{Gammareggammadelta}) and $\Gamma^{(1)}$ is a regular function of time variables statisfying,
\ben
\big\{\Gamma^{(1)}(X_1,X_1^\prime;t'),f(X_1,t_0)\big\}&=&C(X_1,t_0;2)\Gamma^{A}(2,1') \label{gamma1f}\/.
\een
$\Gamma^{(1)}$ can be related to $\Gamma^{\delta}$ and the initial correlation function $C_0$ as follows.
Introducing Eq.(\ref{GammaCappendix}) into (\ref{gammaCCinverse}) yields
\ben
\hspace{-.5cm}C(1,1')&=&\int_{t_0}^{t}{\!\!dt_2\!\!\int_{t_0}^{t'}{dt_3\iint{dX_2dX_3\chi^{R}(1,2)\Gamma^{reg}(23)\chi^{A}(3,1')}}}\nn\\
&+&\int_{t_0}^{t}{\!\!dt_2\!\!\iint{ dX_2dX_3 \chi^{R}(1,2)\Gamma^{\delta}(X_2,X_3;t_2)\chi^{A}(X_3,t_0;1')}}\nn\\
&+&\int{dX_2\int{d3 \chi^{R}(1;X_2,t_0)\Gamma^{(1)}(X_2,X_3;t_3)\chi^{A}(3;1')}}\nn
\een
Setting $t'=t_0$ in the previous expression, the resulting expression is used to calculate the rhs of Eq.(\ref{gamma1f}), which gives
\be
\lefteqn{\big\{\Gamma^{(1)}(X_1,X_1^\prime;t'),f(X_1,t_0)\big\}}&&\\
&=&\big\{\Gamma^{\delta}(X_1^\prime,X_1;t'),f(1,t_0)\big\}+\delta(t'-t_0)\int{d2 \chi^{R}(X_1,t_0;2)\Gamma^{(1)}(X_1^\prime,X_2,t_2)}\,.
\ee
Using $\chi^{R}(X_1,t_0;X_2,t_0)=\big\{f(X_1),\delta(X_1-X_2)\big\}$, the previous relation implies the following expression for $\Gamma^{(1)}$
\ben
\Gamma^{(1)}(X_1,X_1^\prime;t')=\Gamma^{\delta}(X_1^{\prime},X_1;t')+\delta(t'-t_0)\Gamma^{(0)}(X_1,X_1^\prime)\,, \label{Gamma1appendix}
\een
with $\Gamma^{(0)}$ given by Eq.(\ref{Gamma0}).
Combining Eqs.(\ref{Gamma1appendix}) and (\ref{GammaCappendix}) yield the desired result Eq.(\ref{GammaCstructure})\,.

\section{Diagrammatic representation} \label{appendix1}

\begin{figure}[t]
\begin{center}
\includegraphics[scale=.9]{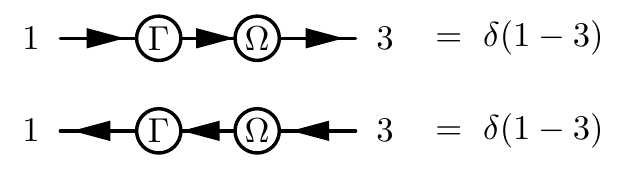}\\
\includegraphics[scale=.9]{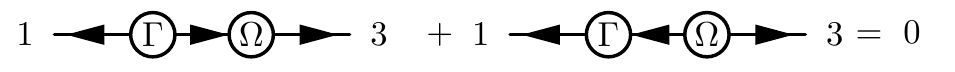}
\end{center}
\caption{Graphical representations of Eqs.(\ref{Gab_1}), (\ref{Gab_2}) and (\ref{Gab_3}).} \label{figure5}
\end{figure}

\subsection{Derivation of Eq.(\ref{gammaij})} \label{appendix1A}

From the chain rule of differentiation, we have
\ben
\frac{\delta f_{p}(3)}{\delta f_{p}(1)}&=&\delta(1-3)=\int{d2\,\Gamma_{12}(12)\Omega_{12}(23)+\Gamma_{11}(12)\Omega_{22}(23)}\nn\\
\frac{\delta f_{\Delta}(3)}{\delta f_{p}(1)}&=&0=\int{d2\,\Gamma_{22}(12)\Omega_{12}(23)+\Gamma_{21}(12)\Omega_{22}(23)}\nn
\een
and
\ben
\frac{\delta f_{\Delta}(3)}{\delta f_{p}(1)}&=&0=\int{d2\,\Gamma_{11}(12)\Omega_{21}(23)}\\
\frac{\delta f_{\Delta}(3)}{\delta f_{\Delta}(1)}&=&\delta(1-3)=\int{d2\,\Gamma_{21}(12)\Omega_{12}(21)}
\een
where we used $\Omega_{11}=0$.
We deduce $\Gamma_{11}=0$, and therefore
\bs \label{Gab}
\ben
\delta(1-3)&=&\int{d2\,\Gamma_{12}(12)\Omega_{12}(23)} \label{Gab_1}\\
\delta(1-3)&=&\int{d2\,\Gamma_{21}(12)\Omega_{12}(21)} \label{Gab_2}\\
0&=&\int{d2\,\Gamma_{22}(12)\Omega_{12}(23)+\Gamma_{21}(12)\Omega_{22}(23)} \label{Gab_3}
\een
\es
Combining equations (\ref{Gab}) with (\ref{omegaij}) leads to Eqs.(\ref{gammaij}).

\subsection{Higher-order derivatives with diagrams}

We define here an economical graphical representation for deriving relations between the correlation and response functions with the vertex functions.
We define the graphical notations as shown in Fig.(\ref{figure3}).

\begin{figure}[t]
\begin{center}
\includegraphics[scale=.8]{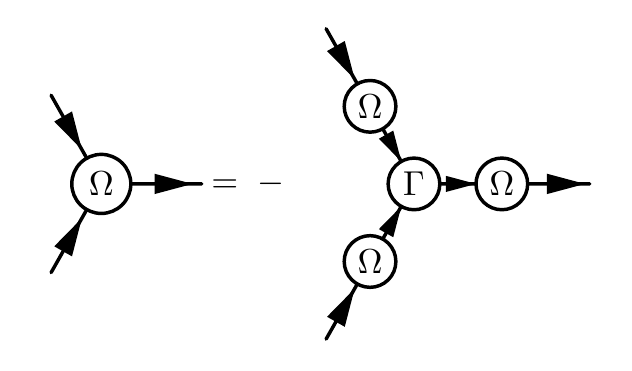}
\end{center}
\caption{$\delta/\delta f_p$ of the first diagram in Fig.(\ref{figure5})}\label{figure6}
\begin{center}
\includegraphics[scale=.8]{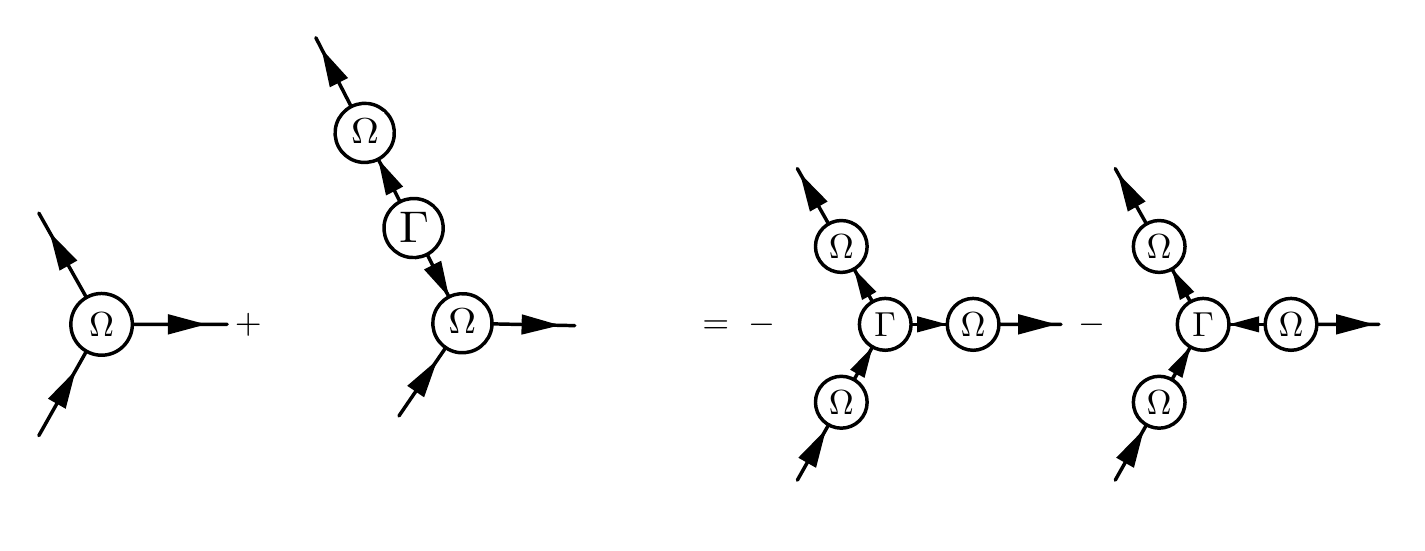}
\end{center}
\caption{$\delta/\delta f_p$ of last diagram in Fig.(\ref{figure5})} \label{figure7}
\end{figure}

\begin{figure}[t]
\begin{center}
\includegraphics[scale=.8]{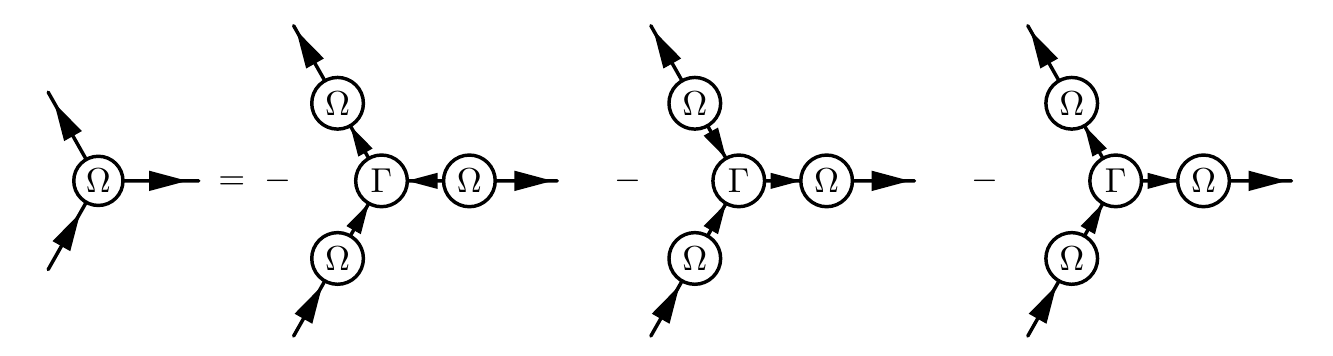}
\end{center}
\caption{Figs.(\ref{figure6}) and (\ref{figure7}) can be combined to lead to this diagram} \label{figure8}
\end{figure}

Using the chain rule, the functional derivative $\delta/\delta f_{p}$ and $\delta/\delta f_{\Delta}$ of $\Omega_{1\dots n 1'\dots n'}$ add an additional two-point vertex to the diagram as illustrated in Figs.\ref{figure4} and \ref{figure5}.
Using $\Gamma_{11}=0$ and the rules in Fig.(\ref{figure4}), we can evaluate the successive derivatives by letting $\delta/\delta f_{p,\Delta}$ act of the diagrams of Fig.(\ref{figure5}).
We illustrate the calculations leading to the third-order derivatives Eq.(\ref{omega221}) used in the body of the paper.
$\delta/\delta f_p$ of the first diagram in Fig.(\ref{figure5}) yields the diagram in Fig.(\ref{figure6}), and $\delta/\delta f_p$ of last diagram in Fig.(\ref{figure5}) yields the diagram in Fig.(\ref{figure7})
Combining the diagrams in Figs. (\ref{figure6}) and (\ref{figure7}) yields the diagrammatic representation of Eq.(\ref{omega221}) shown in Fig.(\ref{figure8}).

\section{Details on the derivation of the real-time formalism} \label{keystepsrecastformalism}

By considering the equation of motion (\ref{dcalNphidt1}) for ${\cal{N}}(\tau)$ in an external potential $\phi$ on both sides on the closed-time consider, it is straigthforward to extract the evolution equation for the classical and non-classical components ${\cal{N}}_{p,\Delta}$ as
\be
\left[\frac{\partial}{\partial t_1}-L_1^p\right]N_p(1)-\frac{1}{4}L_1^{\Delta}N_\Delta(1)&=&\frac{1}{2}\gamma_{3}(1,2,3)\Big(N_p(2)N_p(3)+\frac{1}{4}N_\Delta(2)N_\Delta(3)\Big)\\
\left[\frac{\partial}{\partial t_1}-L_1^{p}\right]N_\Delta(1)-L_1^{\Delta}N_p(1)&=&\gamma_{3}(1,2,3)N_p(2)N_\Delta(3)
\ee
with the single particle Liouville operators,
\be
L_1^p\bullet&=&\big\{h_0(1)+\phi_p(1),\bullet\big\}\\
L_1^\Delta\bullet&=&\big\{\phi_\Delta(1),\bullet\big\}
\ee
By averaging these equations over the initial conditions, we obtain the evolution equation of for the averaged densities $f_{p,\Delta}=\llangle {\cal{N}}_{p,\Delta}\rrangle$,
\ben
\lefteqn{\left[\frac{\partial}{\partial t_1}-L_1^p\right]f_p(1)-\frac{1}{4}L_1^{\Delta}f_\Delta(1)}&&\label{equationforfp}\\
&=&\frac{1}{2}\gamma_{3}(1,2,3)\Big(f_p(2)f_p(3)+\frac{1}{4}f_\Delta(2)f_\Delta(3)\Big)+\frac{is}{2}\gamma_3(123)\Omega_{22}(23)+\delta(t-t_0)f_p(1)\nn
\een
and
\ben
\left[\frac{\partial}{\partial t_1}-L_1^{p}\right]f_\Delta(1)-L_1^{\Delta}f_p(1)&=&\gamma_{3}(1,2,3)f_p(2)f_\Delta(3)is\gamma_3(123)\Omega_{21}(23)\,. \label{equationforfdelta}
\een
As in Sec. \ref{sectionIIIF}, equations of evolutions for the second-order derivatives $\Omega_{ij}$, i.e. for $C$ and $\chi^{R,A}$, can be obtained by differentiation of Eqs.(\ref{equationforfp}-\ref{equationforfdelta}).
The terms involving the third-order derivatives $\Omega_{ijk}$ in the resulting equations (not shown here) can be expressed in terms of memory function kernels $\Sigma^{R,A,C}$ such as
\be
\frac{is}{2}\gamma_{3}(123)\left(
\begin{array}{cc}
0&2\/\Omega_{212}(231')\\
\Omega_{221}(231')&2\/\Omega_{222}(231')
\end{array}
\right)&=&\left(
\begin{array}{ccc}
0 & & \Sigma^{A}\cdot\Omega_{12}\\
\Sigma^{R}\cdot\Omega_{21} & & \frac{2}{is}\left(is\Sigma^R\cdot \Omega_{22}+\Sigma^C\cdot \Omega_{12}\right)
\end{array}
\right)(1,1')\\
&=&\left(
\begin{array}{ccc}
0 & & \Sigma^{A}\cdot\chi^A\\
\Sigma^{R}\cdot\chi^R & & \frac{2}{is}\left(\Sigma^R\cdot C+\Sigma^C\cdot \chi^A\right)
\end{array}
\right)(1,1')\,,
\ee
which leads to equations (\ref{sigmaRAC}) and (\ref{sigmaRACdetail}).
Following the same steps as in Sec. \ref{sectionIIIG}, using Eqs.(\ref{chiRAgammaRA}-\ref{gammaCC}), the evolution equations for $\chi^{R,A}$ and $C$ yield the following (non-trivial) relations between $\Sigma^{R,A,C}$ and $\Gamma^{R,A,C}$,
\be
\big\{\Gamma_{211}(123),f(1)\big\}&=&\big\{\Gamma_{12}(32),\delta(1-2)\big\}-\big\{\Gamma_{21}(12),\delta(1-3)\big\}-\gamma_{123}-\frac{\delta\Sigma^{R}(12)}{\delta f_p(3)}\\
\big\{\Gamma_{121}(123),f(1)\big\}&=&\big\{\Gamma_{12}(32),\delta(1-2)\big\}-\big\{\Gamma_{12}(12),\delta(1-3)\big\}-\gamma_{123}-\frac{\delta\Sigma^{A}(12)}{\delta f_p(3)}\\
\big\{\Gamma_{221}(123),f(1)\big\}&=&-\big\{\Gamma_{22}(12),\delta(1-3)\big\}-\frac{1}{is}\frac{\delta\Sigma^{C}(12)}{\delta f_p(3)}\\
&&\hspace{3cm}-\delta(t_1-t_0)\frac{\delta}{\delta f_p(3)}\big[\Omega_{22}(14)\Gamma_{12}(42)\big]\\
\big\{\Gamma_{222}(123),f(1)\big\}&=&\frac{1}{4}\big\{\Gamma_{21}(32),\delta(1-2)\big\}-\frac{1}{4}\big\{\Gamma_{12}(12),\delta(1-3)\big\}-\frac{1}{4}\gamma_{123}\\
&&-\frac{1}{is}\frac{\delta\Sigma^{C}(12)}{\delta f_\Delta(3)}-\delta(t_1-t_0)\frac{\delta}{\delta f_\Delta(3)}\big[\Omega_{22}(14)\Gamma_{12}(42)\big]\\
\big\{\Gamma_{122}(123),f(1)\big\}&=&\big\{\Gamma_{22}(32),\delta(1-2)\big\}-\big\{\Gamma_{22}(12),\delta(1-3)\big\}-\gamma_{123}-\frac{\delta\Sigma^{A}(12)}{\delta f_\Delta(3)}\\
\big\{\Gamma_{212}(123),f(1)\big\}&=&\big\{\Gamma_{22}(32),\delta(1-2)\big\}-\gamma_{123}-\frac{\delta\Sigma^{R}(12)}{\delta f_\Delta(3)}
\ee
Those relations are then combined with Eq.(\ref{sigmaRACdetail}) to close the hierarchy through the functional differential equations (\ref{SigmaRAdeltaSigma} and \ref{calSdetailed}) for the memory function kernels.


\begin{thebibliography}{}
\bibitem{NoteFluctuationDissipation} With $t_{0}\to -\infty$, all dynamical quantities are time-translational invariant, e.g. $\chi^R(1,1')=\chi^R(X,t-t';X',0)\equiv\chi^R(X,X';t-t')$ and the fluctuation-dissipation theorem reads
\be
\chi^{R}(X,X^{\prime};t)&=&-\beta\theta(t)\frac{d}{dt}C(X,X';t)\,.
\ee
\bibitem{PossibleParametrization} A possible parametrization is given by
\be
t(\tau)=\left\{
\begin{array}{l}
\alpha(\tau-\tau_{i})+t_{0}\quad,\quad \tau_{i}\leq\tau\leq\tau_{M}=\frac{\tau_{i}+\tau_{f}}{2}\\\\
-\alpha(\tau-\tau_{M})+t_{M}\quad,\quad \tau_{M}\leq\tau\leq\tau_{f}
\end{array}
\right.
\ee
with $\alpha=2(t_{M}-t_{0})/(\tau_{f}-\tau_{i})$.
\bibitem{examplederivatives} The situation is analogous to a function, say $f(x,y)=e^x \sin(y)$, which is zero along the x-axis but takes non-zero values when leaving it, i.e. when $y\neq 0$, and therefore has differentials that are generally non-zero along the x-axis.
\bibitem{anotherdistinction} Another distinction between Eqs.(\ref{quantumfieldtheoryactionfunctional}) and (\ref{actionfunctional}) is in the boundary conditions.
In the quantum expression (\ref{quantumfieldtheoryactionfunctional}), the field integral is over all trajectories starting at $\phi(\tau_i)=\phi_1$ at $\tau_i$ and ending $\phi_2$ at $\phi(\tau_f)=\tau_f$, independently of the source $\phi$.
In the classical expression (\ref{actionfunctional}), while initial the value $x(\tau_i)=x(\tau_i)$ is the same for all external potential $\phi$, the final value $\tilde{x}(\tau_f)$ depends on the overall trajectory $\tilde{x}(\tau)$ and therefore on the potential $\phi$.
This saddle difference actually can not be discarded when considering the functional derivatives and the additional term ${\cal{S}}_{b}$ in Eq.(\ref{classicalclosedtimeaction}) was added to absorb such spurious boundary effects (see Sec.\ref{sectionproof}.)
\bibitem{notesymmetryofchi} We remark that, contrary to the functional defined on the real-time axis imagined at the beginning of section \ref{sectionIII}, $\chi$ is fully symmetric in its arguments,
\be
\chi(\tilde{1},\tilde{2})=\chi(\tilde{2},\tilde{1})\quad\forall\,\tilde{1},\tilde{2}\,.
\ee
\bibitem{Heavisidefunction} The Heaviside step function is defined as
\be
\Theta(\tau)=\left\{
\begin{array}{l}
\D 1\quad \tau\geq 0\\
\D 0\quad \tau<0
\end{array}
\right.\,,
\ee
\bibitem{Notesuccessivedifferentiation} Alternatively, of course, the same hierarchy could be obtained from a ``direct calculation'' using Eq.(\ref{dcalNphidt1}) and the explicit expressions of the $\Omega^{(n)}$'s in terms of the microscopic field ${\cal{N}}$. In this direct approach, the secular term $s_2$ in Eq.(\ref{dchi11pdtau}) arises from the time derivative of the Heaviside functions in Eq.(\ref{chichi>chi<}), 
\be
\delta(\tau-\tau')\chi^{>}(1,1')-\delta(\tau'-\tau)\chi^{<}(1,1')&=&\delta(\tau-\tau')\llangle\{{\cal{N}}(1),{\cal{N}}(1')\}\rrangle=s_{2}(1,1')\,.
\ee
The functional derivative of the boundary term in Eq.(\ref{dfphidt1}) yields $\Delta(\tau)\Pi(\tau')C(\tilde 1,\tilde 1')$ as can be seen by applying Eq.(\ref{deltaAdeltaphigeneral}) with $A={\cal{N}}$ and using $\delta {\cal{N}}(\tau_i)/\delta\phi(\tau')=0$ since the initial value is fixed.
\bibitem{Noteonrenormalizationapproach}
Using $\Omega^{(3)}=\delta\chi/\delta\phi$ in Eq.(\ref{dchi11pdtau}), the latter can then be regarded as a functional differential equation for $\chi(1,1')$.
In principle, this equation could be used to generate approximate but closed relations for $\chi$.
Thus, following ordinary perturbation theory, we could for instance use Eq.(\ref{dchi11pdtau}) to expand $\chi$ in powers of the bare interaction vertex $\gamma_{3}$ and of its ``non-interacting'' limit solution $\chi_0$ (solution of Eq.(\ref{dchi11pdtau}) with $\gamma_3=0$).
Lowest order approximation obtained in this way would most certainly suffer form the common divergence and secular problems that often plague ordinary perturbation theory.
These difficulties could in principle be cured by summing infinite classes of terms in the expansion, which would lead to excessively complicated calculations.
In our apporach, following the lessons of field theory, those issues are dealt with more efficiently by working directly with the so-called self-energy and vertex corrections; in the present work, the role of self-energy is played by the memory function.
\bibitem{Noteofadjointness} The equations for $\chi^{R}$ and $\chi^{A}$ are adjoint ($C$ is symmetric so the adjoint equation is equivalent to Eq.(\ref{setCchiRchiA1}).)
\bibitem{equilibriumtheory} Very good introduction to the equilibrium kinetic theory of fluids with references to seminal papers can be found in \cite{BalucaniZoppi,HansenMcDonald2nd}
\bibitem{matrixforgamma} Using the identity $Q\sigma_{z}Q^\dagger=\sigma_{x}$, we find the matrix representation of Eq.(\ref{chigammainverse}),
\ben
\hat\Gamma\hat\sigma_{x}\hat{\chi}\delta(1-1')&=&\hat\sigma_{x}\delta(1-1')\,.
\een
where $\hat\Gamma=Q\Bar{\Bar{\Gamma}} Q^\dagger$ and $\Bar{\Bar{\Gamma}}$ is defined as in Eq.(\ref{barbarchi}).
\bibitem{Noteonmatrixrepresentation} We thus find that the matrix representations $\hat{\chi}$ and $\hat{\Gamma}$ defined in Sec.\ref{SectionIV1} are simply given by
\be
\hat{\chi}=\left(
\begin{array}{cc}
\D 0&\D\Omega_{12}\\
\D \Omega_{21}&\D 2\,\Omega_{22}
\end{array}
\right)
\quad,\quad
\tilde{\Gamma}&=&\left(
\begin{array}{cc}
0&\Gamma_{12}\\
\Gamma_{21}&2\/\Gamma_{22}
\end{array}
\right)\,.
\ee
\end{thebibliography}
\end{document}